\documentclass[11pt]{article}

\usepackage{graphicx}
\usepackage[english]{babel}
\usepackage{amsmath, amssymb, amsfonts,amsthm}
\usepackage{stmaryrd}
\usepackage{dsfont} 
\usepackage[T1]{fontenc}

\usepackage{hyperref}
\usepackage[authoryear]{natbib}
\usepackage{hypernat}

\theoremstyle{plain}
\newtheorem{lem}{Lemma}
\newtheorem{thm}{Theorem}
\newtheorem{prop}{Proposition}

\theoremstyle{definition}

\newtheorem{algo}{Procedure}
\theoremstyle{remark}
\newtheorem{rk}{Remark}
\newcommand{\1}{\mathds{1}}
\newcommand{\paren}[1]{\left( #1 \right)}
\newcommand{\croch}[1]{\left[ #1 \right]}
\newcommand{\acc}[1]{\left\{ #1 \right\}}

\newcommand{\norm}[1]{\left\| #1 \right\|}

\newcommand{\tq}{\mathrm{~s.t.~}} 
\newcommand{\guill}[1]{``#1''}

\newcommand{\defegal}{:=} 

\newcommand{\argmin}{\arg\min}

\newcommand{\R}{\mathbb{R}}

\newcommand{\E}{\mathbb{E}}

\newcommand{\Var}{\mathrm{Var}}

\DeclareMathOperator{\pen}{pen}
\DeclareMathOperator{\crit}{crit}
\DeclareMathOperator{\card}{Card} 
\newcommand{\sh}{\,\widehat{s}\,}
\newcommand{\mh}{\widehat{m}}
\newcommand{\sm}{s_m}
\newcommand{\shm}{\widehat{s}_m}
\newcommand{\Sm}{S_{m}}
\newcommand{\M}{\mathcal{M}}
\newcommand{\Mn}{\mathcal{M}_n}

\newcommand{\Ss}{\mathcal{S}^*}

\newcommand{\Pn}{P_n}

\newcommand{\Png}{P_n\gamma}

\newcommand{\perte}[1]{\norm{ s - #1}_n^2 }

\newcommand{\Rh}{\widehat{R}}

\newcommand{\Il}{I_{\lambda}}

\newcommand{\nl}{n_{\lambda}}
\newcommand{\silr}{\paren{\sigma_{\lambda}^r}^2}

\newcommand{\ERM}{\widehat{s}}
\DeclareMathOperator{\ERMalg}{ERM}
\newcommand{\lamm}{\lambda \in \Lambda_m}

\newcommand{\A}{\mathcal{A}}
\newcommand{\Dh}{\widehat{D}}

\newcommand{\Cor}{\ensuremath{C_{\mathrm{or}}}} 
\newcommand{\CorR}{\ensuremath{\Cor^{(R)}}} 
\newcommand{\Cora}[1]{\ensuremath{\Cor\paren{\alg{#1}{\Id}}}} 
\newcommand{\Corb}[1]{\ensuremath{\Cor\paren{\alg{\Emp}{#1}}}} 

\newcommand{\flapp}{\mapsto}
\newcommand{\flens}{\mapsto}
\newcommand{\grandO}{\ensuremath{\mathcal{O}}}
\newcommand{\set}[1]{\ensuremath{ \left\{ #1 \right\} }}

\newcommand{\mo}{m^{\star}}
\newcommand{\Do}{D^{\star}}
\newcommand{\mM}{m \in \M}

\newcommand{\Lpo}{\mathrm{Lpo}}
\newcommand{\Loo}{\mathrm{Loo}}
\newcommand{\Id}{\mathrm{Id}}
\newcommand{\Emp}{\mathrm{ERM}}
\newcommand{\VF}{\mathrm{VF}}

\newcommand{\mhERM}{\ensuremath{\mh_{\mathrm{ERM}}}}
\newcommand{\mhLoo}{\ensuremath{\mh_{\Loo}}}
\newcommand{\Rhho}{\ensuremath{\Rh_{\mathrm{ho}}}}
\newcommand{\RhLoo}{\ensuremath{\Rh_{\Loo}}}
\newcommand{\mhLpo}{\ensuremath{\mh_{\Lpo_p}}}
\newcommand{\RhLpo}{\ensuremath{\Rh_{\Lpo_p}}}
\newcommand{\mhVFCV}{\ensuremath{\mh_{\VF_V}}}

\newcommand{\BM}{\mathrm{BM}}
\newcommand{\mhBM}{\ensuremath{\mh_{\BM}}}
\newcommand{\DhBM}{\ensuremath{\Dh_{\BM}}}
\newcommand{\penBM}{\ensuremath{\pen_{\BM}}}
\newcommand{\critBM}{\ensuremath{\crit_{\BM}}}

\newcommand{\PML}{\mathrm{PML}}
\newcommand{\critPML}{\ensuremath{\crit_{\PML}}}

\newcommand{\DhVF}{\ensuremath{\Dh_{\VF_V}}}
\newcommand{\RhVF}{\ensuremath{\Rh_{\VF_V}}}
\newcommand{\critVF}{\ensuremath{\crit_{\VF_V}}}

\newcommand{\It}{\ensuremath{I^{(t)}}} 
\newcommand{\Iv}{\ensuremath{I^{(v)}}} 
\newcommand{\Pnt}{\ensuremath{P_n^{(t)}}} 
\newcommand{\Pnv}{\ensuremath{P_n^{(v)}}} 

\newcommand{\alg}[2]{\ensuremath{\llbracket #1 , #2 \rrbracket}} 
\newcommand{\Proc}{\ensuremath{\mathfrak{P}}}

\newcommand{\sigmacst}{\ensuremath{\overline{\sigma}}}
\newcommand{\D}{\mathcal{D}_n}

\newcommand{\Mc}{\ensuremath{\widetilde{\M}}}
\newcommand{\mMc}{\ensuremath{m\in \Mc}}
\newcommand{\pcdeuxOLD}{pc,3}
\newcommand{\pctroisOLD}{pc,2}
\newcommand{\Ch}{\ensuremath{\widehat{C}}}

\newcommand{\refsuppBM}{1}
\newcommand{\refsuppRand}{2}
\newcommand{\refsuppTab}{3}

\newcommand{\largeurfigchr}{.4\linewidth}
\newcommand{\largdeufig}{.45\linewidth}

\begin{document}


\title{Segmentation of the mean of heteroscedastic data via cross-validation}

\author{Sylvain Arlot and Alain Celisse}

\date{\today}

%

\maketitle

\begin{abstract}
This paper tackles the problem of detecting abrupt changes in the
mean of a heteroscedastic signal by model selection, without
knowledge on the variations of the noise.
A new family of change-point detection procedures is proposed,
showing that cross-validation methods can be successful in the
heteroscedastic framework, whereas most existing procedures are not
robust to heteroscedasticity.
The robustness to heteroscedasticity of the proposed procedures is supported
by an extensive simulation study, together with recent theoretical results.
An application to Comparative Genomic Hybridization (CGH) data is provided, showing
that robustness to heteroscedasticity can indeed be required for their analysis.
\end{abstract}




\section{Introduction} \label{sec.intro}

The problem tackled in the paper is the detection of abrupt changes in the
mean of a signal without assuming its variance is constant. Model
selection and cross-validation techniques are used for building
change-point detection procedures that significantly improve on
existing procedures when the variance of the signal is not constant.
Before detailing the approach and the main contributions of the
paper, let us motivate the problem and briefly recall some related
works in the change-point detection literature.

\subsection{Change-point detection} \label{sec.intro.chge-pt}

The change-point detection problem, also called one-dimensional
segmentation, deals with a stochastic process the distribution of
which abruptly changes at some unknown instants. The purpose is to
recover the location of these changes and their number. This problem is motivated by
a wide range of applications, such as voice recognition, financial time-series
analysis \citep{LaTe06} and Comparative Genomic Hybridization (CGH) data analysis \citep{Pica05}.
A large literature exists about change-point detection in many frameworks \citep[see][for a complete bibliography]{BaNi93,BrDa93}.

The first papers on change-point detection were devoted to the
search for the location of a unique change-point, also named
breakpoint \citep[see][for instance]{Pica85}.
Looking for multiple change-points is a harder task and has been studied later.
For instance, Yao \cite{Yao88} used the BIC criterion for detecting multiple change-points in a Gaussian signal, and Miao and Zhao \cite{MiZh93} proposed an approach relying on rank statistics.

%

\medskip

The setting of the paper is the following.
The values $Y_1, \ldots, Y_n \in \R$ of a noisy signal at points $t_1, \ldots, t_n$ are observed, with
\begin{equation}\label{regression model}
Y_i = s(t_i) + \sigma(t_i)\epsilon_i \enspace , \qquad
\E\croch{\epsilon_i}=0 \quad \mbox{and} \quad \Var(\epsilon_i)=1
\enspace .
\end{equation}
The function $s$ is called the {\em regression function} and is assumed to be piecewise-constant, or at least well approximated by piecewise constant functions, that is, $s$ is smooth everywhere except at a few breakpoints.
The noise terms $\epsilon_1, \ldots, \epsilon_n$ are assumed to be
independent and identically distributed. No assumption is made on
$\sigma: [0,1]\mapsto  [0,\infty)$.
Note that all data $(t_i,Y_i)_{1 \leq i \leq n}$ are observed before detecting the change-points, a setting which is called {\em off-line}.

As pointed out by Lavielle \cite{Lavi05}, multiple change-point detection procedures generally tackle one among the following three problems:
\begin{enumerate}
\item Detecting changes in the mean $s$ assuming the standard-deviation $\sigma$ is
constant,
\item Detecting changes in the standard-deviation $\sigma$ assuming the mean $s$ is
constant,
\item Detecting changes in the whole distribution of $Y$, with no distinction between changes in the mean $s$, changes in the standard-deviation~$\sigma$ and changes in the distribution of $\epsilon$.
\end{enumerate}
In applications such as CGH data analysis, changes in the mean $s$
have an important biological meaning, since they correspond to the
limits of amplified or deleted areas of chromosomes. However in the CGH
setting, the standard-deviation $\sigma$ is not always constant, as assumed in problem~1.
See Section~\ref{sec.application.CGH} for more details on CGH data, for
which heteroscedasticity---that is, variations of $\sigma$---
correspond to experimental artefacts or biological nuisance that
should be removed.


Therefore, CGH data analysis requires to solve a fourth problem, which is the purpose of the present article:
\begin{enumerate}
\setcounter{enumi}{3}
\item Detecting changes in the mean $s$ with no constraint on the standard-deviation $\sigma: [0,1] \mapsto [0,\infty)$.
\end{enumerate}
Compared to problem~1, the difference is the presence of an
additional nuisance parameter $\sigma $ making problem~4 harder. Up
to the best of our knowledge, no change-point detection procedure
has ever been proposed for solving problem~4 with {\em no prior
information on $\sigma$}.

\subsection{Model selection} \label{sec.intro.mod-sel}

Model selection is a successful approach for multiple change-point
detection, as shown by Lavielle \cite{Lavi05} and by
Lebarbier \cite{Leba05} for instance. Indeed, a set of
change-points---called a segmentation---is naturally associated with
the set of piecewise-constant functions that may only jump at these
change-points. Given a set of functions (called a model), estimation
can be performed by minimizing the least-squares criterion (or other criteria, see Section~\ref{sec.breakpoint.locations}).
Therefore, detecting changes in the mean of a signal, that is the
choice of a segmentation, amounts to select such a model.

More precisely, given a collection of models
$\acc{\Sm}_{m\in\Mn}$ and the associated collection of least-squares estimators
$\acc{\shm}_{m\in \Mn}$, the purpose of model selection is to
provide a model index $\mh$ such that $\sh_{\mh}$ reaches the ``best
performance'' among all estimators $\acc{\shm}_{m\in\Mn}$.

\medskip

Model selection can target two different goals.
On the one hand, a model selection procedure is {\em efficient} when
 its quadratic risk is smaller than the smallest quadratic risk of
 the estimators $\acc{\shm}_{m\in\Mn}$, up to a constant factor $C_n \geq 1$.
Such a property is called an {\em oracle inequality} when it holds
for every finite sample size. The procedure is said to be
\textit{asymptotic efficient} when the previous property holds with
$C_n\to 1$ as $n$ tends to infinity.
%
%
%
Asymptotic efficiency is the goal of AIC \citep{Akai69,Akai73} and
Mallows'~$C_p$ \citep{Mall73}, among many others.

\sloppy On the other hand, assuming that $s$ belongs to one of the
models $\acc{\Sm}_{m\in\Mn}$, a procedure is {\em model consistent}
when it chooses the smallest model containing $s$ asymptotically
with probability one. Model consistency is the goal of BIC
\citep{Schw78} for instance. See also the article by Yang \cite{Yang03}
about the distinction between efficiency and model consistency.

In the present paper as in \cite{Leba05}, the quality
of a multiple change-point detection procedure is assessed by the
quadratic risk; hence, {\em a change in the mean hidden by the noise
should not be detected}.
This choice is motivated by applications where {\em the
signal-to-noise ratio may be small}, so that exactly recovering
every true change-point is hopeless.
Therefore, {\em efficient} model selection procedures will be used in order to detect the change-points.

\medskip

Without prior information on the locations of the change-points, the natural collection of models for change-point detection depends on the sample size $n$.
Indeed, there exist ${n-1 \choose D-1}$ different partitions of the $n$ design points into $D$ intervals, each partition corresponding to a set of $(D-1)$ change-points.
Since $D$ can take any value between 1 and $n$, $2^{n-1}$ models can be considered.
Therefore, model selection procedures used for multiple change-point detection have to satisfy {\em non-asymptotic} oracle inequalities: the collection
of models cannot be assumed to be fixed with the sample size $n$ tending to infinity.
(See Section~\ref{sec.framework.models} for a precise definition of the collection $\set{S_m}_{\mM_n}$ used for change-point detection.)

\medskip

Most model selection results consider ``polynomial'' collections of
models $\set{S_m}_{\mM_n}$, that is $\card(\M_n) \leq C n^{\alpha}$
for some constants $C,\alpha \geq 0$. For polynomial collections,
procedures like AIC or Mallows' $C_p$ are proved to satisfy oracle
inequalities in various frameworks
\citep{Bara00,BiMa01,Bara02,BiMa06}, assuming that data are {\em
homoscedastic}, that is, $\sigma(t_i)$ does not depend on~$t_i$.

However as shown in \cite{Arlo08a}, Mallows' $C_p$ is suboptimal
when data are {\em heteroscedastic}, that is the variance is
non-constant. Therefore, other procedures must be used. For
instance, resampling penalization is optimal with heteroscedastic
data \citep{Arlo08}.
Another approach has been explored by Gendre \cite{Gend08}, which
consists in simultaneously estimating the mean and the variance,
using a particular polynomial collection of models.

\medskip

However in change-point detection, the collection of models is
''exponential'', that is $\card(\M_n)$ is of order $\exp(\alpha n)$
for some $\alpha>0$. For such large collections, especially larger
than polynomial, the above penalization procedures fail.
Indeed, Birg\'e and Massart \cite{BiMa06} proved that the minimal
amount of penalization required for a procedure to satisfy an oracle
inequality is of the form
\begin{equation} \label{pen.BM}
\pen(m)=c_1\frac{\sigma^2 D_m}{n}+c_2\frac{\sigma^2 D_m}{n}\log\paren{\frac{n}{D_m}}
\enspace ,
\end{equation}
where $c_1$ and $c_2$ are positive constants and $\sigma^2$ is the variance of the noise, assumed to be constant.
Lebarbier \cite{Leba05} proposed $c_1 = 5$ and $c_2 = 2$ for optimizing the penalty \eqref{pen.BM} in the context of change-point detection.
Penalties similar to \eqref{pen.BM} have been introduced independently by other authors
\citep{Riss83,ABDJ06,BaBM99,TiKn99} and are shown to provide
satisfactory results.

\medskip

Nevertheless, all these results assume that data are homoscedastic.
Actually, the model selection problem with heteroscedastic data and an exponential collection of models has never been
considered in the literature, up to the best of our
knowledge.

Furthermore, penalties of the form \eqref{pen.BM} are very close to be proportional to $D_m$, at least for small values of $D_m$.
Therefore, the results of \cite{Arlo08a} lead to conjecture that
the penalty \eqref{pen.BM} is suboptimal for model selection over an exponential collection of models, when data are heteroscedastic.
The suggest of this paper is to use cross-validation methods instead.

\subsection{Cross-validation} \label{sec.intro.CV}

Cross-validation (CV) methods allow to estimate (almost) unbiasedly the quadratic risk of any estimator, such as $\ERM_m$ (see Section~\ref{sec.breakpoints.locations.CV} about the heuristics underlying CV).
%
%
Classical examples of CV methods are the leave-one-out
\citep[Loo,][]{LaMi68,Ston74} and $V$-fold cross-validation
\citep[VFCV,][]{Geis74,Geis75}. 
More references on cross-validation can be found in \cite{Arlo08b,Celi08} for instance.

CV can be used for model selection, by choosing the model $S_m$ for which the CV estimate of the risk of $\ERM_m$ is minimal.
The properties of CV for model selection with a polynomial collection of models and homoscedastic data have been widely studied.
In short, CV is known to adapt to a wide range of statistical settings,
from density estimation \citep{Ston84,CeRo08} to regression
\citep{Ston77,Yang07} and classification \citep{KMNR97,Yang06}.
In particular, Loo is asymptotically equivalent to AIC or Mallows' $C_p$ in several frameworks where they are asymptotically optimal, and other CV methods have similar performances, provided the size of the training sample is close enough to the sample size \citep[see for instance][]{Li87,Shao97,DuLa05}.
In addition, CV methods are robust to heteroscedasticity of data \citep{Arlo08,Arlo08b}, as well as several other resampling methods \citep{Arlo08a}.
Therefore, CV is a natural alternative to penalization procedures assuming homoscedasticity.


\medskip

Nevertheless, nearly nothing is known about CV for model selection with an exponential collection of models, such as in the change-point detection setting.
The literature on model selection and CV \citep{BiMa97,Shao97,BiMa06,Cel:2008:phd} only suggests that minimizing directly the Loo estimate of the risk over $2^{n-1}$ models would lead to overfitting.

In this paper, a remark made by Birg\'e and Massart \cite{BiMa06} about penalization procedure is used for solving this issue in the context of change-point detection.
Model selection is perfomed in two steps:
First, choose a segmentation given the number of change-points; second, choose the number of change-points.
CV methods can be used at each step, leading to Procedure~\ref{algo.general} (Section~\ref{sec.2steps}).
The paper shows that such an approach is indeed successful for detecting changes in the mean of a heteroscedastic signal.

\subsection{Contributions of the paper}
The main purpose of the present work is to design a CV-based
model selection procedure (Procedure~\ref{algo.general}) that can be
used for detecting multiple changes in the mean of a heteroscedastic
signal.
Such a procedure experimentally adapts to heteroscedasticity when the collection of models is exponential, which has never been obtained before.
In particular, Procedure~\ref{algo.general} is a reliable alternative to Birg\'e and Massart's penalization procedure \citep{BiMa01} when data can be heteroscedastic.

Another major difficulty tackled in this paper is the computational
cost of resampling methods when selecting among $2^n$ models. Even
when the number $(D-1)$ of change-points is given, exploring the
${n-1\choose D-1}$ partitions of $[0,1]$ into $D$ intervals and
performing a resampling algorithm for each partition is not feasible
when $n$ is large and $D>0$.
An implementation of Procedure~\ref{algo.general} with a tractable
computational complexity is proposed in the paper, using closed-form
formulas for Leave-$p$-out (Lpo) estimators of the risk, dynamic
programming, and $V$-fold cross-validation.

The paper also points out that least-squares estimators are not
reliable for change-point detection when the number of breakpoints
is given, although they are widely used to this purpose in the
literature. Indeed, experimental and theoretical results detailed in
Section~\ref{sec.breakpoints.locations.ERM} show that least-squares
estimators suffer from {\em local overfitting} when the variance of
the signal is varying over the sequence of observations. On the
contrary, minimizers of the Lpo estimator of the risk do not suffer
from this drawback, which emphasizes the interest of using
cross-validation methods in the context of change-point detection.

\medskip

The paper is organized as follows.
The statistical framework is described in Section~\ref{sec.framework}.
First, the problem of selecting the \guill{best} segmentation given the number of
change-points is tackled in Section~\ref{sec.breakpoint.locations}.
Theoretical results and an extensive simulation study show that the
usual minimization of the least-squares criterion can be misleading
when data are heteroscedastic, whereas cross-validation-based procedures
provide satisfactory results in the same framework.

Then, the problem of choosing the number of
breakpoints from data is addressed in Section~\ref{sec.number.breakpoints}.
As supported by an extensive simulation study, $V$-fold
cross-validation (VFCV) leads to a computationally feasible and
statistically efficient model selection procedure when data are
heteroscedastic, contrary to procedures implicitly assuming
homoscedasticity.

The resampling methods of Sections~\ref{sec.breakpoint.locations}
and~\ref{sec.number.breakpoints} are combined in
Section~\ref{sec.2steps}, leading to a family of resampling-based
procedures for detecting changes in the mean of a heteroscedastic
signal.
A wide simulation study shows they perform well with both
homoscedastic and heteroscedastic data, significantly improving the
performance of procedures which implicitly assume homoscedasticity.

Finally, Section~\ref{sec.application.CGH} illustrates on a real
data set the promising behaviour of the proposed procedures for
analyzing CGH microarray data, compared to procedures previously
used in this setting.

\section{Statistical framework}\label{sec.framework}
%
In this section, the statistical framework of change-point detection
via model selection is introduced, as well as some notation.
%
\subsection{Regression on a fixed design} \label{sec.framework.reg}
%
%
Let $\Ss$ denote the set of measurable functions $[0,1] \flens \R$.
Let $t_1 < \cdots < t_n \in [0,1]$ be some deterministic design points, $s \in \Ss$ and $\sigma: [0,1] \flens [0,\infty)$ be some functions and define
\begin{equation}\label{model.rupt}
\forall i \in \set{1 , \ldots, n}, \qquad Y_i = s(t_i) + \sigma(t_i) \epsilon_i \enspace ,
\end{equation}
where $\epsilon_1, \ldots, \epsilon_n$ are independent and identically distributed random variables with $\E\croch{\epsilon_i}=0$ and $\E\croch{\epsilon_i^2}=1$.

As explained in Section~\ref{sec.intro.chge-pt}, the goal is to find
from $(t_i,Y_i)_{1 \leq i \leq n}$
a piecewise-constant function $f \in \Ss$ close to $s$ in terms of
the quadratic loss
\[ \perte{f} \defegal \frac{1}{n} \sum_{i=1}^n \paren{ f(t_i) - s(t_i)}^2 \enspace . \]

\subsection{Least-squares estimator} \label{sec.framework.lse}
A classical estimator of $s$ is the {\em least-squares estimator}, defined as follows.
For every $f\in \Ss$, the least-squares criterion at $f$ is defined by
\[ P_n \gamma(f) \defegal \frac{1}{n} \sum_{i=1}^n \paren{ Y_i - f(t_i) }^2 \enspace . \]
\sloppy The notation $P_n \gamma(f)$ means that the function $(t,Y)
\flapp \gamma(f;(t,Y)) \defegal \paren{Y - f(t)}^2 $ is integrated
with respect to the empirical distribution $P_n \defegal n^{-1}
\sum_{i=1}^n \delta_{(t_i,Y_i)}$. $P_n \gamma(f)$ is also called the
{\em empirical risk} of $f$.

Then, given a set $S \subset \Ss$ of functions $[0,1] \flens \R$ (called a {\em
model}),
the least-squares estimator on model $S$ is
\[ \ERMalg (S ; P_n) \defegal \argmin_{f \in S}
\acc{ P_n \gamma(f) } \enspace . \]
The notation $\ERMalg (S ; P_n)$ stresses that the least-squares estimator is the output of the empirical risk minimization algorithm over $S$, which takes a model $S$ and a data sample as inputs.
%
When a collection of models $\acc{\Sm}_{m\in\Mn}$ is given, $\ERM_m(P_n)$ or $\ERM_m$ are shortcuts for $\ERMalg (S_m ; P_n)$.
%
%

\subsection{Collection of models} \label{sec.framework.models}
Since the goal is to detect jumps of $s$, every model considered in this article is the set of piecewise constant functions with respect to some partition of $[0,1]$.

For every $K \in \set{ 1, \ldots, n-1}$ and every sequence of integers $\alpha_0 = 1 < \alpha_1 < \alpha_2 < \cdots < \alpha_{K} \leq n$ (the breakpoints),
$\paren{\Il}_{\lambda\in\Lambda_{(\alpha_1 , \ldots \alpha_K)}}$
denotes the partition
\[ [t_{\alpha_0} ; t_{\alpha_1}), \, \ldots \, , \, [t_{\alpha_{K-1}} ; t_{\alpha_K}), \,  [t_{\alpha_K}; 1]  \]
of $[0,1]$ into $(K+1)$ intervals.
Then, the model $S_{(\alpha_1 , \ldots \alpha_K)}$ is defined as the set of piecewise constant
functions that can only jump at $t= t_{\alpha_j}$ for some $j \in \set{1, \ldots, K}$.

For every $K \in \set{ 1, \ldots, n-1}$, let $\Mc_n(K+1)$ denote the
set of such sequences $(\alpha_1 , \ldots \alpha_K)$ of length $K$,
so that $\acc{S_m}_{m \in \Mc_n(K+1)}$ is the collection of models of
piecewise constant functions with $K$ breakpoints. When $K=0$,
$\Mc_n(1) \defegal \set{\emptyset}$ and the model $S_{\emptyset}$ is the
linear space of constant functions on $[0,1]$.
Remark that for every $K$ and $m \in \Mc_n(K+1)$, $S_m$ is a vector space of dimension $D_m = K+1$.
In the rest of the paper, the relationship between the number of breakpoints $K$ and the dimension $D = K+1$ of the model $S_{(\alpha_1 , \ldots \alpha_K)}$ is used repeatedly; in particular, estimating of the number of breakpoints (Section~\ref{sec.number.breakpoints}) is equivalent to choosing the dimension of a model.
In addition, since a model $\Sm$ is uniquely defined by $m$, the index $m$ is also called a model.

The classical collection of models for change-point detection can now be
defined as $\acc{S_m}_{m \in \Mc_n}$, where $\Mc_n = \bigcup_{D \in \D} \Mc_n(D)$ and $\D = \set{1, \ldots, n}$.
This collection has a cardinality $2^{n-1}$.

In this paper, a slightly smaller collection of models is considered, that is, all $\mMc_n$ such that each element of the partition $\paren{\Il}_{\lamm}$ contains at least two design points $(t_j)_{1 \leq j \leq n}$.
Indeed, when nothing is known about the noise-level $\sigma(\cdot)$, one cannot hope to distinguish two consecutive change-points from a local variation of $\sigma$.
For every $D \in \set{1, \ldots, n}$, let $\M_n(D)$ denote the set of $\mMc_n(D)$ satisfying this property.
Then, the collection of models used in this paper is defined as $\acc{S_m}_{m \in \M_n}$ where $\M_n = \bigcup_{D \in \D} \M_n(D)$ and $\D \subset \set{ 1, \ldots, n/2}$.
Finally, in all the experiments of the paper, $\D = \set{1, \ldots,
4n/10}$ for reasons detailed in
Section~\ref{sec.number.breakpoints.CV}, in particular
Remark~\ref{rk.CV.SD.design}.

\subsection{Model selection} \label{sec.framework.mod-sel}
Among $\acc{S_m}_{m \in \M_n}$, the best model is
defined as the minimizer of the {\em quadratic loss}
$\perte{\shm}$ over $m\in\Mn$ and called the {\em oracle} $\mo$.
Since the oracle depends on $s$, one can only expect to select
$\mh(P_n)$ from the data such that the quadratic loss of
$\ERM_{\mh}$ is close to that of the oracle with high probability,
that is,
\begin{equation}
\label{eq.oracle.rupt}
\perte{\ERM_{\mh}} \leq C \inf_{m\in\Mn} \acc{ \perte{\ERM_{m}} } + R_n
\end{equation}
where $C$ is close to 1 and $R_n$ is a small remainder term
(typically of order $n^{-1}$). Inequality \eqref{eq.oracle.rupt} is
called an {\em oracle inequality}.


\section{Localization of the breakpoints} \label{sec.breakpoint.locations}

A usual strategy for multiple change-point detection
\citep{Lavi05,Leba05} is to dissociate the search for the best
segmentation given the number of breakpoints from the choice of the
number of breakpoints.

In this section, the number $K=D-1$ of breakpoints is fixed and the goal is to
localize them.
In other words, the goal is to select a model among $\acc{\Sm}_{\mM_n (D)}$.


\subsection{Empirical risk minimization's failure with heteroscedastic data}\label{sec.breakpoints.locations.ERM}

As explained by many authors such as Lavielle~\cite{Lavi05}, minimizing the least-squares criterion over $\set{ \ERM_m}_{m \in \M(D)}$ is a classical way of estimating the best segmentation with $(D-1)$ change-points.
This leads to the following procedure:
\begin{algo} \label{alg.ERMD}
\begin{gather*}
\mhERM(D) \defegal \argmin_{m\in\M_n(D)} \set{ \Png(\shm) } = \ERMalg\paren{\widetilde{S}_D ; \Pn} \enspace  , \\
\mbox{where} \qquad
\widetilde{S}_{D} \defegal \cup_{\mM_n(D)}\Sm
\end{gather*}
is the set of piecewise constant functions with exactly $(D-1)$
change-points, chosen among $t_2, \ldots, t_{n}$ (see Section~\ref{sec.framework.models}).
\end{algo}
\begin{rk}
Dynamic programming \citep{BeDr62} leads to an efficient implementation of Procedure~\ref{alg.ERMD} with computational complexity $\grandO\paren{n^2}$.
\end{rk}

\begin{figure}
\begin{center}
\includegraphics[width=\largdeufig]{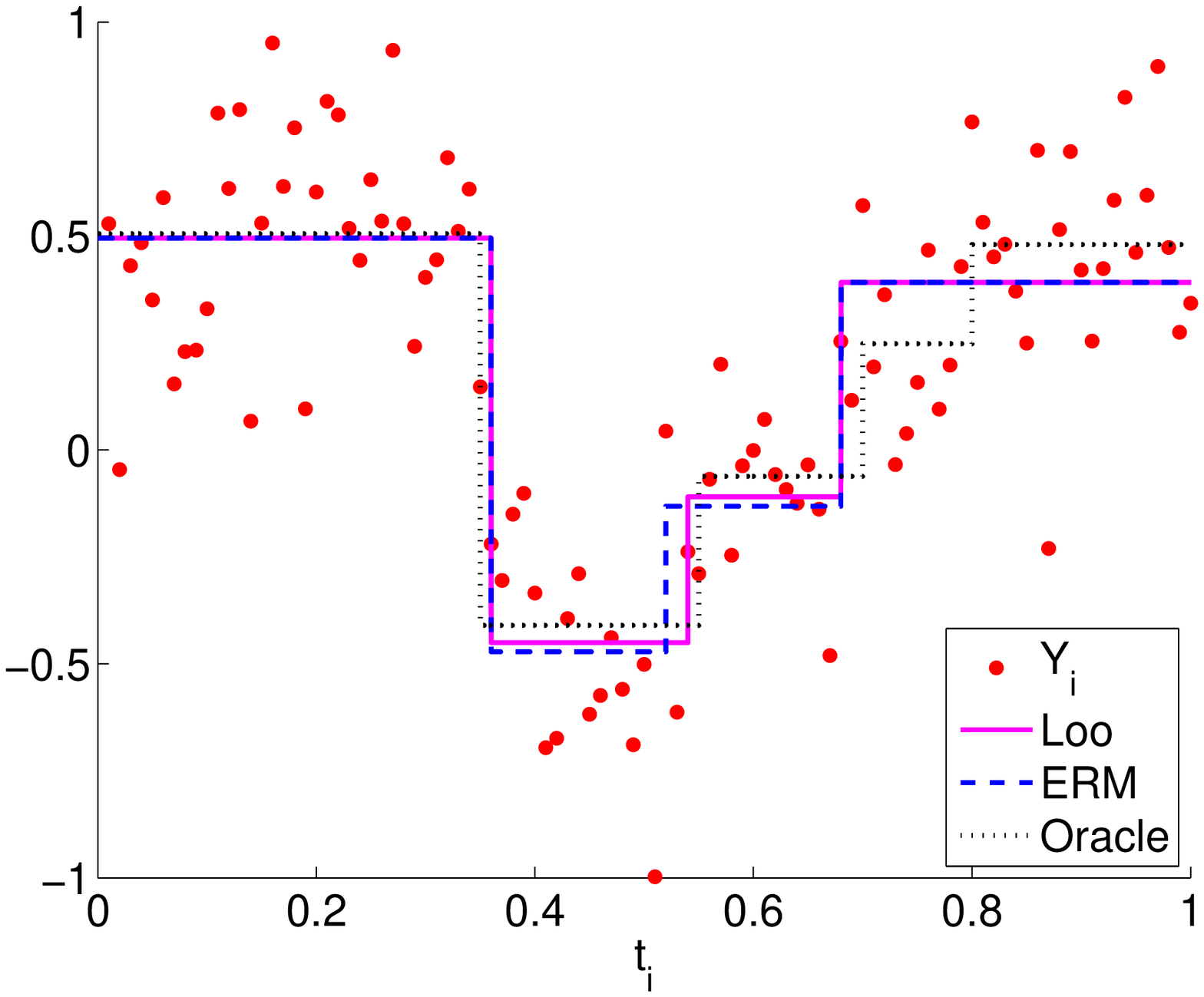}
\hspace{.025\linewidth}
\includegraphics[width=\largdeufig]{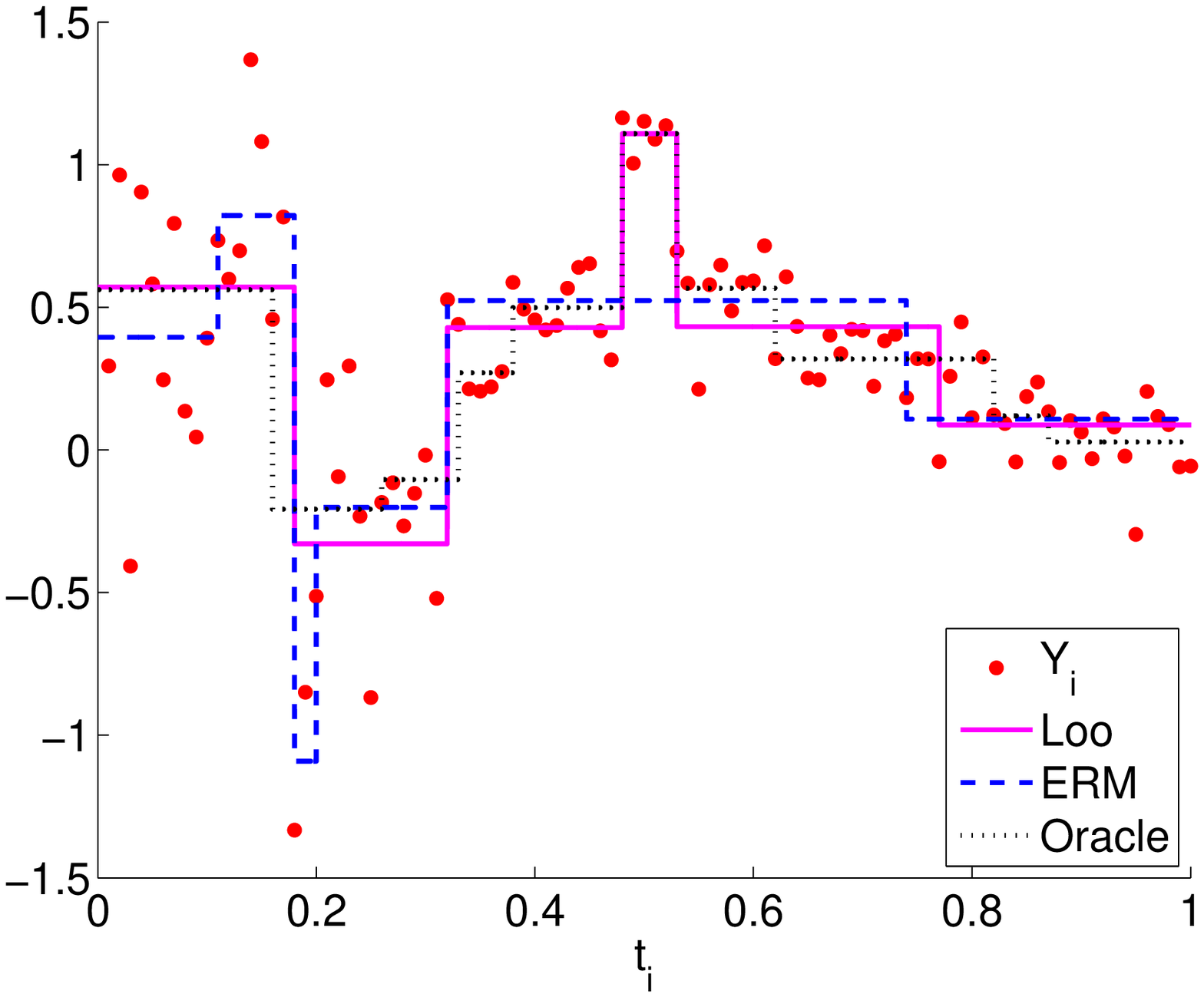}
\vspace{-.8cm}
\end{center}
\caption{ Comparison of $\ERM_{\mo(D)}$ (dotted black line), $\ERM_{\mhERM(D)}$ (dashed blue line)  and $\ERM_{\mhLoo(D)}$ (plain magenta line, see Section~\ref{sec.breakpoints.locations.CV.def}), $D$ being the ``optimal'' dimension (see Figure~\ref{fig.segmentation.quality.homo.hetero}).
Data are generated as described in Section~\ref{sec.breakpoint.locations.simu.setting} with $n=100$
data points.
Left: homoscedastic data $(s_2,\sigma_c)$, $D=4$. Right: heteroscedastic data $(s_3,\sigma_{\pcdeuxOLD})$, $D=6$.
\label{fig.homosced.heterosced.ERM.segmentation}}
\end{figure}

Among models corresponding to segmentations with $(D-1)$
change-points, the oracle model can be defined as
\[ \mo(D) \defegal \argmin_{m\in\M_n(D)} \set{ \perte{\shm }  } \enspace . \]
Figure~\ref{fig.homosced.heterosced.ERM.segmentation} illustrates how far $\mhERM(D)$ typically is from $\mo(D)$ according to variations of the standard-deviation $\sigma$.
On the one hand, when data are homoscedastic, empirical risk
minimization yields a segmentation close to the oracle
(Figure~\ref{fig.homosced.heterosced.ERM.segmentation}, left).
On the other hand, when data are heteroscedastic, empirical risk
minimization introduces artificial breakpoints in areas where the
noise-level is above average, and misses breakpoints in areas where
the noise-level is below average
(Figure~\ref{fig.homosced.heterosced.ERM.segmentation}, right). In
other words, when data are heteroscedastic, {\em empirical risk
minimization over $\widetilde{S}_{D}$ locally overfits in high-noise
areas}, and locally underfits in low-noise areas.

The failure of empirical risk minimization with heteroscedastic data
observed on Figure~\ref{fig.homosced.heterosced.ERM.segmentation} is general \citep[Chapter~7]{Cel:2008:phd} and
can be explained by Lemma~\ref{le.exp.ERM} below.
Indeed, the criteria $\Png(\shm)$ and
$\perte{\shm }$, respectively minimized by $\mhERM(D)$ and $\mo(D)$
over $\M_n(D)$, are close to their respective expectations, as proved by the
concentration inequalities of \cite[Proposition~9]{Arlo08b} for
instance.
Lemma~\ref{le.exp.ERM} enables to compare these expectations.
\begin{lem} \label{le.exp.ERM}
Let $\mM_n$ and define $\sm \defegal \argmin_{f \in \Sm} \perte{f}$.
Then, 
\begin{align}
\label{exp.ERM.expectation.hetero}
\E\croch{\Png\paren{\shm}} &=
\perte{s_m} - V(m)
 + \frac{1}{n}\sum_{i=1}^n\sigma(t_i)^2 \\
\E\croch{\perte{\shm} } & =
\perte{\sm} + V(m)
\label{exp.risk.expectation.hetero}
\end{align}
where
\begin{equation} \label{eq.varterm}
\hspace{-.3cm} V(m) \defegal \frac{\sum_{\lambda \in \Lambda_m} \silr}{n}
\mbox{ and } \forall \lamm, \quad
 \silr \defegal \frac{\sum_{i=1}^n \sigma(t_i)^2 \1_{t_i \in \Il} }{\card\paren{\acc{k\mid t_k\in \Il}}}  \enspace .
\end{equation}
\end{lem}
Lemma~\ref{le.exp.ERM} is proved in \cite{Cel:2008:phd}.
As it is well-known in the model selection literature, the
expectation of the quadratic loss
\eqref{exp.risk.expectation.hetero} is the sum of two terms:
$\perte{\sm}$ is the bias of model $S_m$, and
$V(m)$ is a variance term, measuring the difficulty of estimating the
$D_m$ parameters of model $S_m$.
Up to the term $n^{-1} \sum_{i=1}^n\sigma(t_i)^2 $ which does not
depend on $m$, the empirical risk underestimates the quadratic risk
(that is, the expectation of the quadratic loss), as shown by
\eqref{exp.ERM.expectation.hetero}, because of the sign in front of $V(m)$.

Nevertheless, when data are homoscedastic, that is when $\forall i$, $\sigma(t_i) = \sigmacst$,
$V(m) = D_m \sigmacst^2 n^{-1}$
is the same for all $m \in \M_n(D)$.
%
Therefore, \eqref{exp.ERM.expectation.hetero} and \eqref{exp.risk.expectation.hetero} show that for every $D \geq 1$,
when data are homoscedastic
\[ \argmin_{m \in\M_n (D)} \set{ \E\croch{\Png\paren{\shm}} } = \argmin_{m \in\M_n (D)} \set{ \E\croch{\perte{\shm}} } \enspace . \]
%
Hence, $\mhERM(D)$ and $\mo(D)$ tend to be close to one another, as
on the left of
Figure~\ref{fig.homosced.heterosced.ERM.segmentation}.

On the contrary, when data are heteroscedastic, the variance term
$V(m)$ can be quite different among models $\mM_n(D)$, even though they have the same dimension $D$.
Indeed, $V(m)$ increases when a breakpoint is moved from an area
where $\sigma$ is small to an area where $\sigma$ is large.
Therefore, the empirical risk minimization algorithm rather puts
breakpoints in noisy areas in order to minimize $-V(m)$ in
\eqref{exp.ERM.expectation.hetero}. This is illustrated in the right
panel of Figure~\ref{fig.homosced.heterosced.ERM.segmentation}, where the
oracle segmentation $\mo(D)$ has more breakpoints in areas where
$\sigma$ is small.

\subsection{Cross-validation}\label{sec.breakpoints.locations.CV}

Cross-validation (CV) methods are natural candidates for fixing the failure of empirical risk minimization when data are
heteroscedastic, since CV methods are naturally adaptive to heteroscedasticity (see Section~\ref{sec.intro.CV}).
The purpose of this section is to properly define how CV can be used for selecting $\mh \in \M_n(D)$ (Procedure~\ref{algo.mhLoo}), and to recall theoretical results showing why this procedure adapts to heteroscedasticity (Proposition~\ref{prop.RhLpo.heterosc}).

\subsubsection{Heuristics} \label{sec.breakpoints.locations.CV.heur}
The cross-validation heuristics \citep{Alle74, Ston74} relies on a
data splitting idea: For each candidate algorithm---say $\ERMalg(S_m;
\cdot)$ for some $\mM_n (D)$---, part of the data---called training
set---is used for training the algorithm. The remaining part---called
validation set---is used for estimating the risk of the algorithm.
This simple strategy is called {\em validation} or {\em hold-out}.
One can also split data several times and average the estimated
values of the risk over the splits. Such a strategy is called {\em
cross-validation} (CV).
CV with general repeated splits of data has been introduced by Geisser \cite{Geis74,Geis75}.

In the fixed-design setting, $(t_i, Y_i)_{1 \leq i
\leq n}$ are not identically distributed so that CV estimates a quantity slightly different from the usual prediction error.
Let $T$ be uniformly distributed over $\set{t_1, \ldots, t_n}$
and $Y = s(T) + \sigma(T) \epsilon$, where $\epsilon$ is independent
from $\epsilon_1, \ldots, \epsilon_n$ with the same distribution.
Then, the CV estimator of the risk of $\ERM(P_n)$ estimates
\begin{align*} \E_{(T,Y)} \croch{ \paren{ \ERM(T) - Y}^2 }
&= \frac{1}{n} \sum_{i=1}^n \E_{\epsilon} \croch{ \paren{s(t_i) + \sigma(t_i) \epsilon_i - \ERM(t_i)}^2 } \\
&= \perte{\ERM} + \frac{1}{n} \sum_{i=1}^n \sigma(t_i)^2 \enspace . \end{align*}
Hence, minimizing the CV estimator of $\E_{(T,Y)} \croch{ \paren{
\ERM_m(T) - Y}^2 }$ over $m$ amounts to minimize
$\perte{\ERM_m}$, up to estimation errors.

Even though the use of CV in a fixed-design setting is not usual,
theoretical results detailed in
Section~\ref{sec.breakpoints.locations.CV.theory} below show that CV
actually leads to a good estimator of the quadratic risk
$\perte{\ERM_m}$. This fact is confirmed by all the experimental
results of the paper.


\subsubsection{Definition}  \label{sec.breakpoints.locations.CV.def}

Let us now formally define how CV is used for selecting some $m \in
\M_n(D)$ from data.
A (statistical) algorithm $\A$ is defined as any measurable function
$P_n \flapp \A(P_n) \in \Ss$. For any $t_i \in [0,1]$, $\A(t_i;P_n)$
denotes the value of $\A(P_n)$ at point~$t_i$.

%
For any $\It \subset \set{1, \ldots, n}$, define $\Iv \defegal
\set{1, \ldots, n} \backslash \It$,
\[ \Pnt \defegal \frac{1}{\card(\It)} \sum_{i  \in \It} \delta_{(t_i,Y_i)} \quad \mbox{and} \quad \Pnv \defegal \frac{1}{\card(\Iv)} \sum_{i  \in \Iv} \delta_{(t_i,Y_i)} \enspace . \]
Then, the {\em hold-out estimator of the risk} of any algorithm $\A$ is defined as
\[ \Rhho (\A, P_n, \It) \defegal \Pnv \gamma\paren{ \A\paren{\Pnt}} = \frac{1} {\card(\Iv)} \sum_{i \in \Iv} \paren{ \A(t_i; \Pnt) - Y_i}^2 \enspace . \]

The {\em cross-validation estimators of the risk} of $\A$ are then
defined as the average of $\Rhho (\A, P_n, \It_j)$ over $j = 1,
\ldots, B$ where $\It_1, \ldots, \It_B$ are chosen in a predetermined
way \citep{Geis75}.
Leave-one-out, leave-$p$-out and $V$-fold cross-validation are among
the most classical examples of CV procedures. They differ one another
by the choice of $\It_1, \ldots, \It_B$.
\begin{itemize}
\item {\em Leave-one-out} ($\Loo$), often called {\em ordinary CV} \citep{Alle74, Ston74},
consists in training with the whole sample except one point, used for
testing, and repeating this for each data point: $\It_j = \set{1 ,
\ldots, n} \setminus \set{j}$ for $j = 1, \ldots, n$.
The Loo estimator of the risk of $\A$ is defined by
\begin{equation*} 
\RhLoo (\A, P_n) \defegal \frac{1}{n} \sum_{j=1}^n \croch{ \paren{Y_j - \A \paren{t_j; P_n^{(-j)}} }^2} \enspace ,
\end{equation*}
where $P_n^{(-j)} = (n-1)^{-1} \sum_{i,\,i \neq j} \delta_{(t_i,Y_i)} \enspace $.
\item {\em Leave-$p$-out} ($\Lpo_p$, with any $p \in \set{1, \ldots, n-1}$) generalizes Loo. Let $\mathcal{E}_p$
denote the collection of all possible subsets of $\set{1, \ldots, n}$
with cardinality $n-p$. Then, Lpo consists in considering every
$\It\in \mathcal{E}_p$ as training set indices:
%
%
\begin{equation} \label{eq.def.lpo}
\hspace{-.3cm}\RhLpo (\A, P_n) \defegal {n\choose p}^{-1}
\sum_{\It\in\mathcal{E}_{p}} \croch{ \frac{1}{p} \sum_{j \in \Iv}
\croch{
\paren{Y_j - \A
\paren{t_j; P_n^{(t)}} }^2} } \enspace .
\end{equation}
%
\item {\em $V$-fold cross-validation} (VFCV) is a computationally efficient alternative to $\Lpo$ and $\Loo$.
The idea is to first partition the data into $V$ blocks, to use all
the data but one block as a training sample, and to repeat the
process $V$ times. In other words, VFCV is a blockwise Loo, so that
its computational complexity is $V$ times that of $\A$.
Formally, let $B_1, \ldots, B_V$ be a partition of $\set{1, \ldots,
n}$ and $P_n^{(\overline{B_k})} \defegal (n-\card(B_k))^{-1} \sum_{i \notin B_k}
\delta_{(t_i,Y_i)}$ for every $k \in \set{1, \ldots, V}$.
The VFCV estimator of the risk of $\A$ is defined by
\begin{equation} \label{eq.def.vfcv}
\hspace{-.7cm} \RhVF \paren{ \A, P_n} \defegal \frac{1}{V}
\sum_{k=1}^V \croch{ \frac{1}{\card(B_k)} \sum_{j \in B_k}
\croch{ \paren{Y_j - \A \paren{t_j; P_n^{(\overline{B_k})} } }^2} } \enspace .
\end{equation}
The interested reader will find theoretical and experimental results
on VFCV and the best way to use it in
\cite{Arlo08b,Cel:2008:phd} and references therein, in particular
\cite{Burm89}.
\end{itemize}

\medskip

Given the Loo estimator of the risk of each algorithm $\A$ among
$\set{\ERMalg(S_m;\cdot)}_{\mM_n(D)}$, the segmentation with $(D-1)$
breakpoints chosen by Loo is defined as follows.
\begin{algo} \label{algo.mhLoo}
\begin{align*}
\mhLoo(D)\defegal \argmin_{\mM_n(D)} \set{ \RhLoo  \paren{
\ERMalg\paren{\Sm;\cdot} , P_n }} \enspace .
\end{align*}
The segmentations chosen by $\Lpo$ and VFCV are defined similarly and denoted respectively by
$\mhLpo(D)$ and by $\mhVFCV(D)$.
\end{algo}

\medskip


As illustrated by Figure~\ref{fig.homosced.heterosced.ERM.segmentation}, when data are heteroscedastic, $\mhLoo(D)$ is often closer to the oracle segmentation $\mo(D)$ than $\mhERM(D)$.
%
This improvement will be explained by theoretical results in Section~\ref{sec.breakpoints.locations.CV.theory} below.

\subsubsection{Computational tractability} \label{sec.breakpoints.locations.CV.comput}
The computational complexity of $\ERMalg(S_m;P_n)$ is $\grandO(n)$
since for every $\lamm$, the value of $\ERM_m(P_n)$ on $\Il$ is
equal to the mean of $\set{Y_i}_{t_i \in \Il}$.
Therefore, a naive implementation of $\Lpo_p$ has a computational
complexity $\grandO\paren{ n {n \choose p}}$, which can be
intractable for large $n$ in the context of model selection, even when $p=1$.
In such cases, only VFCV with a small $V$ would work
straightforwardly, since its computational complexity is $\grandO(n
V)$.

Nevertheless, closed-form formulas for the $\Lpo$ estimator of the risk
have been derived in the density estimation \citep{CeRo08,Celi08} and
regression \citep{Cel:2008:phd} frameworks. Some of these closed-form
formulas apply to regressograms $\ERM_m$ with $\mM_n$.
%
The following theorem gives a closed-form expression for $\RhLpo(m)
\defegal \RhLpo (\ERMalg(S_m;\cdot), P_n)$ which can be computed with
$\grandO(n)$ elementary operations.

\begin{thm}[Corollary~3.3.2 in \cite{Cel:2008:phd}] \label{th.formule.lpo}
Let $m\in\Mn$, $S_m$ and $\ERM_m=\ERMalg(S_m;\cdot)$ be defined as in Section~\ref{sec.framework}.
For every $(t_1, Y_1), \ldots, (t_n,Y_n) \in \R^2$ and $\lamm$, define
\[ S_{\lambda,1} \defegal \sum_{j=1}^n Y_j \1_{\set{t_j \in I_{\lambda}}} \quad \mbox{and} \quad S_{\lambda,2} \defegal \sum_{j=1}^n Y_j^2 \1_{\set{t_j \in I_{\lambda}}} \enspace . \]
Then, for every $p \in \set{1, \ldots, n-1}$, the $\Lpo_p$ estimator of the risk of $\ERM_m$ defined by \eqref{eq.def.lpo} is given by
\begin{align*}
\RhLpo(m) &= \sum_{\lamm} \frac{1}{pN_{\lambda}} \left[ \acc{ \paren{ A_{\lambda} - B_{\lambda}}  S_{\lambda,2}
+ B_{\lambda} S_{\lambda,1}^2} \1_{\set{n_{\lambda}\geq
2}}
+ \acc{+\infty} \1_{\set{n_{\lambda}= 1}} \right] \enspace ,
\end{align*}
where for every $\lamm$,
\begin{gather*}
n_{\lambda}\defegal\card\paren{\acc{i\mid t_i\in\Il}} \qquad
N_{\lambda} \defegal 1-\1_{\set{p\geq
n_{\lambda}}}{n-n_{\lambda}\choose p-n_{\lambda}}/{n\choose p} \\
A_{\lambda} \defegal V_{\lambda}(0) \paren{ 1 - \frac{1}{\nl}} - \frac{V_{\lambda}(1)}{n_{\lambda}}  + V_{\lambda}(-1) \\
B_{\lambda}
\defegal 
V_{\lambda}(1) \frac{ 2 - \1_{\nl \geq 3}} { \nl (\nl - 1)} + \frac{V_{\lambda}(0)} {\nl - 1} \croch{ \paren{ 1 + \frac{1}{\nl}} \1_{\nl \geq 3} - 2} - \frac{ V_{\lambda}(-1) \1_{\nl \geq 3}} {\nl - 1} \\
\mbox{and} \quad \forall k\in\acc{-1,0,1}, \quad V_{\lambda}(k)
\defegal \sum_{r= \max\set{ 1, (p-n_{\lambda}) } }^{ \min\set{
n_{\lambda} , (n-p) }} r^k \frac{{n-p\choose r}{p\choose
n_{\lambda}-r}}{{n\choose n_{\lambda}}} \enspace .
\end{gather*}
\end{thm}
\begin{rk}
$V_{\lambda}(k)$ can also be written as $\E\croch{ Z^k \1_{Z> 0}}$
where $Z$ has hypergeometric distribution with parameters $( n ,
n-p, \nl)$.
\end{rk}%

An important practical consequence of Theorem~\ref{th.formule.lpo}
is that for every $D$ and $p$, $\mhLpo (D)$ can be computed
with the same computational complexity as $\mh_{\ERMalg} (D)$, that
is $\mathcal{O}\paren{n^2}$. Indeed, Theorem~\ref{th.formule.lpo}
shows that $\RhLpo(m)$ is a sum over $\lamm$ of terms
depending only on $\set{Y_i}_{t_i \in \Il}$, so that dynamic
programming \citep{BeDr62} can be used for computing the minimizer
$\mhLpo(D)$ of $\RhLpo(m)$ over $\mM_n$.
Therefore, {\em $\Lpo$ and $\Loo$ are computationally tractable for change-point detection}
when the number of breakpoints is given.

Dynamic programming also applies to $\mhVFCV$ with a computational
complexity  $\mathcal{O}\paren{V n^2}$, since each term appearing in
$\RhVF(m)$ is the average over $V$ quantities that must be
computed, except when $V=n$ since VFCV then becomes Loo.
Since VFCV is mostly an approximation to $\Loo$ or $\Lpo$ but has a larger
computational complexity, $\mhLpo$ 
will be preferred to $\mhVFCV(D)$ in the following.

\subsubsection{Theoretical guarantees} \label{sec.breakpoints.locations.CV.theory}
In order to understand why CV indeed works for change-point detection with a given number of breakpoints, let us recall a straightforward consequence of Theorem~\ref{th.formule.lpo} which is proved in details in \cite[Lemma~7.2.1 and Proposition~7.2.3]{Cel:2008:phd}.
\begin{prop} \label{prop.RhLpo.heterosc}
Using the notation of Lemma~\ref{le.exp.ERM}, for any
$m\in\Mn$,
\begin{equation}
\E\croch{ \RhLpo(m) } \approx \perte{s_m} + \frac{1}{n-p} \sum_{\lamm} \silr + \frac{1}{n} \sum_{i=1}^n \sigma(t_i)^2 \enspace ,
\label{exp.Lpo.expectation.hetero}
\end{equation}
where the approximation holds as soon as $\min_{\lamm} \nl$ is large enough (in particular larger than $p$).
\end{prop}
%

The comparison of \eqref{exp.risk.expectation.hetero} and
\eqref{exp.Lpo.expectation.hetero} shows that $\Lpo_p$ yields an almost
unbiased estimator of $\perte{\ERM_m}$: The only
difference is that the factor $1/n$ in front of the variance term
$V(m)$ has been changed into $1/(n-p)$.
Therefore, minimizing the $\Lpo_p$ estimator of the risk instead of the
empirical risk allows to automatically take into account
heteroscedasticity of data.

\subsection{Simulation study} \label{sec.breakpoint.locations.simu}
%
The goal of this section is to experimentally assess, for several
values of $p$, the performance of $\Lpo_p$ for detecting a given number of
changes in the mean of a heteroscedastic signal. This performance is
also compared with that of empirical risk minimization.

\subsubsection{Setting} \label{sec.breakpoint.locations.simu.setting}
The setting described in this section is used in all the experiments of the paper.

\begin{figure}
\hspace*{-.5cm}
\begin{center}
\includegraphics[width=.31\linewidth]{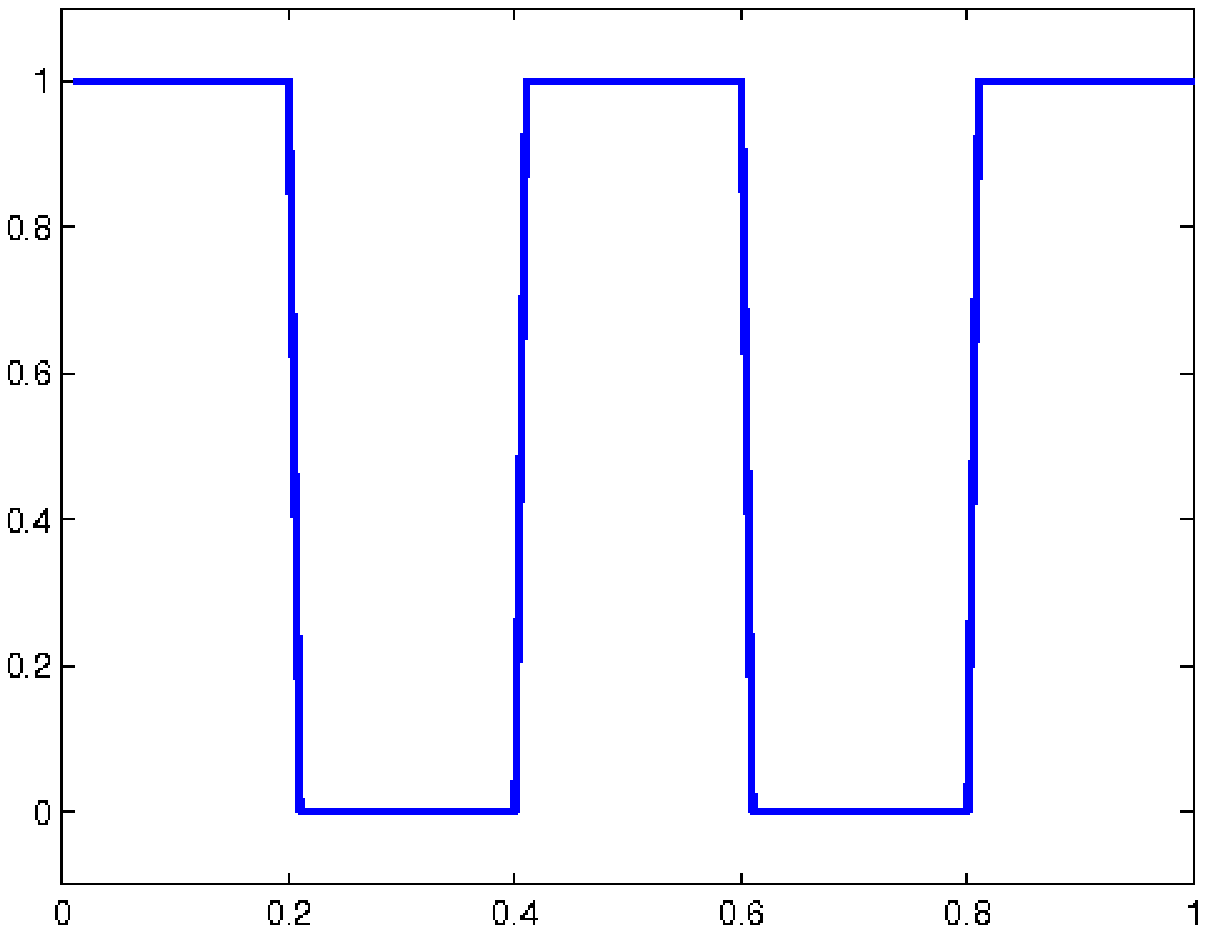} \hspace{.015\linewidth}
\includegraphics[width=.31\linewidth]{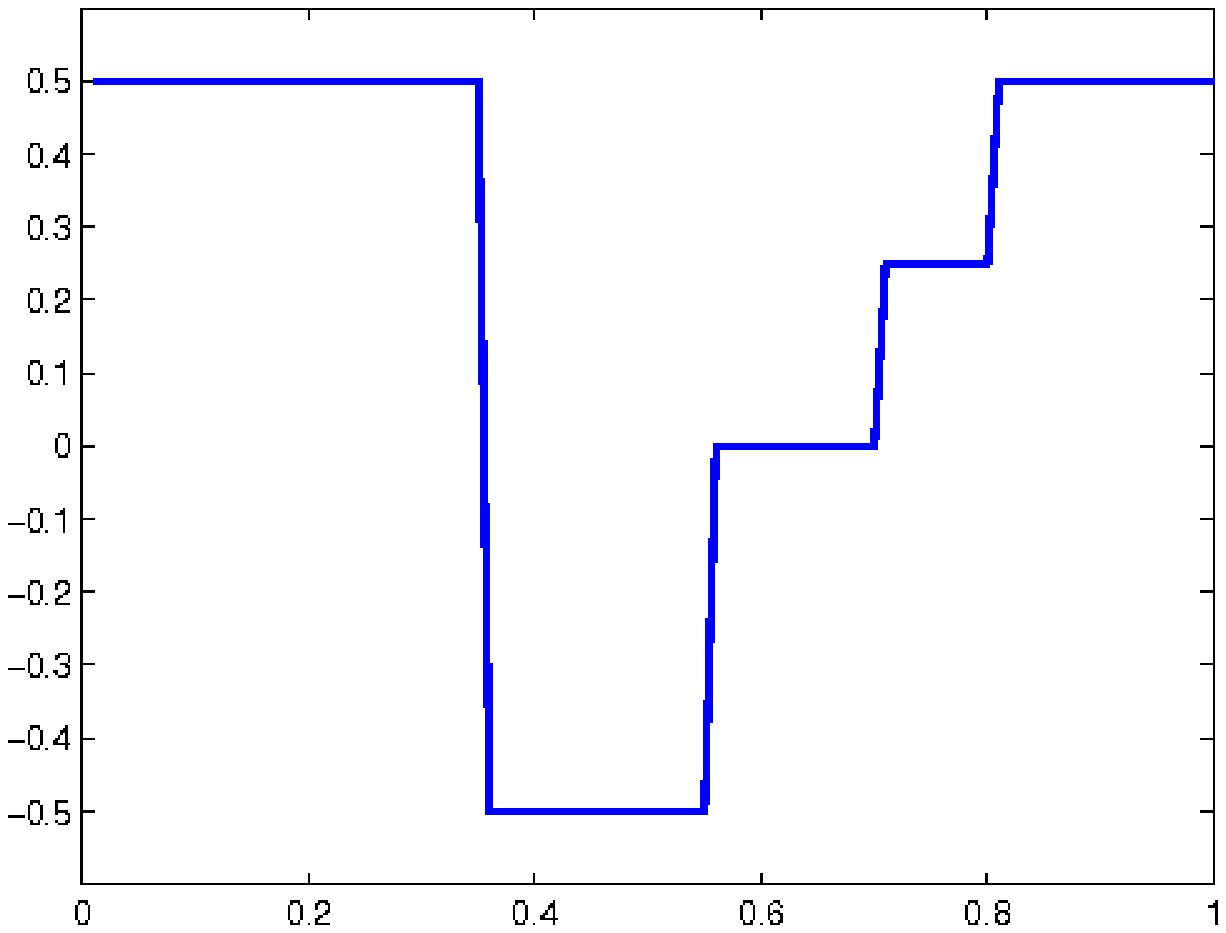} \hspace{.015\linewidth}
\includegraphics[width=.31\linewidth]{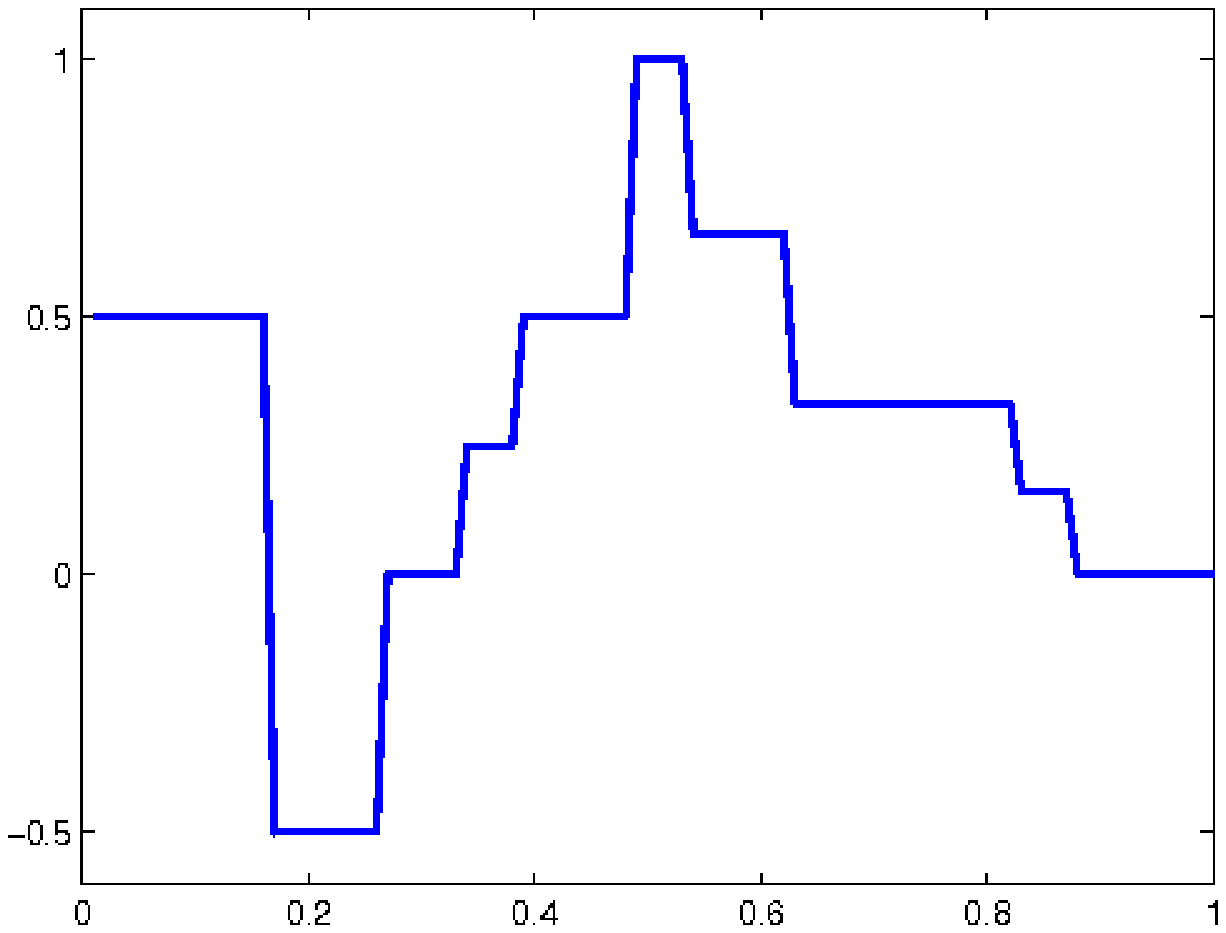}
\end{center}
\caption{Regression functions $s_1, s_2, s_3$; $s_1$ and $s_2$ are piecewise constant with 4 jumps; $s_3$ is piecewise constant with 9 jumps. \label{fig.example.regression.functions}}
\end{figure}

Data are generated according to \eqref{model.rupt} with $n=100$. For
every $i$, $t_i=i/n$ and $\epsilon_i$ has a standard Gaussian
distribution.
The regression function $s$ is chosen among three piecewise constant functions $s_1, s_2, s_3$ plotted on Figure~\ref{fig.example.regression.functions}.
The model collection described in Section~\ref{sec.framework.models}
is used with $\D=\acc{1,\ldots,4n/10}$.
The noise-level function $\sigma(\cdot)$ is chosen among the following functions:
\begin{enumerate}
\item Homoscedastic noise: $\sigma_c=0.25\,\1_{[0,1]}$,
\item Heteroscedastic piecewise constant noise:
$\sigma_{pc,1}=0.2\,\1_{[0,1/3]}+0.05\1_{[1/3,1]}$,
$\sigma_{\pctroisOLD}=2\sigma_{pc,1}$ or $\sigma_{\pcdeuxOLD}=2.5 \sigma_{pc,1} \enspace $.
\item Heteroscedastic sinusoidal noise:
$\sigma_{s}=0.5\sin\paren{t\pi/4}$.
\end{enumerate}

All combinations between the regression functions $(s_i)_{i=1,2,3}$
and the five noise-levels $\sigma_{\cdot}$ have been considered, each
time with $N=10\,000$ independent samples.
Results below only report a small part of the entire simulation
study but intend to be representative of the main observed
behaviour. A more complete report of the results, including other
regression functions $s$ and noise-level functions $\sigma$, is given in the second
authors' thesis \citep[Chapter~7]{Cel:2008:phd}; see also Section~\refsuppTab\ of the supplementary material.

%
%
%

\subsubsection{Results: Comparison of segmentations for each dimension} \label{sec.breakpoint.locations.simu.courbes}
\begin{figure}
\begin{center}
\includegraphics[width=\largdeufig]{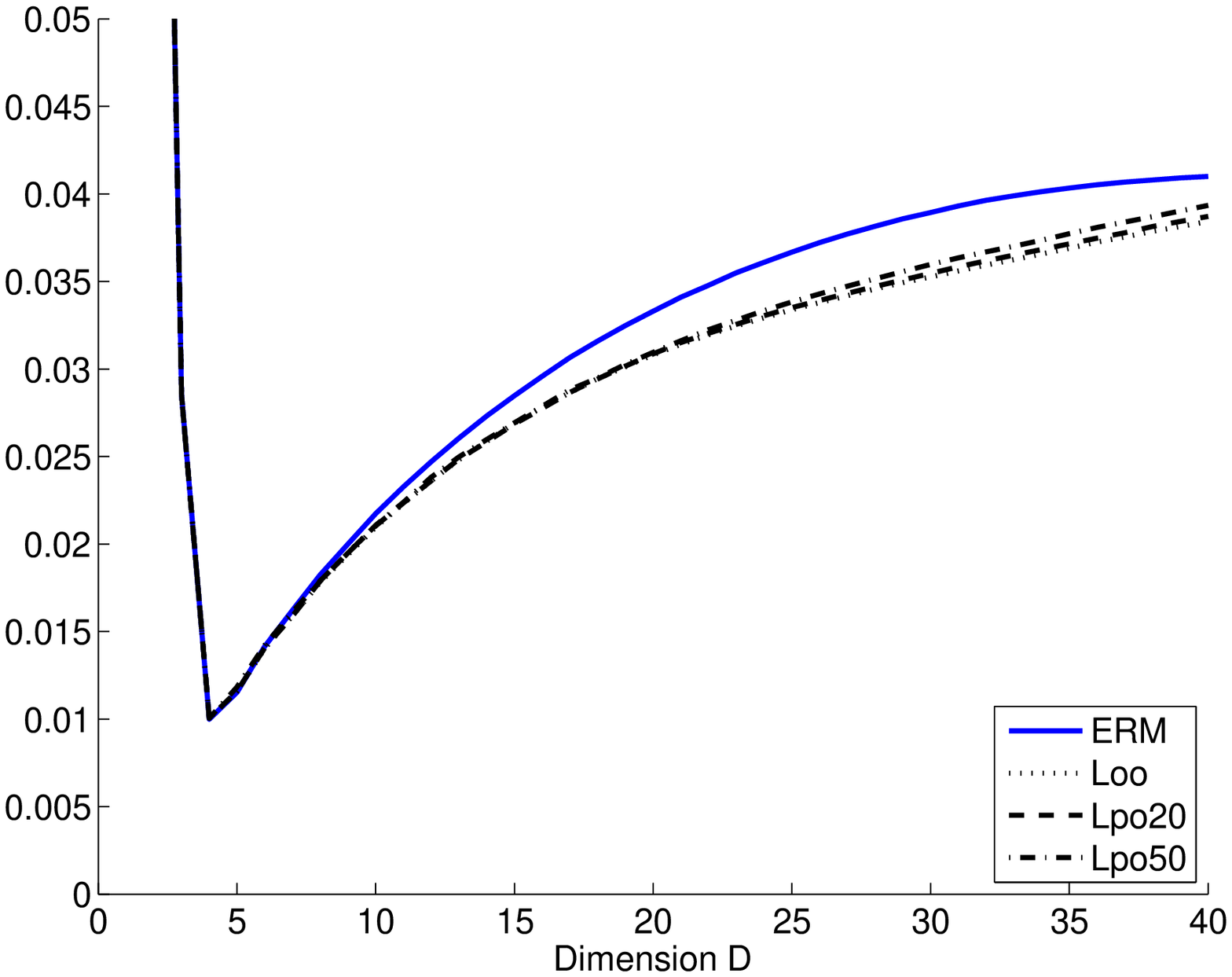}
\hspace{.025\linewidth}
\includegraphics[width=\largdeufig]{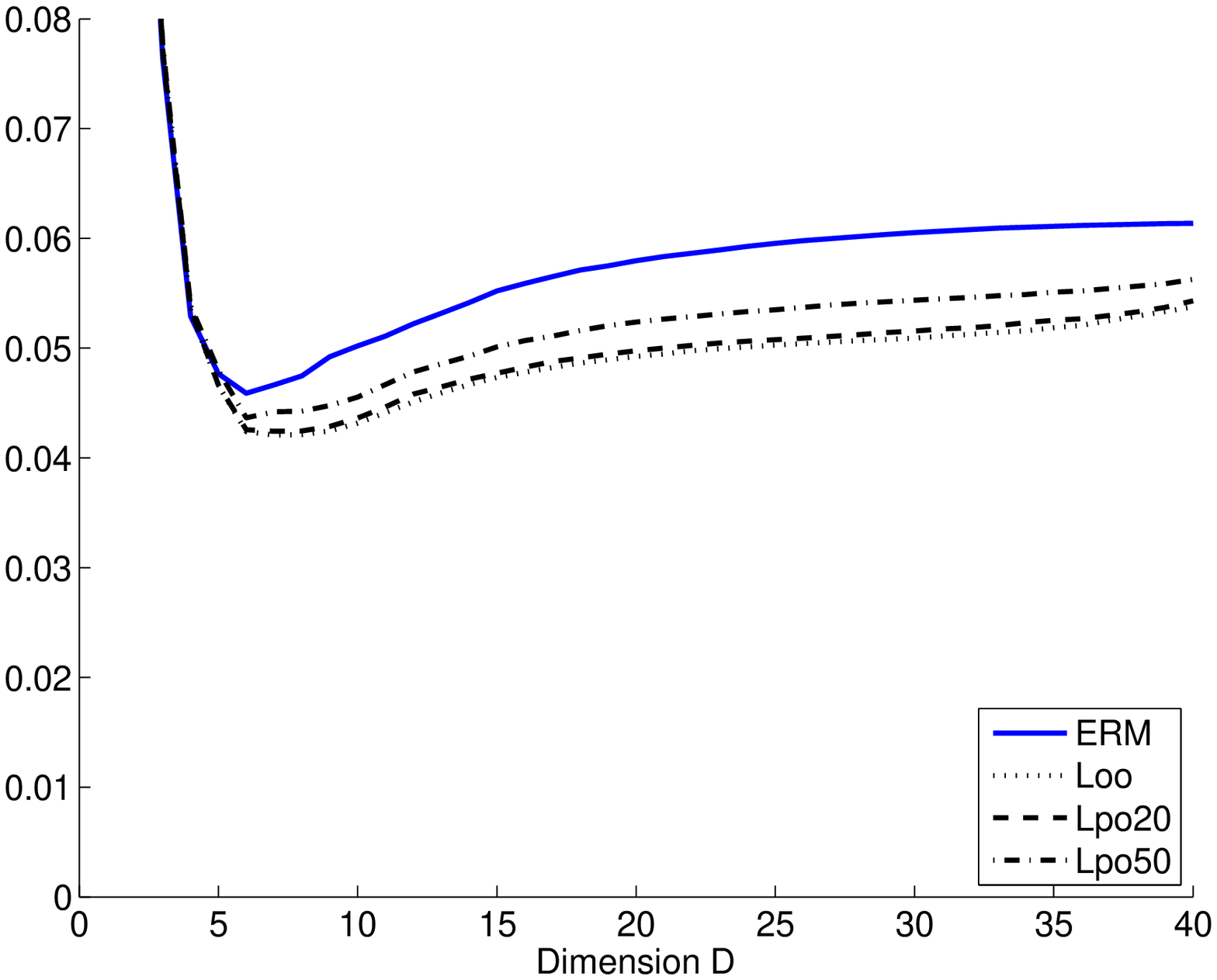}
\vspace{-.8cm}
\end{center}
\caption{ $\E\croch{ \perte{\ERM_{\mh_{\Proc} (D)}} }$ as a function of
$D$ for $\Proc$ among `ERM' (empirical risk minimization), `Loo'
(Leave-one-out), `$\Lpo(20)$' ($\Lpo_p$ with $p=20$) and `$\Lpo(50)$' ($\Lpo_p$ with
$p=50$). Left: homoscedastic $(s_2, \sigma_c)$.
Right: heteroscedastic $(s_3, \sigma_{\pcdeuxOLD})$.
{\small All curves have been estimated from $N =10\,000$ independent
samples; error bars are all negligible in front of visible
differences (the larger ones are smaller than $8.10^{-5}$ on the
left, and smaller than $2.10^{-4}$ on the right).
The curves $D \flapp \perte{\ERM_{\mh_{\Proc} (D)}}$ behave similarly to their expectations.
}
\label{fig.segmentation.quality.homo.hetero}}
\end{figure}

%
The segmentations of each dimension $D \in \D$ obtained by empirical
risk minimization (`$\Emp$', Procedure~\ref{alg.ERMD}) and $\Lpo_p$
(Procedure~\ref{algo.mhLoo}) for several values of $p$ are compared
on Figure~\ref{fig.segmentation.quality.homo.hetero}, through the
expected values of the quadratic loss $\E\croch{\perte{\ERM_{\mh_{\Proc} (D)}} }$ for procedure $\Proc$.

On the one hand, when data are homoscedastic (Figure~\ref{fig.segmentation.quality.homo.hetero}, left), all procedures
yield similar performances for all dimensions up to twice the best
dimension; $\Lpo_p$ performs significantly better for larger dimensions.
Therefore, unless the dimension is strongly overestimated (whatever
the way $D$ is chosen), all procedures are equivalent with
homoscedastic data.

On the other hand, when data are heteroscedastic (Figure~\ref{fig.segmentation.quality.homo.hetero}, right), $\Emp$ yields
significantly worse performance than $\Lpo$ for dimensions larger than
half the true dimension.
%
As explained in Sections~\ref{sec.breakpoints.locations.ERM}
and~\ref{sec.breakpoints.locations.CV.theory}, $\mhERM(D)$ often puts
breakpoints inside pure noise for dimensions $D$ smaller than the
true dimension, whereas $\Lpo$ does not have this drawback.
Therefore, whatever the choice of the dimension (except $D \leq 4$, that is for detecting the obvious jumps), $\Lpo$ should be prefered to empirical risk minimization as soon as data are heteroscedastic.

\subsubsection{Results: Comparison of the ``best'' segmentations} \label{sec.breakpoint.locations.simu.tab}
%
This section focuses on the segmentation obtained with the best
possible choice of $D$, that is the one corresponding to the minimum
of $D \flapp \perte{\ERM_{\mh_{\Proc} (D)}}$ (plotted on
Figure~\ref{fig.segmentation.quality.homo.hetero}) for procedures
$\Proc$ among $\Emp$, $\Loo$, and $\Lpo_p$ with $p=20$ and $p=50$.
Therefore, the performance of a procedure $\Proc$ is defined by
\[ \Cora{\Proc} \defegal \frac{ \E \croch{ \inf_{1 \leq D \leq n} \set{ \perte{\ERM_{\mh_{\Proc} (D)} } } } } { \E \croch{ \inf_{\mM_n} \set{ \perte{\ERM_m } }} } \enspace , \]
which measures what is lost compared to the oracle when
selecting one segmentation $\mh_{\Proc} (D)$ per dimension.
Even if the choice of $D$ is a real practical problem---which will be tackled in
the next sections---, $\Cora{\Proc}$ helps to understand which is the best
procedure for selecting a segmentation of a given dimension.
The notation $\Cora{\Proc}$ has been chosen for consistency with notation used in the next sections (see Section~\ref{sec.2steps.def}).
%

Table~\ref{table.losses.1star2id} confirms the results of Section~\ref{sec.breakpoint.locations.simu.courbes}.
On the one hand, when data are homoscedastic, $\Emp$ performs slightly better than $\Loo$ or $\Lpo_p$.
On the other hand, when data are heteroscedastic, $\Lpo_p$ often performs better than $\Emp$ (whatever $p$), and the improvement can be large (more than 20\% in the setting  $(s_2,\sigma_{\pcdeuxOLD})$).
Overall, when homoscedasticity of the signal is questionable,
$\Lpo_p$ appears much more reliable than $\Emp$ for localizing a
given number of change-points of the mean.

The question of choosing $p$ for optimizing the performance of $\Lpo_p$ remains a widely open problem.
The simulation experiment summarized with Table~\ref{table.losses.1star2id} only shows that $\Lpo_p$ improves $\Emp$ whatever $p$, the optimal value of $p$ depending on $s$ and $\sigma$.

\begin{table}
\begin{center}
{\small
\begin{tabular}{cc|r@{ $\pm$ }lr@{ $\pm$ }lr@{ $\pm$ }lr@{ $\pm$ }l}
$s_{\cdot}$ &   $\sigma_{\cdot}$   & \multicolumn{2}{c}{$\Emp$} & \multicolumn{2}{c}{$\Loo$} & \multicolumn{2}{c}{$\Lpo_{20}$} & \multicolumn{2}{c}{$\Lpo_{50}$} \\
\hline
 2 & c            & {\bf 2.88} & 0.01 & 2.93       & 0.01 & 2.93       & 0.01 & 2.94       & 0.01 \\
   & pc,1         & 1.31       & 0.02 & 1.16       & 0.02 & 1.14       & 0.02 & {\bf 1.11} & 0.01 \\
   & \pcdeuxOLD\  & 3.09       & 0.03 & 2.52       & 0.03 & 2.48       & 0.03 & {\bf 2.32} & 0.03 \\
\hline
 3 & c            & {\bf 3.18} & 0.01 & 3.25       & 0.01 & 3.29       & 0.01 & 3.44       & 0.01 \\
   & pc,1         & 3.00       & 0.01 & {\bf 2.67} & 0.02 & {\bf 2.68} & 0.02 & 2.77       & 0.02 \\
   & \pcdeuxOLD\  & 4.41       & 0.02 & {\bf 3.97} & 0.02 & 4.00       & 0.02 & 4.11       & 0.02 \\
\end{tabular}}
\caption{\label{table.losses.1star2id} Average performance
$\Cora{\Proc}$ for change-point detection procedures $\Proc$ among $\Emp$, $\Loo$
and $\Lpo_p$ with $p=20$ and $p=50$. Several regression functions $s$ and
noise-level functions $\sigma$ have been considered, each time with
$N=10\,000$ independent samples.
Next to each value is indicated the corresponding empirical standard
deviation divided by $\sqrt{N}$, measuring the uncertainty of the
estimated performance.}
\end{center}
\end{table}
%

\section{Estimation of the number of breakpoints}\label{sec.number.breakpoints}

In this section, the number of breakpoints is no longer fixed or known {\it a priori}. The goal is precisely the estimation of this number, as often needed with real data.

Two main procedures are considered.
First, a penalization procedure introduced by Birg\'e and Massart \cite{BiMa01} is analyzed in Section~\ref{sec.number.breakpoints.BM}; this procedure is successful for change-point detection when data are homoscedastic \citep{Lavi05,Leba05}.
On the basis of this analysis, $V$-fold cross-validation (VFCV) is then proposed as an alternative to Birg\'e and Massart's penalization procedure (BM) when data can be heteroscedastic.

In order to enable the comparison between BM and VFCV when
focusing on the question of choosing the number of breakpoints, VFCV
is used for choosing among the same segmentations as BM, that is
$\set{\mhERM(D)}_{D \in \D}$.
The combination of VFCV for choosing $D$ with the new procedures proposed in Section~\ref{sec.breakpoint.locations} will be studied in Section~\ref{sec.2steps}.

\subsection{Birg\'e and Massart's penalization}  \label{sec.number.breakpoints.BM}
%
First, let us define precisely the penalization procedure proposed
by Birg\'e and Massart \cite{BiMa01} successfully used for
change-point detection in \cite{Lavi05,Leba05}.


\begin{algo}[Birg\'e and Massart \cite{BiMa01}]\label{algo.BM1} \hfill \\ \vspace{-.5cm}
\begin{enumerate}
\item $ \forall \mM_n$, $\shm\defegal\ERMalg\paren{\Sm ; \Pn}\enspace $.
\item $ \mhBM \defegal \argmin_{m\in\Mn, \, D_m \in \D} \acc{\Png(\shm)+\penBM(m)}\enspace $,
where for every $\mM_n$, the penalty $\penBM(m)$ only depends on $S_m$ through its dimension:
\begin{equation} \label{def.penBM}
\penBM (m) = \penBM (D_m) \defegal \frac{\Ch D_m}{n} \paren{5 + 2 \log\paren{\frac{n}{D_m}}} \enspace ,
\end{equation}
where $\Ch$ is estimated from data using Birg\'e and
Massart's {\em slope heuristics} \citep{BiMa06,ArMa08}, as proposed
by Lebarbier \cite{Leba05} and by Lavielle \cite{Lavi05}. See Section~\refsuppBM\ of the supplementary material for a detailed discussion about $\Ch$.
\item $\widetilde{s}_{\mathrm{BM}} \defegal \sh_{\mhBM} $.
\end{enumerate}
\end{algo}
\medskip

All $\mM_n(D)$ are penalized in the same way by $\penBM(m)$, so that Procedure~\ref{algo.BM1} actually selects a segmentation among $\set{\mhERM(D)}_{D \in \D}$.
Therefore, Procedure~\ref{algo.BM1} can be reformulated as follows, as noticed in \cite[Section~4.3]{BiMa06}.
\begin{algo}[Reformulation of Procedure~\ref{algo.BM1}]\label{algo.BM2} \hfill \\ \vspace{-.5cm}
\begin{enumerate}
\item $ \forall D\in\D$, $\sh_{\mhERM(D)}\defegal\ERMalg\paren{\widetilde{S}_D ; \Pn}$ where $\widetilde{S}_D\defegal \bigcup_{\mM_n(D)}\Sm\enspace $.
\item $ \DhBM \defegal \argmin_{D \in \D} \acc{\Png(\sh_{\mhERM(D)})+\penBM(D)} $ where $\penBM(D)$ is defined by \eqref{def.penBM}.
\item $\widetilde{s}_{\mathrm{BM}} \defegal \sh_{\mhERM(\DhBM)} \enspace $.
\end{enumerate}
\end{algo}
In the following, `BM' denotes Procedure~\ref{algo.BM2} and
\[ \critBM(D) \defegal \Png(\sh_{\mhERM(D)})+\penBM(D) \] is called the BM criterion.

\medskip

Procedure~\ref{algo.BM2} clarifies the reason why \penBM\ must be larger than Mallows' $C_p$ penalty. Indeed, for every $\mM_n$, Lemma~\ref{le.exp.ERM} shows that when data are homoscedastic, $\Png(\sh_{m})+\pen(m)$ is an unbiased estimator of $\perte{\shm}$ when $\pen(m)=2 \sigma^2 D_m n^{-1}$, that is Mallows' $C_p$ penalty.
When $\card(\M_n)$ is at most polynomial in $n$, Mallows' $C_p$ penalty leads to an efficient model selection procedure, as proved in several regression frameworks \citep{Shib81,Li87,Bara02}. Hence, Mallows' $C_p$ penalty is an adequate measure of the ``capacity'' of any vector space $S_m$ of dimension $D_m$, at least when data are homoscedastic.

On the contrary, in the change-point detection framework, $\card(\M_n)$ grows exponentially with $n$.
The formulation of Procedure~\ref{algo.BM2} points out that $\penBM(D)$
has been built so that $\critBM(D)$ estimates unbiasedly
$\perte{\sh_{\mhERM(D)}}$ for every $D$, where
$\sh_{\mhERM(D)}$ is the empirical risk minimizer over
$\widetilde{S}_D$. Hence, $\penBM(D)$ measures the ``capacity'' of
$\widetilde{S}_D$, which is much bigger than a vector space of
dimension $D$.
Therefore, $\penBM$ should be larger than Mallows' $C_p$, as confirmed by the results of Birg\'e and Massart \cite{BiMa06} on minimal penalties for exponential collections of models.


\medskip

\begin{figure}
\begin{center}
\includegraphics[width=\largdeufig]{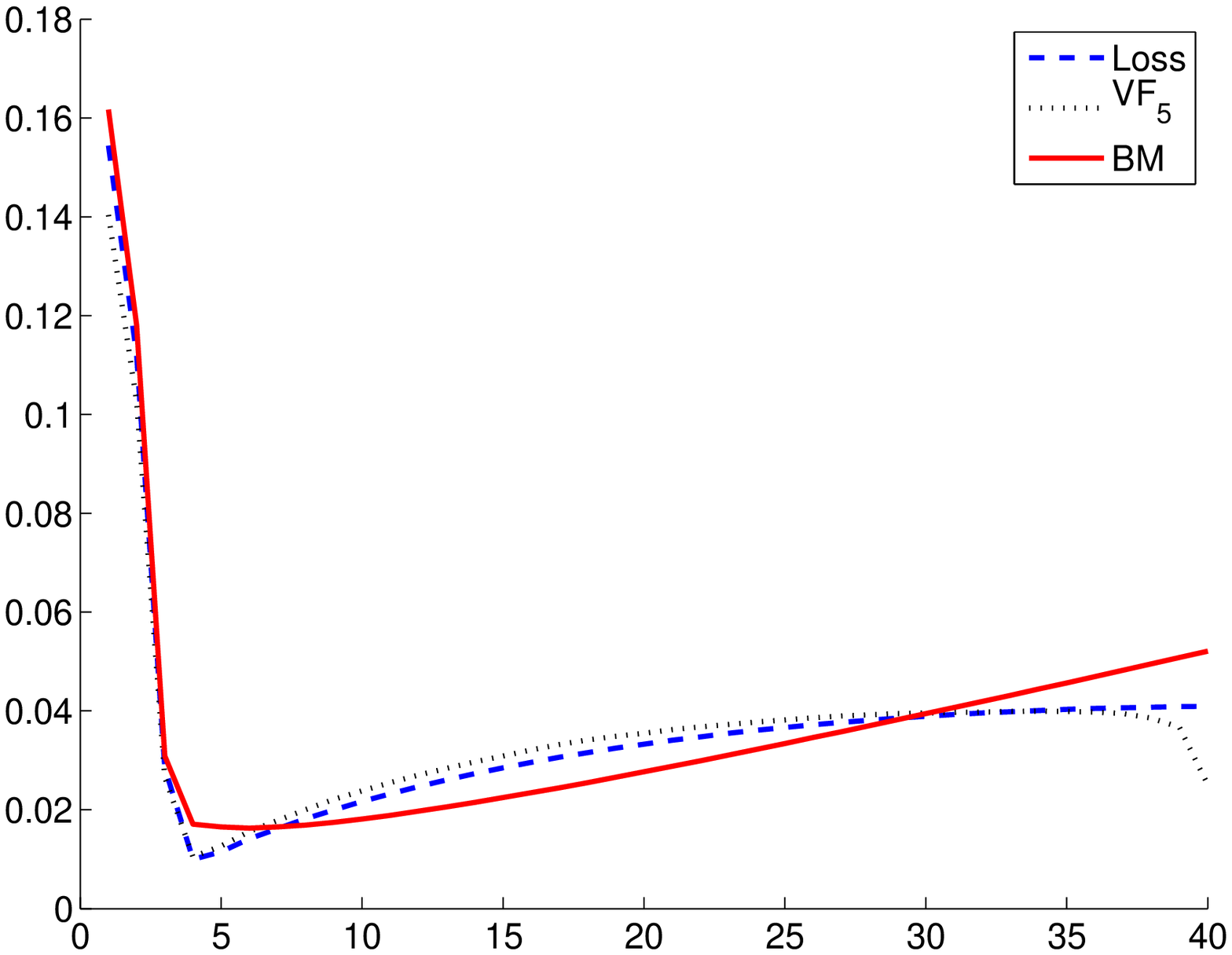}
\hspace{.025\linewidth}
\includegraphics[width=\largdeufig]{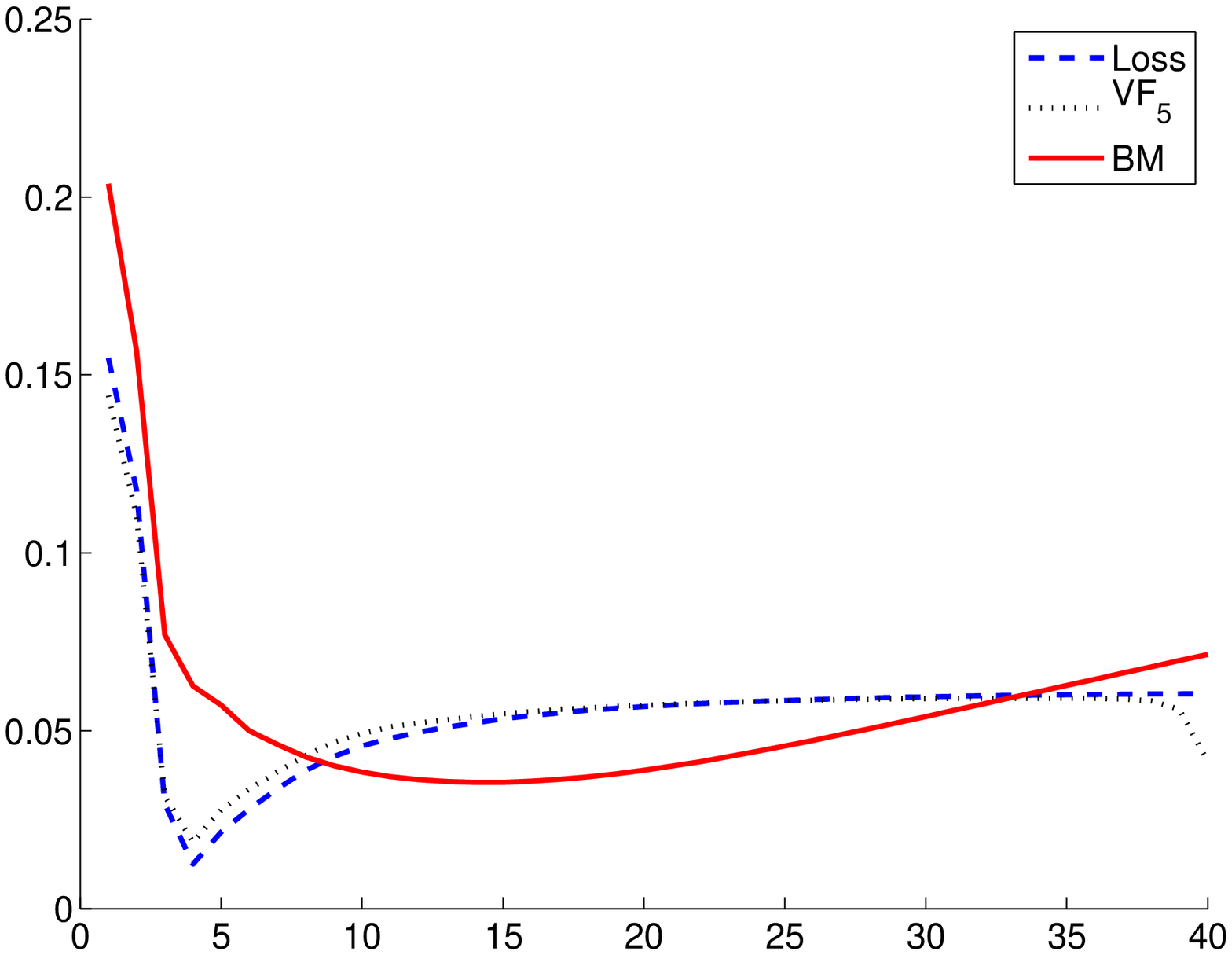}
\vspace{-.8cm}
\end{center}
\caption{
Comparison of the expectations of $\perte{\sh_{\mh(D)}}$ (`Loss'), $\critVF(D)$ (`$\VF_5$') and $\critBM(D)$ (`$\BM$'). Data are generated as explained in Section~\ref{sec.breakpoint.locations.simu.setting}. Left: homoscedastic $(s_2, \sigma_{c})$. Right: heteroscedastic $(s_2, \sigma_{\pcdeuxOLD})$.
{\small Expectations have been estimated from $N =10\,000$ independent samples; error bars are all negligible in front of visible differences (the larger ones are smaller than $5.10^{-4}$ on the left, and smaller than $2.10^{-3}$ on the right).
Similar behaviours are observed for every single sample, with slightly larger fluctuations for $\critVF(D)$ than for $\critBM(D)$.
The curves `$\BM$' and `$\VF_5$' have been shifted in order to make comparison with `Loss' easier, without changing the location of the minimum.}
\label{fig.capacity.BM.VFCV}}
\end{figure}

Simulation experiments support the fact that $\critBM(D)$ is an unbiased estimator of $\perte{\sh_{\mh(D)}}$ for every $D$ (up to an additive constant) when data are homoscedastic (Figure~\ref{fig.capacity.BM.VFCV} left).
However, when data are heteroscedastic, theoretical results proved
by Birg\'e and Massart \cite{BiMa01,BiMa06} no longer apply, and
simulations show that $\critBM(D)$ does not always estimate
$\perte{\sh_{\mhERM(D)}}$ well
(Figure~\ref{fig.capacity.BM.VFCV} right).
This result is consistent with Lemma~\ref{le.exp.ERM}, as well as
the suboptimality of penalties
proportional to $D_m$ for model selection among a
polynomial collection of models when data are heteroscedastic \citep{Arlo08a}.

Therefore, $\penBM(D)$ is not an adequate capacity measure of
$\widetilde{S}_D$ in general when data are heteroscedastic, and
another capacity measure is required.


\subsection{Cross-validation} \label{sec.number.breakpoints.CV}

As shown in Section~\ref{sec.breakpoints.locations.CV.def}, CV can
be used for estimating the quadratic loss $\perte{\A(P_n)}$ for any
algorithm $\A$. In particular, CV was successfully used in
Section~\ref{sec.breakpoint.locations} for estimating the quadratic
risk of $\ERMalg(S_m ; \cdot)$ for all segmentations $\mM_n (D)$
with a given number $(D-1)$ of breakpoints
(Procedure~\ref{algo.mhLoo}), even when data are heteroscedastic.

Therefore, CV methods are natural candidates for fixing BM's failure.
The proposed procedure---with VFCV---is the following.
\begin{algo} \label{alg.2CV} \hfill \\ \vspace{-.5cm}
\begin{enumerate}
\item $ \forall D\in\D$, $\sh_{\mhERM(D)}\defegal\ERMalg\paren{\widetilde{S}_D ; \Pn} \enspace $,
\item $ \DhVF \defegal \argmin_{D\in\D} \acc{ \critVF(D) }$ \\
where $\critVF(D) \defegal  \RhVF \paren{ \ERMalg\paren{\widetilde{S}_D(\cdot) ; \cdot}, \cdot}$ and $\RhVF$ is defined by \eqref{eq.def.vfcv}.
\end{enumerate}
\end{algo}
\begin{rk}\label{rk.CV.SD.design}
In algorithm $(t_i,Y_i)_{1 \leq i \leq n} \flapp \ERMalg\paren{\widetilde{S}_D ; \Pn}$, the model $\widetilde{S}_D$ depends on the design points.
When the training set is $(t_i,Y_i)_{i \notin B_k}$, the model $\widetilde{S}_D$ is the union of the $S_m$ such that $\forall\lamm$, $\Il$ contains at least two elements of $\set{ t_i \tq i \notin B_k}$.
Such an $m$ exists as soon as $D \leq (n-\max_k \set{\card(B_k)})/2$ and two consecutive design points $t_i, t_{i+1}$ always belong to different blocks $B_k$, which is always assumed in this paper.
Note that the dynamic programming algorithms \citep{BeDr62} quoted in Section~\ref{sec.breakpoints.locations.CV.comput} can straightforwardly take into account such constraints when minimizing the empirical risk over $\widetilde{S}_D$.

The dependence of $\widetilde{S}_D$ on the design explains why $\critVF(D)$ decreases for $D$ close to $n(V-1)/(2V)$, as observed on Figure~\ref{fig.capacity.BM.VFCV}.
Indeed, when $D$ is close to $n_t/2$ (where $n_t$ is the size of the design), only a few $\set{S_m}_{\mM_{n_t}(D)}$ remain in $\widetilde{S}_D$; for instance, when $D=n_t/2$, $\widetilde{S}_D$ is equal to one of the $\set{S_m}_{\mM_{n_t}(D)}$.
Therefore, the ``capacity'' of $\widetilde{S}_D$ decreases in the neighborhood of $D = n_t/2$.
\end{rk}

Similar procedures can be defined with $\Loo$ and $\Lpo_p$ instead of VFCV.
The interest of VFCV is its reasonably small computational
cost---taking $V \leq 10$ for instance---, since no closed-form
formula exists for CV estimators of the risk of
$\ERMalg\paren{\widetilde{S}_D ; \Pn}$.

\subsection{Simulation results} \label{sec.number.breakpoints.res}
%
A simulation experiment was performed in the setting presented in
Section~\ref{sec.breakpoint.locations.simu.setting}, for comparing $\BM$ and $\VF_V$ with $V=5$
blocks.
A representative picture of the results is given by Figure~\ref{fig.capacity.BM.VFCV} and by Table~\ref{table.losses.capacity}
\citep[see][Chapter~7, and Section~\refsuppTab\ of the supplementary material for additional results]{Cel:2008:phd}.

\begin{table}
\begin{center}
\begin{tabular}{cc|r@{ $\pm$ }lr@{ $\pm$ }lr@{ $\pm$ }l}
$s_{\cdot}$ & $\sigma_{\cdot}$ & \multicolumn{2}{c}{Oracle} & \multicolumn{2}{c}{$\VF_5$} & \multicolumn{2}{c}{BM}  \\
\hline
 2& c            & 2.88 & 0.01 & {\bf 4.51}  & 0.03 &  5.27      & 0.03 \\
  & \pctroisOLD\ & 2.88 & 0.02 & {\bf 6.58}  & 0.06 & 19.82      & 0.07 \\
  & s            & 3.01 & 0.01 & {\bf 5.21}  & 0.04 &  9.69      & 0.40 \\
\hline
 3& c            & 3.18 & 0.01 &  4.41       & 0.02 & {\bf 4.39} & 0.01 \\
  & \pctroisOLD\ & 4.06 & 0.02 & {\bf 5.99}  & 0.02 & 7.86       & 0.03 \\
  & s            & 4.02 & 0.01 & {\bf 5.97}  & 0.03 & 7.59       & 0.03 \\
\end{tabular}
\caption{Performance $\Corb{\Proc}$ for $\Proc=\Id$ (that is, choosing the dimension $\Do \defegal \argmin_{D \in \D}\set{
\perte{\ERM_{\mhERM(D)}}}$), $\Proc=\VF_V$ with $V=5$ or $\Proc = \BM$.
Several regression functions $s$ and noise-level functions $\sigma$
have been considered, each time with $N=10\,000$ independent samples.
Next to each value is
indicated the corresponding empirical standard deviation divided by
$\sqrt{N}$, measuring the uncertainty of the estimated performance.
\label{table.losses.capacity}}
\end{center}
\end{table}

As illustrated by Figure~\ref{fig.capacity.BM.VFCV}, $\critVF(D)$ can be used for measuring the capacity of $\widetilde{S}_D$.
Indeed, VFCV correctly estimates the risk of empirical risk minimizers over
$\widetilde{S}_D$ for every $D$ and for both homoscedastic and
heteroscedastic data; $\critVF(D)$ only underestimates $\perte{\sh_{\mh(D)}}$ for dimensions $D$ close to $n(V-1)/(2V)$, for reasons explained at the end of Remark~\ref{rk.CV.SD.design}.
On the contrary, $\critBM(D)$ is a poor estimate of $\perte{\sh_{\mh(D)}}$ when data are heteroscedastic.

Subsequently, VFCV yields a much smaller performance index
\[ \Corb{\Proc} \defegal \frac{ \E\croch{ \perte{\ERM_{\mhERM(\Dh_{\Proc})}}}} {\E\croch{ \inf_{\mM_n} \set{ \perte{\ERM_m(P_n)}}} } \]
than BM when data are heteroscedastic (Table~\ref{table.losses.capacity}); see also the supplementary material (Section~\refsuppBM) for details about the performances of BM and possible ways to improve them.
When data are homoscedastic, VFCV and BM have similar performances (maybe with a slight advantage for BM), which is not surprising since BM uses the knowledge that data are homoscedastic. Moreover, BM has been proved to be optimal in the homoscedastic setting \citep{BiMa01,BiMa06}.

Overall, VFCV appears to be a reliable alternative to BM when no prior knowledge guarantees that data are homoscedastic.

\section{New change-point detection procedures via cross-validation} \label{sec.2steps}

Sections~\ref{sec.breakpoint.locations} and~\ref{sec.number.breakpoints} showed that when data are heteroscedastic, CV can be used successfully instead of penalized criteria for detecting breakpoints given their number, as well as for estimating the number of breakpoints.
Nevertheless, in Section~\ref{sec.number.breakpoints}, the
segmentations compared by CV were obtained by empirical risk
minimization, so that they can be suboptimal according to the results
of Section~\ref{sec.breakpoint.locations}.

The next step for obtaining reliable change-point detection procedures for heteroscedastic data is to combine the two ideas, that is, to use CV twice.
The goal of the present section is to properly define such procedures (with various kinds of CV) and to assess their performances.

\subsection{Definition of a family of change-point detection procedures} \label{sec.2steps.def}
%
The general strategy used in this article for change-point detection
relies on two steps: First, detect where $(D-1)$ breakpoints should be
located for every $D \in \D$; second, estimate the number $(D-1)$ of
breakpoints.
This strategy can be summarized with the following procedure:
\begin{algo}[General two-step change-point detection procedure]
\label{algo.general} \hfill \\ \vspace{-.5cm}
\begin{enumerate}
\item $ \forall D \in \D$, $\A_D (P_n) \defegal \ERM_{\mh(D)} =
\argmin_{\mM_n(D)} \set{ \crit_1(S_m, P_n)}$ where for every model $S$, $\crit_1(S,P_n) \in \R$ estimates $\perte{\ERMalg(S ; P_n)}$ and $\ERM_m = \ERMalg(S_m ; P_n)$ is defined as in Section~\ref{sec.breakpoints.locations.ERM}.
\item $\Dh = \argmin_{D\in\D} \set{ \crit_2( \A_D,P_n) }$, where for every algorithm $\A_D$, $\crit_2( \A_D,P_n) \in \R$ estimates $\perte{\A_D(P_n)}$.
\item Output: the segmentation $\mh(\Dh)$ and the corresponding estimator $\ERM_{\mh(\Dh)}$ of $s$.
\end{enumerate}
\end{algo}

\medskip

Let us now detail which are the candidate criteria $\crit_1$ and $\crit_2$ for being used in Procedure~\ref{algo.general}.
For the first step:
\begin{itemize}
\item The empirical risk (`$\Emp$') is \[ \crit_{1,\Emp} (S, P_n) \defegal P_n \gamma \paren{ \ERMalg \paren{S ; P_n} } \]
\item The Leave-$p$-out estimator of the risk (`$\Lpo_p$') is, for every $p \in \set{1, \ldots, n-1}$,
\[ \crit_{1,\Lpo} (S, P_n,p) \defegal \RhLpo (\ERMalg(S;\cdot),P_n)\]
\item For comparison, the ideal criterion (`Id') is defined by $\crit_{1,Id} (S,P_n) \defegal \perte{\ERMalg(S ; P_n)}$.
\end{itemize}
As in Section~\ref{sec.breakpoint.locations}, $\Loo$ denotes $\Lpo_1$.
The VFCV estimator of the risk $\RhVF$ could also be used as $\crit_1$; it will not be considered in the following because it is computationally more expensive and more variable than $\Lpo$ (see Section~\ref{sec.breakpoints.locations.CV}).

For the second step:
\begin{itemize}
\item Birg\'e and Massart's penalization criterion (`BM') is
\[ \crit_{2,\BM} (\A_D , P_n ) \defegal P_n \gamma \paren{ \A_D \paren{P_n} } + \penBM(D) \enspace , \]
where $\penBM(D)$ is defined by \eqref{def.penBM} with $c_1=5$, $c_2=2$ and $\Ch$ is chosen by the slope heuristics (see Section~\refsuppBM\ of the supplementary material).
\item The $V$-fold cross-validation estimator of the risk (`$\VF_V$') is, for every $V \in \set{1, \ldots, n}$,
\[ \crit_{2,\VF_V} (\A_D, P_n) \defegal \RhVF (\A_D, P_n) \enspace , \]
where $\RhVF$ is defined by \eqref{eq.def.vfcv} and the blocks $B_1, \ldots, B_V$ are chosen as in Procedure~\ref{alg.2CV} (see Remark~\ref{rk.CV.SD.design}).
\item For comparison, the ideal criterion (`Id') is defined by $\crit_{2,Id} (\A_D,P_n) \defegal \perte{\A_D(P_n)}$.
\end{itemize}
\begin{rk} For $\crit_2$, definitions using $\Lpo$ could theoretically be
considered. They are not investigated here because they are
computationally intractable.
\end{rk}

\medskip

In the following, the notation $\alg{\alpha}{\beta}$ is used as a
shortcut for ``Procedure~\ref{algo.general} with $\crit_{1,\alpha}$ and
$\crit_{2,\beta}$'', and the outputs of $\alg{\alpha}{\beta}$ are denoted by $\mh_{\alg{\alpha}{\beta}} \in \M_n$ and $\widetilde{s}_{\alg{\alpha}{\beta}} \in \Ss$.
For instance, BM coincides with $\alg{\Emp}{\BM}$; Procedures
$\alg{\alpha}{Id}$ are compared for several $\alpha$ in
Section~\ref{sec.breakpoint.locations};  Procedures $\alg{\Emp}{\beta}$
are compared for $\beta \in \set{ \Id, \BM, \VF_5}$ in
Section~\ref{sec.number.breakpoints}.


\subsection{Simulation study} \label{sec.2steps.simus}
%
A simulation experiment compares procedures $\alg{\alpha}{\VF_5}$
for several $\alpha$ and $\alg{\Emp}{\BM}$,
in the setting described in
Section~\ref{sec.breakpoint.locations.simu.setting}.
A representative picture of the results is given by Table~\ref{table.1*2VF} \cite[see][Chapter~7, for additional results]{Cel:2008:phd}.
The (statistical) performance of each competing procedure $\Proc$ is measured by
\[ \Cor(\Proc) \defegal  \frac{ \E\croch{\perte{\widetilde{s}_{\Proc} (P_n)}}} {\E\croch{\inf_{\mM_n} \set{ \perte{\ERM_m(P_n)}}} } \enspace , \]
both expectations being evaluated by averaging over $N=10\,000$ independent samples.
\begin{table}
\begin{center}
\begin{tabular}{cc|r@{ $\pm$ }lr@{ $\pm$ }lr@{ $\pm$ }lr@{ $\pm$ }l}
$s_{\cdot}$ &   $\sigma_{\cdot}$   & \multicolumn{2}{c}{$\alg{\Emp}{\VF_5}$}  & \multicolumn{2}{c}{$\alg{\Loo}{\VF_5}$} & \multicolumn{2}{c}{$\alg{\Lpo_{20}}{\VF_5}$} & \multicolumn{2}{c}{$\alg{\Emp}{\BM}$} \\
\hline
 1  & c            &  5.40 & 0.05 & 5.03        & 0.05 &  5.10      & 0.05 & {\bf 3.91} & 0.03 \\
    & pc,1         & 11.96 & 0.03 & {\bf 10.25} & 0.03 & 10.28      & 0.03 & 12.85      & 0.04 \\
    & \pcdeuxOLD\  &  4.96 & 0.05 & {\bf 4.82}  & 0.04 & {\bf 4.79} & 0.05 & 13.08      & 0.04 \\
    & s            &  7.33 & 0.06 & {\bf 6.82}  & 0.05 &  6.99      & 0.06 &  9.41      & 0.04 \\
\hline
 2  & c            & {\bf 4.51} & 0.03 & {\bf  4.55} & 0.03 & {\bf  4.50} & 0.03 &  5.27 & 0.03 \\
    & pc,1         & 11.67      & 0.09 & {\bf 10.26} & 0.08 & {\bf 10.29} & 0.08 & 19.36 & 0.07 \\
    & \pcdeuxOLD\  &  6.66      & 0.06 & {\bf  5.81} & 0.06 & {\bf  5.74} & 0.06 & 20.12 & 0.06 \\
    & s    & {\bf 5.21} & 0.04 & {\bf  5.19} & 0.03 & {\bf  5.17} & 0.03 &  9.69 & 0.04 \\
\hline
 3  & c            & 4.41 & 0.02 & 4.54       & 0.02 & 4.62       & 0.02 & {\bf 4.39} & 0.01 \\
    & pc,1         & 4.91 & 0.02 & {\bf 4.40} & 0.02 & {\bf 4.44} & 0.02 & 6.50       & 0.02 \\
    & \pcdeuxOLD\  & 6.32 & 0.02 & {\bf 5.74} & 0.02 & 5.81       & 0.02 & 8.47       & 0.03 \\
    & s            & 5.97 & 0.02 & {\bf 5.72} & 0.02 & 5.86       & 0.02 & 7.59       & 0.03 \\
\end{tabular}
\caption{\label{table.1*2VF} Performance
$\Cor(\Proc)$ for several change-point detection procedures $\Proc$ in several settings $(s,\sigma)$.
Each time, $N=10\,000$ independent samples have been generated.
Next to each value is indicated the
corresponding empirical standard deviation divided by $\sqrt{N}$.}
\end{center}
\end{table}

\begin{rk} \label{rk.BM}
Birg\'e and Massart's penalization procedure is the only classical change-point detection procedure considered in this experiment for two reasons.
First, change-point detection procedure looking for changes in the distribution of $Y_i$ would clearly fail to detect changes in the mean of the signal, as soon as the noise-level $\sigma$ varies inside areas where the mean is constant.
Second, among procedures detecting changes in the mean of a signal in a setting comparable to the setting of the paper (that is, frequentist, parametric, off-line, with no information on the number of change-points), BM appears to be the most reliable procedure according to recent papers \citep{Lavi05,Leba05}.
The question of the calibration of $\Ch$ is addressed in Section~\refsuppBM\ of the supplementary material.
\end{rk}

\medskip

%
First, BM is consistently outperformed by the other procedures, except in the homoscedastic settings in which it confirms its strength.

Second, empirical risk minimization ($\Emp$) slightly outperforms CV ($\Loo$ and $\Lpo_{20}$) when data are homoscedastic.
On the contrary, when data are heteroscedastic, $\Loo$ and $\Lpo_{20}$ clearly outperform $\Emp$,
often by a margin larger than $10\%$ (for instance, when $\sigma = \sigma_{pc,1}$).
Therefore, the results of Section~\ref{sec.breakpoint.locations} are
confirmed when using $\VF_5$ (instead of $\Id$) for choosing the dimension.

Third, the comparison between $\alg{\Lpo_{p}}{\VF_5}$ for several values of $p$ is less clear.
Even though $p=1$ (that is, $\Loo$) mostly outperforms $p=20$ (as well as $p=50$, see the supplementary material), differences are small and often not significant despite the large number of samples generated.
The conclusion of the simulation experiment on this question is that all values of $p$ between $1$ and $n/2$ all perform almost equally well, with a small advantage to $p=1$ which may not be general.
Let us mention here that the choice of $p$ for $\Lpo_p$ is usually related to overpenalization \citep[see for instance][]{Arlo08,Celi08,Cel:2008:phd}, but it seems difficult to characterize the settings for which overpenalization is needed for detecting change-points given their number.

\subsection{Random frameworks} \label{sec.2steps.simusrand}
In order to assess the generality of the results of Table~\ref{table.1*2VF}, the procedures considered in Section~\ref{sec.2steps.simus} have been compared in three random settings.
The following process has been repeated $N=10,\,000$ times.
First, piecewise constant functions $s$ and $\sigma$ are randomly chosen (see Section~\refsuppRand\ of the supplementary material for details).
Then, given $s$ and $\sigma$, a data sample $(t_i,Y_i)_{1 \leq i \leq n}$ is generated as described in Section~\ref{sec.breakpoint.locations.simu.setting}, and the same collection of models is used.
Finally, each procedure $\Proc$ is applied to the sample $(t_i,Y_i)_{1 \leq i \leq n}$, and its loss $\perte{ \widetilde{s}_{\Proc} (P_n)} $ is measured, as well as the loss of the oracle $\inf_{\mM_n} \set{ \perte{\ERM_m} }$.

To summarize the results, the quality of each procedure is measured by the ratio
\[ \CorR(\Proc) =  \frac{ \E_{s,\sigma, \epsilon_1, \ldots, \epsilon_n} \croch{ \perte{ \widetilde{s}_{\Proc} (P_n)} }} {\E_{s,\sigma, \epsilon_1, \ldots, \epsilon_n} \croch{\inf_{\mM_n} \set{ \perte{\ERM_m} } } } \enspace .\]
The notation $\CorR(\Proc)$ differs from $\Cor(\Proc)$ to emphasize that each expectation includes the randomness of $s$ and $\sigma$, in addition to the one of $\paren{\epsilon_i}_{1 \leq i \leq n}$.

\begin{table}
\begin{center}
\begin{tabular}{c|r@{ $\pm$ }lr@{ $\pm$ }lr@{ $\pm$ }l}
Framework       & \multicolumn{2}{c}{A}   & \multicolumn{2}{c}{B}   & \multicolumn{2}{c}{C} \\
\hline
 $\alg{\Emp}{\BM}$                    & 6.82       & 0.03 & 7.21       & 0.04 & 13.49      & 0.07 \\
 $\alg{\Emp}{\VF_5}$                  & 4.78       & 0.03 & 5.09       & 0.03 &  7.17      & 0.05 \\
 $\alg{\Loo}{\VF_5}$                  & {\bf 4.65} & 0.03 & {\bf 4.88} & 0.03 &  6.61      & 0.05 \\
 $\alg{\Lpo_{20}}{\VF_5}$             & 4.78       & 0.03 & {\bf 4.91} & 0.03 & {\bf 6.49} & 0.05 \\
 $\alg{\Lpo_{50}}{\VF_5}$             & 4.97       & 0.03 & 5.18       & 0.04 &  6.69      & 0.05 \\
\end{tabular}
\caption{\label{table.random}
Performance $\CorR(\Proc)$ of several model selection procedures $\Proc$ in  frameworks A, B, C with sample size $n=100$.
In each framework, $N=10,\,000$ independent samples have been considered.
Next to each value is indicated the corresponding empirical standard deviation divided by $\sqrt{N}$.}
\end{center}
\end{table}
The results of this experiment---which are reported in Table~\ref{table.random}---mostly confirm the results of the previous section (except that all the frameworks are heteroscedastic here), that is, whatever $p$, $\alg{\Lpo_p}{\VF_5}$ outperforms $\alg{\Emp}{\VF_5}$, which strongly outperforms $\alg{\Emp}{\BM}$.
Similar results---not reported here---have been obtained with a sample size $n=200$ and $N=1\,000$ samples.

Moreover, the difference between the performances of $\alg{\Lpo_p}{\VF_5}$ and $\alg{\Emp}{\VF_5}$ is the largest in setting C and the smallest in setting~A.
This fact confirms the interpretation given in Section~\ref{sec.breakpoint.locations} for the failure of $\Emp$ for localizing a given number of change-points.
Indeed, the main differences between frameworks A, B and C---which are precisely defined in Section~\refsuppRand\ of the supplementary material--- can be sketched as follows:
\begin{itemize}
\item[A] the partitions on which $s$ is built is often close to regular, and $\sigma$ is chosen independently from $s$.
\item[B] the partitions on which $s$ is built are often irregular, and $\sigma$ is chosen independently from $s$.
\item[C] the partitions on which $s$ is built are often irregular, and $\sigma$ depends on $s$, so that the noise-level is smaller where $s$ jumps more often.
\end{itemize}
In other words, frameworks A, B and C have been built so that for any $D\in\D$, the largest variations over $\M_n(D)$ of $V(m)$ (defined by \eqref{eq.varterm}) occur in framework~C, and the smallest variations occur in framework~A.
As a consequence, variations of the performance of $\alg{\Emp}{\VF_5}$ compared to $\alg{\Lpo_p}{\VF_5}$ according to the framework certainly come from the local overfitting phenomenon presented in Section~\ref{sec.breakpoint.locations}.

\section{Application to CGH microarray data}\label{sec.application.CGH}
%
In this section, the new change-point detection procedures proposed in the paper are applied to CGH microarray data.

\subsection{Biological context} \label{sec.application.CGH.bio}
%
The purpose of Comparative Genomic Hybridization (CGH) microarray experiments is to detect and map chromosomal aberrations.
For instance, a piece of chromosome can be {\em amplified}, that is appear several times more than usual, or {\em deleted}.
Such aberrations are often related to cancer disease.

Roughly, CGH profiles give the log-ratio of the DNA copy number along the chromosomes, compared to a reference DNA sequence
%
\citep[see][for details about the biological context of CGH data]{Pica05,PRLV05,PRLD07}.

The goal of CGH data analysis is to detect abrupt changes in the mean of a signal (the log-ratio of copy numbers), and to estimate the mean in each segment.
Hence, change-point detection procedures are needed.

Moreover, assuming that CGH data are homoscedastic is often
unrealistic. Indeed, changes in the chemical composition of the
sequence are known to induce changes in the variance of the observed
CGH profile, possibly independently from variations of the true copy
number. Therefore, procedures robust to heteroscedasticity, such as
the ones proposed in Section~\ref{sec.2steps}, should yield better
results---in terms of detecting changes of copy number---than
procedures assuming homoscedasticity.

\medskip

The data set considered in this section is based on the Bt474 cell
lines, which denote epithelial cells obtained from
human breast cancer tumors of a sixty-year-old woman \citep{PRLV05}. A test genome of
Bt474 cell lines is compared to a normal reference male genome.
Even though several chromosomes are studied in these cell lines,
this section focuses on chromosomes~1 and~9. Chromosome~1
exhibits a putative heterogenous variance along the CGH profile,
and chromosome~9 is likely to meet the homoscedasticity assumption.
Log-ratios of copy numbers have been measured
at 119 locations for chromosome~1 and at 93 locations for chromosome~9.

\subsection{Procedures used in the CGH literature} \label{sec.application.CGH.litt}
%
Before applying Procedure~\ref{algo.general} to the analysis of Bt474 CGH data, let us recall the definition of two change-point detection procedures, which were the most successful for analyzing the same data according to the literature \citep{PRLV05}.

The first procedure is a simplified version of BM proposed by Lavielle \cite[Section~2]{Lavi05} and first used on CGH data in \citep{PRLV05}.
Note that BM would give similar results on the data of Figure~\ref{fig.change.points.Chr9.rupt}.

%
%

\medskip

The second procedure---denoted by `PML' for penalized maximum likelihood---aims at detecting changes
in either the mean or the variance, that is breakpoints for
$(s,\sigma)$. The selected model is defined as the
minimizer over $\mM_n$ of
\begin{equation*} 
\critPML (m) \defegal \sum_{\lamm} \nl
\log\paren{\frac{1}{\nl}\sum_{t_i \in \Il} (Y_i-\ERM_m(t_i;P_n))^2} + \Ch^{\prime \prime} D_m \enspace ,
\end{equation*}
where $\nl = \card\set{t_i \in \Il}$ and $\Ch^{\prime \prime}$ is estimated from data by the
slope heuristics algorithm \citep{Lavi05,Leba05}.

\subsection{Results} \label{sec.application.CGH.res}
%
Results obtained with BMsimple, PML, $\alg{\Emp}{\VF_5}$ and $\alg{\Lpo_{20}}{\VF_5}$ on the Bt474 data set are reported on Figure~\ref{fig.change.points.Chr9.rupt}.

\medskip

\newcommand{\verticalCGHa}{\vspace{-.52cm}}
\newcommand{\verticalCGHb}{\vspace{-.38cm}}

\begin{figure}
\begin{center}
\begin{minipage}[b]{\largeurfigchr}
\includegraphics[width=\linewidth]{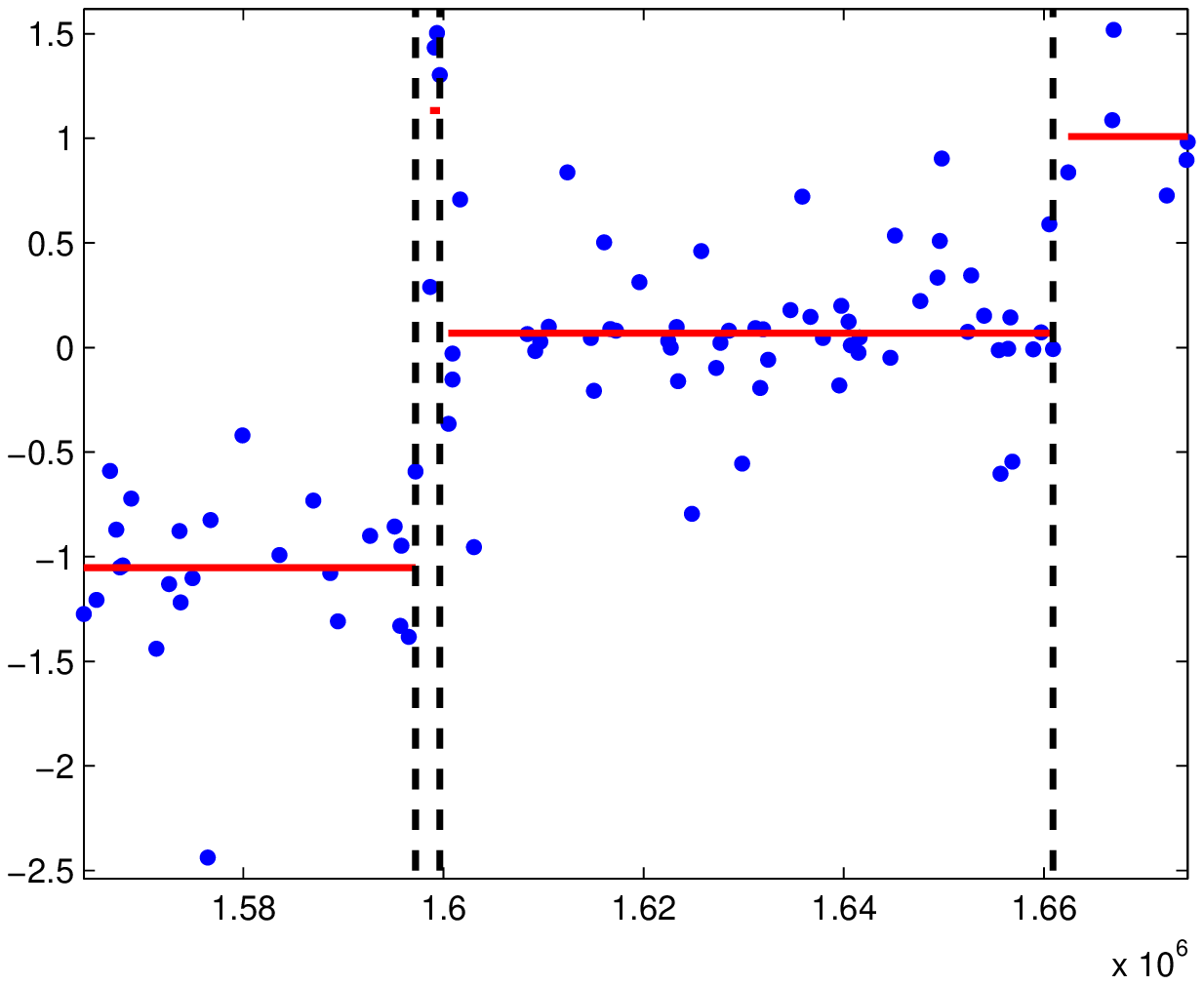}
\begin{center} \verticalCGHa
(a) BMsimple
\end{center} \verticalCGHb
\includegraphics[width=\linewidth]{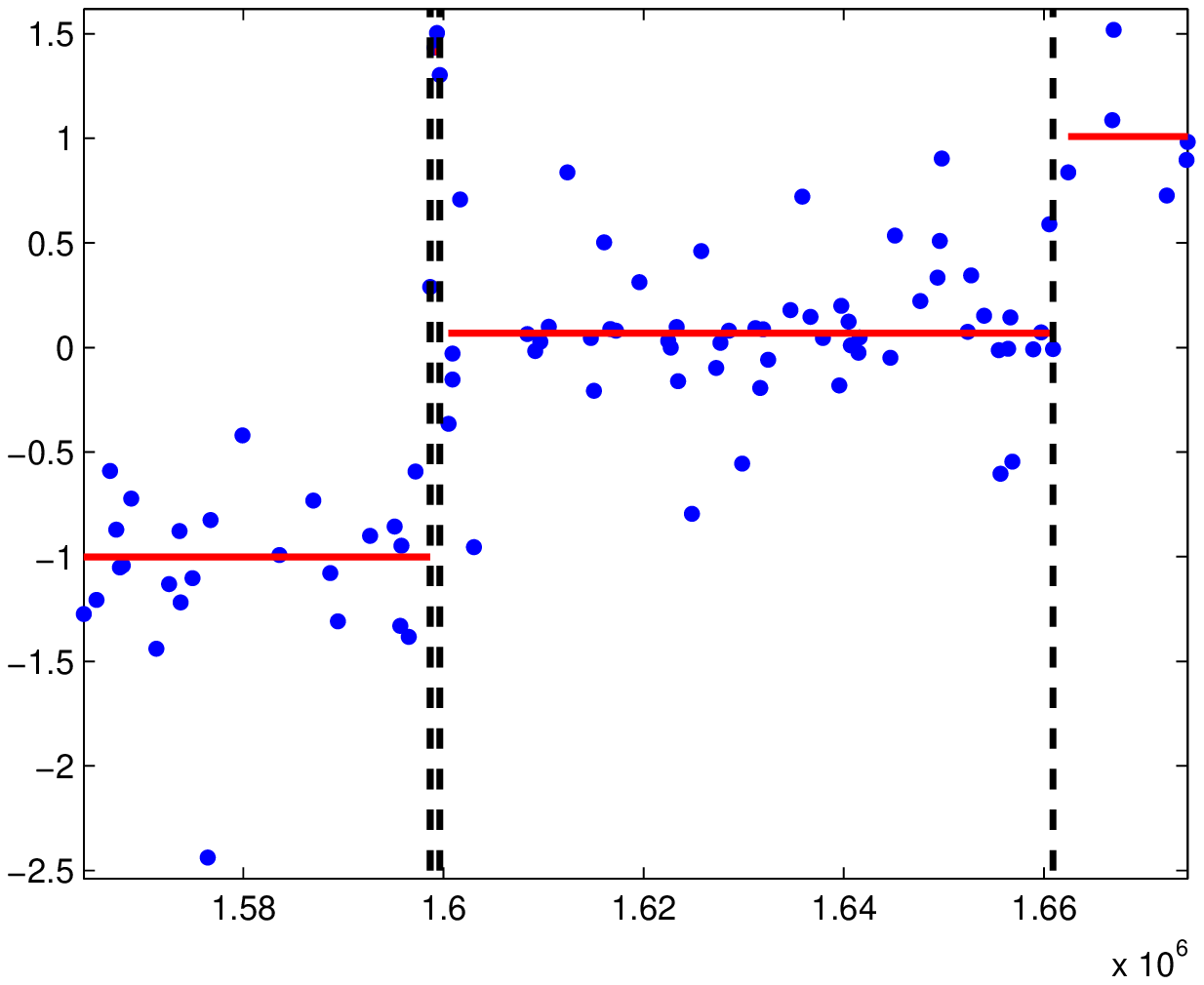}
\begin{center} \verticalCGHa
(b) PML
\end{center} \verticalCGHb
\includegraphics[width=\linewidth]{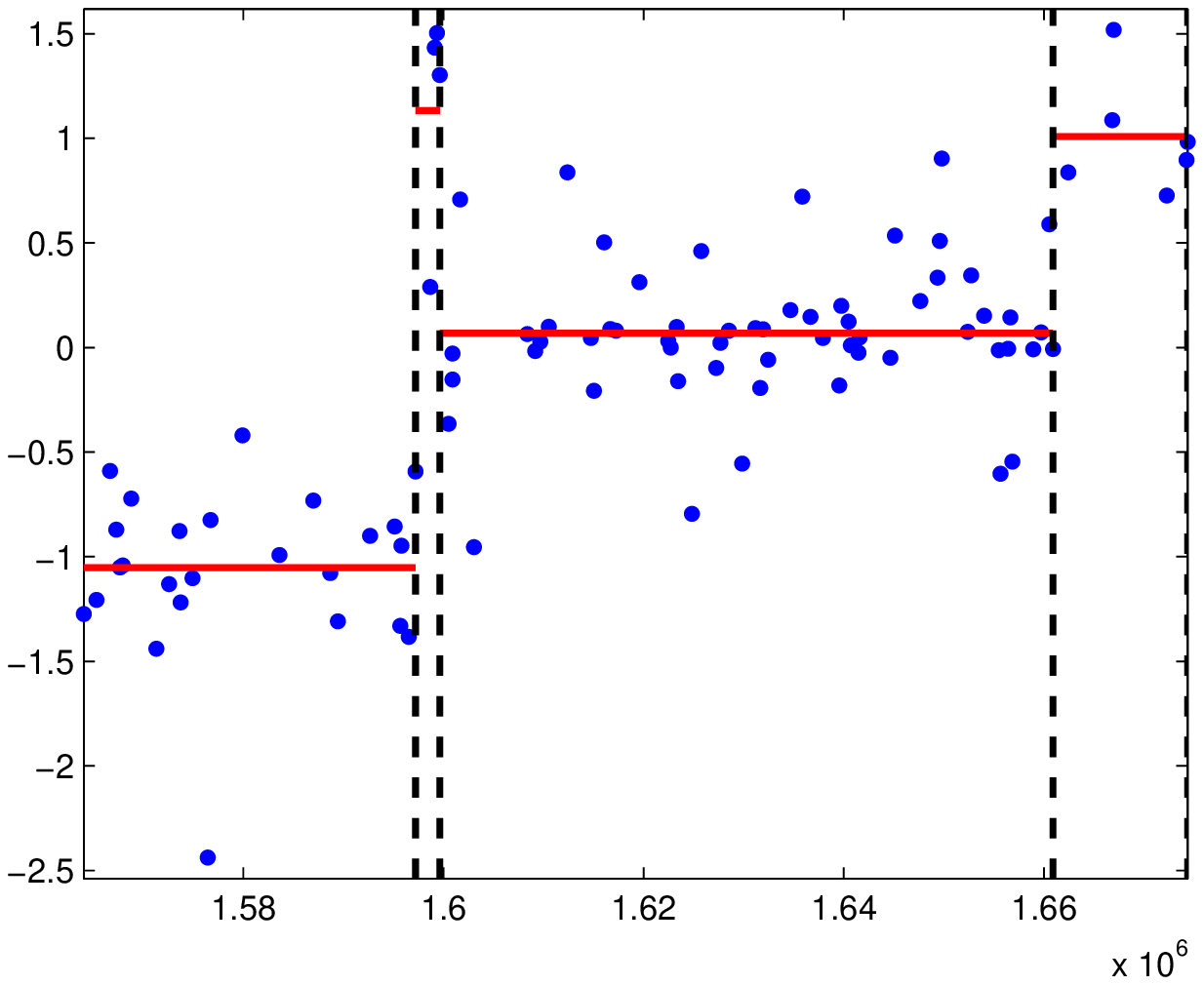}
\begin{center} \verticalCGHa
 (c) $\alg{\Emp}{\VF_5}$
\end{center} \verticalCGHb
\includegraphics[width=\linewidth]{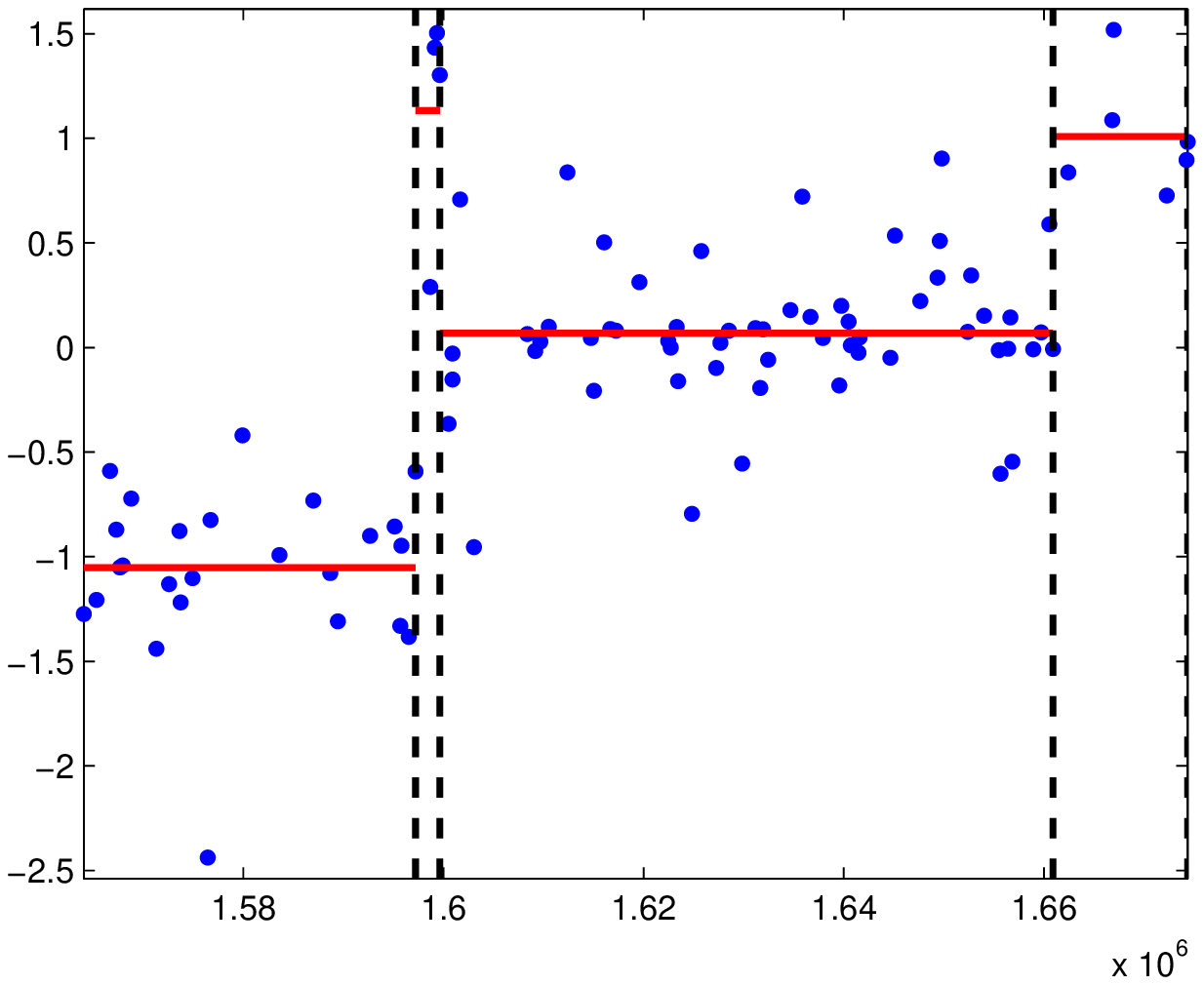}
\begin{center} \verticalCGHa
(d) $\alg{\Lpo_{20}}{\VF_5}$
\end{center} \verticalCGHb
\end{minipage}
\begin{minipage}[b]{\largeurfigchr}
\includegraphics[width=\linewidth]{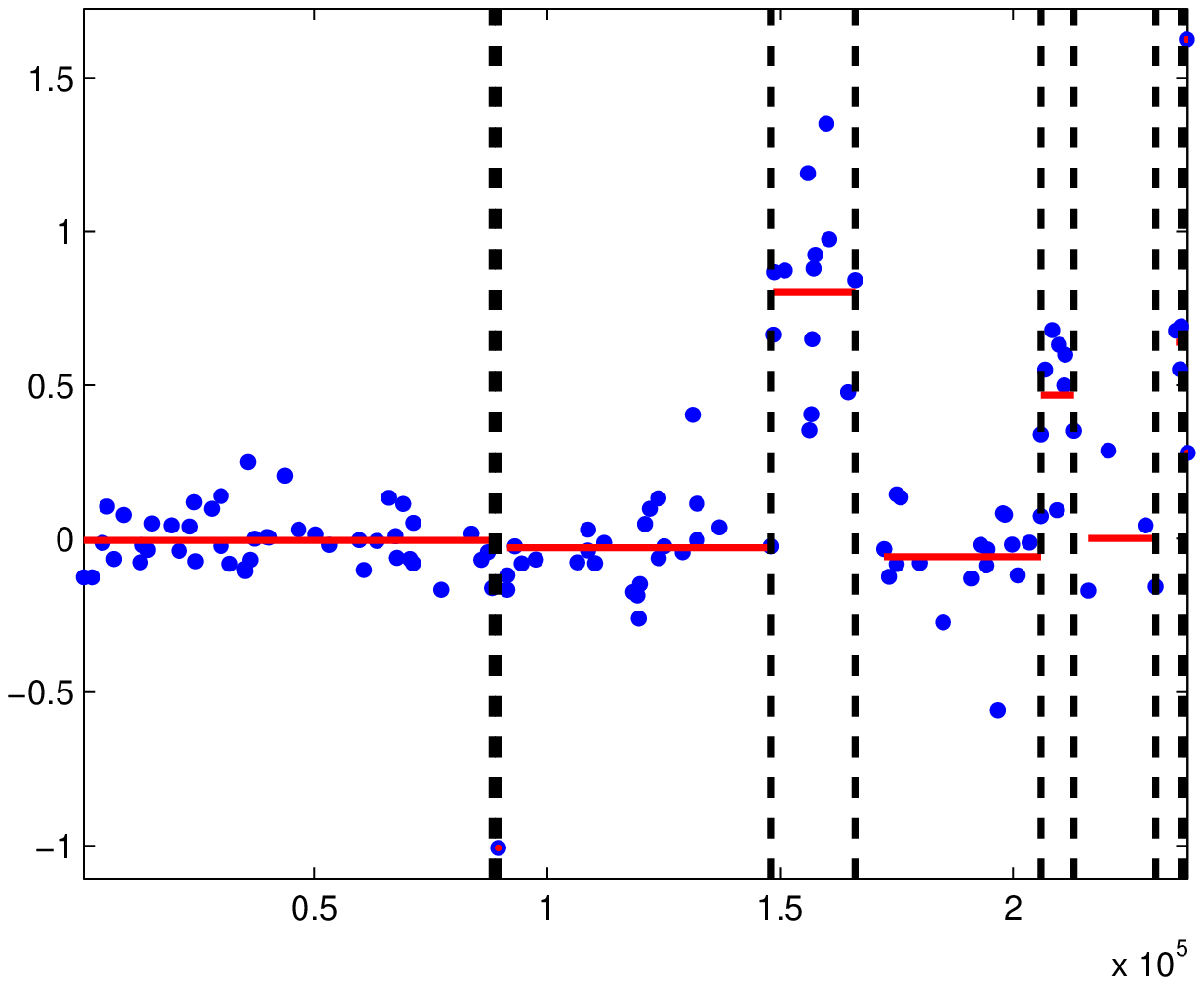}
\begin{center} \verticalCGHa
(e) BMsimple
\end{center} \verticalCGHb
\includegraphics[width=\linewidth]{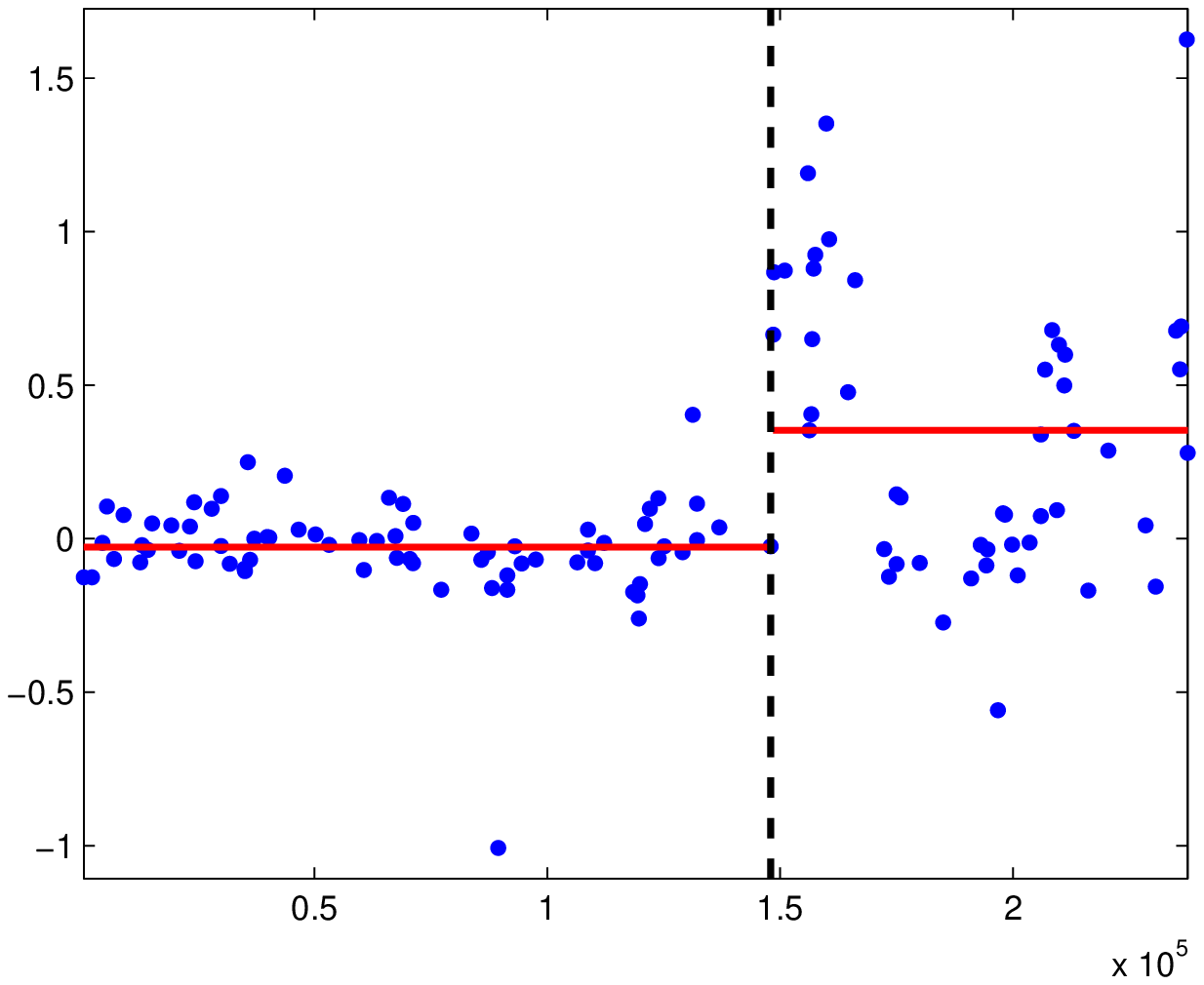}
\begin{center} \verticalCGHa
(f) PML
\end{center} \verticalCGHb
\includegraphics[width=\linewidth]{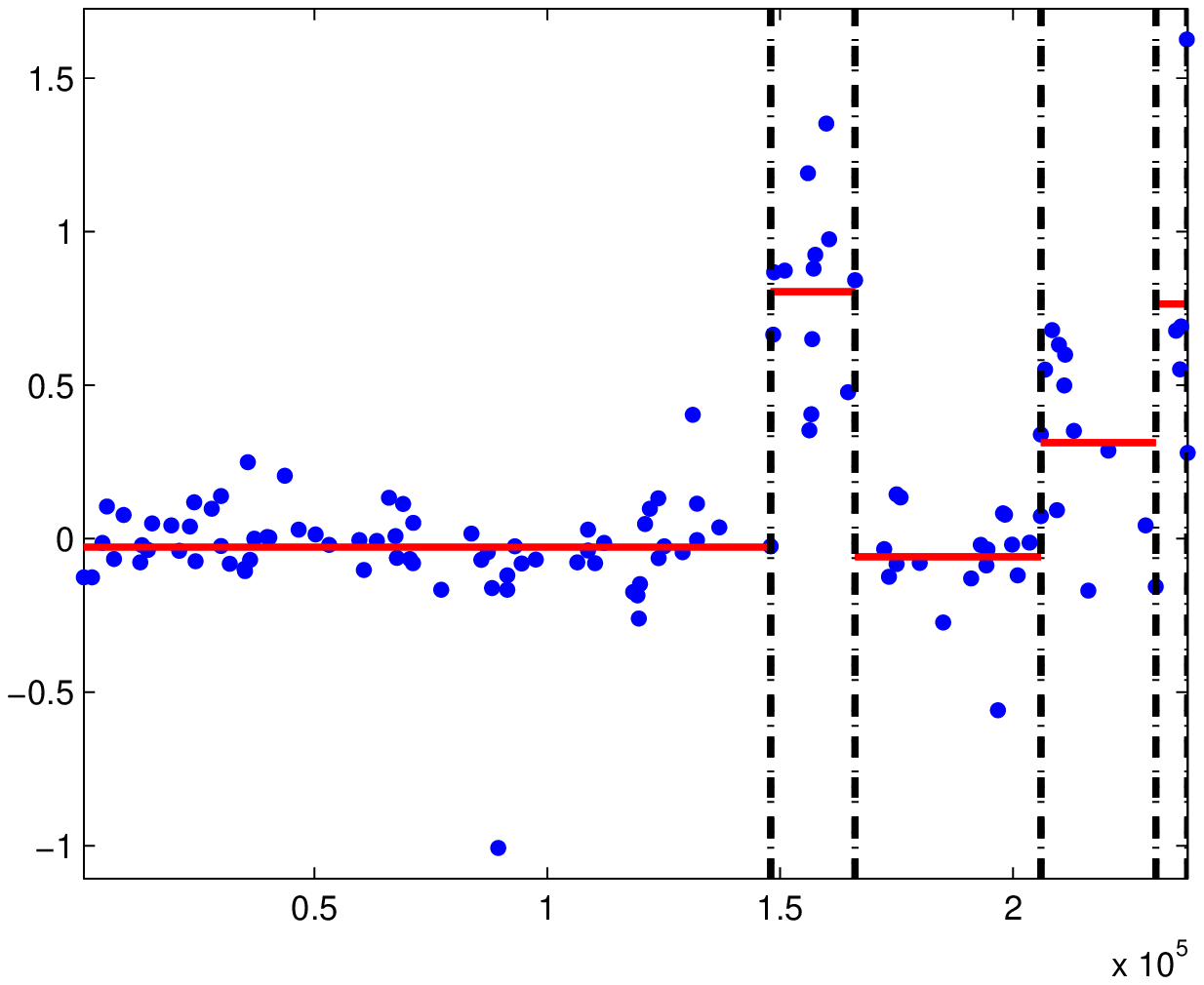}
\begin{center} \verticalCGHa
(g) $\alg{\Emp}{\VF_5}$
\end{center} \verticalCGHb
\includegraphics[width=\linewidth]{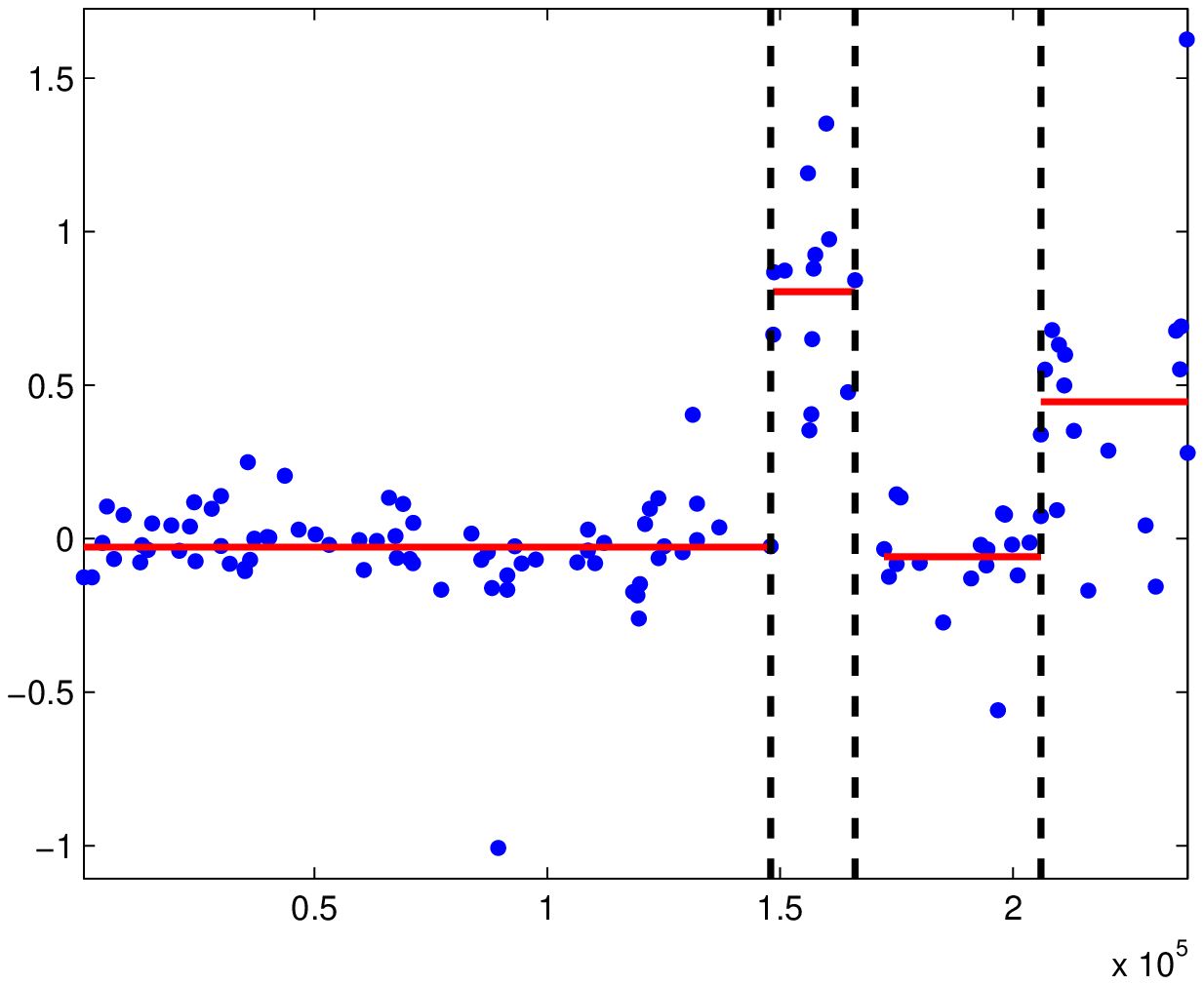}
\begin{center} \verticalCGHa
(h) $\alg{\Lpo_{20}}{\VF_5}$
\end{center} \verticalCGHb
\end{minipage}
\end{center}
\caption{\label{fig.change.points.Chr9.rupt} Change-points locations
along Chromosome~9 (Left) and Chromosome~1 (Right). The mean on each homogeneous region is indicated by plain horizontal lines.
}
\end{figure}
%

For chromosome~9, BMsimple and PML yield (almost) the same segmentation, so that the homoscedasticity assumption is certainly not much violated.
As expected, $\alg{\Emp}{\VF_5}$ and $\alg{\Lpo_{20}}{\VF_5}$ also yield very similar segmentations, which confirms the reliability of these procedures for homoscedastic signal \citep[see][Section~7.6 for details]{Cel:2008:phd}.
\medskip

%

The picture is quite different for chromosome~1.
Indeed, as shown by Figure~\ref{fig.change.points.Chr9.rupt} (right),
BMsimple selects a segmentation with 7 breakpoints, whereas PML
selects a segmentation with only one breakpoint.
The major difference between BMsimple and PML supports at least the
idea that these data must be heteroscedastic.

Nevertheless, none of the segmentations chosen by BMsimple and PML are entirely satisfactory:
BMsimple relies on an assumption which is certainly violated;
PML may use a change in the estimated variance for explaining several changes in the mean.

CV-based procedures $\alg{\Emp}{\VF_5}$ and $\alg{\Lpo_{20}}{\VF_5}$ yield two other segmentations, with a medium number of breakpoints, respectively 4 and 3.
In view of the simulation experiments of the previous sections, the segmentation obtained via $\alg{\Lpo_{20}}{\VF_5}$ should be the most reliable one since data are heteroscedastic.
Therefore, the right of Figure~\ref{fig.change.points.Chr9.rupt} can be
interpretated as follows: The noise-level is small in the first part
of chromosome~1, then higher, but not as high as estimated by PML. In
particular, the copy number changes twice inside the second part of
chromosome~1 (as defined by the segmentation obtained with PML),
indicating that two putative amplified regions of chromosome~1 have been detected.

Note however that choosing among the segmentations obtained with
$\alg{\Emp}{\VF_5}$ and $\alg{\Lpo_{20}}{\VF_5}$ is not an easy
task without additional data. A definitive answer would need further
biological experiments.



\section{Conclusion}\label{sec.conclusion}

\subsection{Results summary}
%
Cross-validation (CV) methods have been used to build reliable procedures (Procedure~\ref{algo.general}) for detecting changes in the mean of a signal whose variance may not be constant.

First, when the number of breakpoints is given, empirical risk minimization has been proved to fail for some heteroscedastic problems, from both theoretical and experimental points of view.
On the contrary, the Leave-$p$-out ($\Lpo_p$) remains robust to
heteroscedasticity while being computationally efficient thanks to
closed-form formulas given in
Section~\ref{sec.breakpoints.locations.CV.comput}
(Theorem~\ref{th.formule.lpo}).

Second, for choosing the number of breakpoints, the commonly used
penalization procedure proposed by Birg\'e and Massart in the
homoscedastic framework should not be applied to
heteroscedastic data.
$V$-fold cross-validation (VFCV) turns out to
be a reliable alternative---both with homoscedastic and
heteroscedastic data---, leading to much better segmentations in
terms of quadratic risk when data are heteroscedastic.
Furthermore, unlike usual deterministic penalized criteria, VFCV
efficiently chooses among segmentations obtained by either $\Lpo$ or
empirical risk minimization, without any specific change in the
procedure.

To conclude, the combination of $\Lpo$ (for choosing a segmentation for
each possible number of breakpoints) and VFCV yields the most
reliable procedure for detecting changes in the mean of a signal which
is not {\it a priori} known to be homoscedastic.
The resulting procedure is computationally tractable for small values of $V$,
since its computational
complexity is of order $\grandO(V n^2)$, which is similar to many comparable change-point detection procedures.
The influence of $V$ on the statistical performance of the procedure is not studied specifically in this paper; nevertheless, considering $V=5$ only was sufficient to obtain a better statistical performance than Birg\'e and Massart's penalization procedure when data are heteroscedastic.
When applied to real data (CGH profiles in
Section~\ref{sec.application.CGH}), the proposed procedure turns out
to be quite useful and effective, for a data set on which existing
procedures highly disagree because of heteroscedasticity.

\subsection{Prospects}
%
The general form of Procedure~\ref{algo.general} could be used with several other criteria, at both steps of the change-point detection procedure.
For instance, resampling penalties \citep{Arlo08} could be used at the first step, for localizing the change-points given their number.
At the second step, $V$-fold penalization \citep{Arlo08a} could also be used instead of VFCV, with the same computational cost and possibly an improved statistical performance.

Comparing precisely these resampling-based criteria for optimizing the performance of Procedure~\ref{algo.general} would be of great interest and deserves further works.
Simultaneously, several values of $V$ should be compared for the second step of Procedure~\ref{algo.general}, and the precise influence of $p$ when $\Lpo_p$ is used at the first step should be further investigated.
Preliminary results in this direction can already be found in \cite[Chapter~7]{Cel:2008:phd}.

\bibliographystyle{plain}
\bibliography{biblio}

%

\end{document}



\title{Supplementary material for Segmentation of the mean of heteroscedastic data via cross-validation}

\author{Sylvain Arlot and Alain Celisse}

\date{\today}

%

\maketitle

\section{Calibration of Birg\'e and Massart's penalization} \label{app.calibration.BM}
Birg\'e and Massart's penalization makes use of the penalty
\begin{equation*} 
\penBM (D) \defegal \frac{\Ch D}{n} \paren{5 + 2 \log\paren{\frac{n}{D}}} \enspace .
\end{equation*}
In a previous version of this work \citep[Chapter~7]{Cel:2008:phd}, $\Ch$ was defined as suggested in \cite{Lavi05,Leba05}, that is, $\Ch = 2 \widehat{K}_{\mathrm{max. jump}}$ with the notation below.
This yielded poor performances, which seemed related to the definition of $\Ch$.
Therefore, alternative definitions for $\Ch$ have been investigated, leading to the choice $\Ch = 2 \widehat{K}_{\mathrm{thresh.}}$ throughout the paper, where $\widehat{K}_{\mathrm{thresh.}}$ is defined by \eqref{eq.Kthresh} below.
The present appendix intends to motivate this choice.

\medskip

Two main approaches have been considered in the literature for defining $\Ch$ in the penalty $\penBM$:
\begin{itemize}
\item Use $\Ch = \sighsq$ any estimate of the noise-level, for instance,
\begin{equation} \label{eq.sighat}
\sighsq \defegal \frac{1}{n} \sum_{i=1}^{n/2} \paren{ Y_{2i} - Y_{2i-1} }^2 \enspace ,
\end{equation}
assuming $n$ is even and $t_1 < \cdots < t_n$.
\item Use Birg\'e and Massart's {\em slope heuristics}, that is, compute the sequence \[ \Dh(K) \defegal \argmin_{D \in \D} \acc{\Png(\sh_{\mhERM(D)}) + \frac{K D}{n} \paren{5 + 2 \log\paren{\frac{n}{D}}} } \enspace , \]
find the (unique) $K = \widehat{K}_{\mathrm{jump}}$ at which $\Dh(K)$ jumps from large to small values, and define $\Ch= 2 \widehat{K}_{\mathrm{jump}}$.
\end{itemize}

The first approach follows from theoretical and experimental results \citep{BiMa01,Leba05} which show that $\Ch$ should be close to $\sigma^2$ when the noise-level is constant; \eqref{eq.sighat} is a classical estimator of the variance used for instance by Baraud \cite{Bara02} for model selection in a different setting.

The optimality (in terms of oracle inequalities) of the second approach has been proved for regression with homoscedastic Gaussian noise and possibly exponential collections of models \citep{BiMa06}, as well as in a heteroscedastic framework with polynomial collections of models \citep{ArMa08}.
In the context of change-point detection with homoscedastic data, Lavielle \cite{Lavi05} and Lebarbier \cite{Leba05} showed that $\Ch= 2 \widehat{K}_{\mathrm{max. jump}}$ can even perform better than $\Ch = \sigma^2$ when $\widehat{K}_{\mathrm{max. jump}}$ corresponds to the highest jump of $\Dh(K)$.

Alternatively, it was proposed in \cite{ArMa08} to define $\Ch= 2 \widehat{K}_{\mathrm{thresh.}}$ where
\begin{equation} \label{eq.Kthresh}
\widehat{K}_{\mathrm{thresh.}} \defegal \min\set{ K \tq \Dh(K) \leq D_{\mathrm{thresh.}} \defegal \left\lfloor \frac{ n } { \ln(n) } \right\rfloor } \enspace . \end{equation}

\begin{table}
\begin{center}
\begin{tabular}{cc|r@{ $\pm$ }lr@{ $\pm$ }lr@{ $\pm$ }lr@{ $\pm$ }l}
$s_{\cdot}$ &   $\sigma_{\cdot}$   & \multicolumn{2}{c}{$2 \widehat{K}_{\mathrm{max. jump}}$}  & \multicolumn{2}{c}{$2 \widehat{K}_{\mathrm{thresh.}}$} & \multicolumn{2}{c}{$\sighsq$} & \multicolumn{2}{c}{$\sigtrue^2$} \\
\hline
 1  & c                 &  6.85 & 0.12 &  3.91 & 0.03 &  1.74 & 0.02 &  2.05 & 0.02 \\
    & \pcdeuxOLD\       & 17.56 & 0.15 & 13.08 & 0.04 &  4.42 & 0.04 & 10.43 & 0.05 \\
    & s                 & 20.07 & 0.31 &  9.41 & 0.04 &  2.18 & 0.03 &  1.66 & 0.02 \\
\hline
2  & c                 &  6.02 & 0.03 &  5.27 & 0.03 &  3.58 & 0.02 &  3.54 & 0.02 \\
   & \pcdeuxOLD\       & 17.76 & 0.10 & 20.12 & 0.07 & 10.58 & 0.07 & 16.64 & 0.08 \\
   & s                 & 10.17 & 0.05 &  9.69 & 0.04 &  5.28 & 0.03 & 10.95 & 0.02 \\
\hline
 3  & c                 &  4.97 & 0.02 &  4.39 & 0.01 &  4.62 & 0.01 &  4.21 & 0.01 \\
    & \pcdeuxOLD\       &  8.66 & 0.03 &  8.47 & 0.03 &  6.64 & 0.02 &  8.00 & 0.03 \\
    & s                 &  8.50 & 0.04 &  7.59 & 0.03 &  5.94 & 0.02 & 15.50 & 0.04 \\
\hline
 \multicolumn{2}{c|}{A} &  7.52 & 0.04 &  6.82 & 0.03 &  4.86 & 0.03 &  5.55 & 0.03 \\
 \multicolumn{2}{c|}{B} &  7.89 & 0.04 &  7.21 & 0.04 &  5.18 & 0.03 &  5.77 & 0.03 \\
 \multicolumn{2}{c|}{C} & 12.81 & 0.08 & 13.49 & 0.07 &  8.93 & 0.06 & 12.44 & 0.07 \\
 \end{tabular}
\caption{\label{table.BM} Performance
$\Cor(\BM)$ with four different definitions of $\Ch$ (see text), in some of the simulation settings considered in the paper.
In each setting, $N=10\,000$ independent samples have been generated.
Next to each value is indicated the corresponding empirical standard deviation divided by $\sqrt{N}$.}
\end{center}
\end{table}

\medskip

These three definitions of $\Ch$ have been compared with $\Ch=\sigtrue^2 \defegal n^{-1} \sum_{i=1}^n \sigma(t_i)^2$ in the settings of the paper.
A representative part of the results is reported in Table~\ref{table.BM}.
The main conclusions are the following.
\begin{itemize}
\item $2 \widehat{K}_{\mathrm{thresh.}}$ almost always beats $2 \widehat{K}_{\mathrm{max. jump}}$, even in homoscedastic settings.
This confirms some simulation results reported in \cite{ArMa08}.
\item $\sigtrue^2$ often beats slope heuristics-based definitions of $\Ch$, but not always, as previously noticed by Lebarbier \cite{Leba05}.
Differences of performance can be huge (in particular when $\sigma = \sigma_{s}$), but not always in favour of $\sigtrue^2$ (for instance, when $s=s_3$).
\item $\sighsq$ yields significantly better performance than $\sigtrue^2$ in most settings (but not all), with huge margins in some heteroscedastic settings.
\end{itemize}

The latter result actually comes from an artefact, which can be explained as follows.
First, \[
\E\croch{ \sighsq } = \frac{1}{n} \sum_{i=1}^n \sigma(t_i)^2 + \frac{1}{n} \sum_{i=1}^n \paren{ s(t_{2i}) - s(t_{2i-1}) }^2 \geq \frac{1}{n} \sum_{i=1}^n \sigma(t_i)^2=\sigtrue^2 \enspace .
\]
The difference between these expectations is not negligible in all the settings of the paper. For instance, when $n=100$, $t_i=i/n$ and $s=s_1$, $n^{-1} \sum_i ( s(t_{2i}) - s(t_{2i-1}) )^2 = 0.04$ whereas $\sigtrue^2$ varies between $0.015$ (when $\sigma=\sigma_{pc,1}$) to $0.093$ (when $\sigma=\sigma_{\pcdeuxOLD}$).
%
%
%
Nevertheless, $\sighsq$ would not overestimate $\sigtrue^2$ at all in a very close setting: Shifting the jumps of $s_1$ by $1/100$ is sufficient to make $n^{-1} \sum_i ( s(t_{2i}) - s(t_{2i-1}) )^2$ equal to zero, and the performances of BM with $\Ch=\sighsq$ would then be very close to the performances of BM with $\Ch=\sigtrue$.

Second, overpenalization turns out to improve the results of BM in most of the heteroscedastic settings considered in the paper.
The reason for this phenomenon is illustrated by the right panel of Figure~\reffigcrit.
Indeed, $\penBM$ is a poor penalty when data are heteroscedastic, underpenalizing dimensions close to the oracle but overpenalizing the largest dimensions (remember that $\Ch = 2 \widehat{K}_{\mathrm{thresh.}}$ on Figure~\reffigcrit).
Then, in a setting like $(s_2,\sigma_{\pcdeuxOLD}$) multiplying $\penBM$ by a factor $C_{\mathrm{over}}>1$ helps decreasing the selected dimension; the same cause has different consequences in other settings, such as $(s_1,\sigma_{s}$ or $(s_3,\sigma_c)$.
Nevertheless, even choosing $\Ch$ using both $P_n$ and $s$, $\paren{\critBM(D)}_{D>0}$ remains a poor estimate of $\paren{ \perte{\ERM_{\mhERM(D)}} }_{D>0}$ in most heteroscedastic settings (even up to an additive constant).

\medskip

To conclude, $\penBM$ with $\Ch = \sighsq$ is not a reliable change-point detection procedure, and the apparently good performances observed in Table~\ref{table.BM} could be misleading.
This leads to the remaining choice $\Ch =2 \widehat{K}_{\mathrm{thresh.}}$ which has been used throughout the paper, although this calibration method may certainly be improved.

Results of Table~\ref{table.BM} for $\Ch = \sigtrue^2$ indicate how far the performances of $\penBM$ could be improved without overpenalization.
According to Tables~\ref{table.1*2VF} and~\ref{table.random}, BM with $\Ch=\sigtrue^2$ only has significantly better performances than $\alg{\Emp}{\VF_5}$ or $\alg{\Loo}{\VF_5}$ in the three homoscedastic settings and in setting $(s_1,\sigma_s)$.

Finally, overpenalization could be used to improve BM, but choosing the overpenalization factor from data is a difficult problem, especially without knowing {\it a priori} whether the signal is homoscedastic or heteroscedastic.
This question deserves a specific extensive simulation experiment.
To be completely fair with CV methods, such an experiment should also compare BM with overpenalization to $V$-fold penalization \citep{Arlo08b} with overpenalization, for choosing the number of change-points.
%
%
%
%
%

\section{Random frameworks generation}
\label{app.random}

The purpose of this appendix is to detail how piecewise constant functions $s$ and $\sigma$ have been generated in the frameworks A, B and C of Section~\refsecrandom.
In each framework, $s$ and $\sigma$ are of the form
\begin{align*}
s(x) &= \sum_{j=0}^{K_s - 1} \alpha_j \1_{[a_j ; a_{j+1} )} + \alpha_{K_s} \1_{[a_{K_s} ; a_{{K_s}+1} ]} \quad &\mbox{with }  a_0 = 0 < a_1 < \cdots < a_{K_s} = 1 \\
\sigma(x) &= \sum_{j=0}^{K_{\sigma} - 1} \beta_j \1_{[b_j ; b_{j+1} )} + \beta_{K_{\sigma}} \1_{[b_{K_{\sigma}} ; b_{{K_{\sigma}}+1} ]} \quad &\mbox{with } b_0 = 0 < b_1 < \cdots < b_{K_{\sigma}} = 1
\end{align*}
for some positive integers $K_s, K_{\sigma}$ and real numbers $\alpha_0, \ldots, \alpha_{K_s} \in \R$ and $\beta_0, \ldots, \beta_{K_{\sigma}} > 0$.

\begin{rk}
The frameworks A, B and C depend on the sample size $n$, through the distribution of $K_s$, $K_{\sigma}$, and of the size of the intervals $[a_j; a_{j+1})$ and $[b_j; b_{j+1})$.
This ensures that the signal-to-noise ratio remains rather small, so that the quadratic risk remains an adequate performance measure for change-point detection.

When the signal-to-noise ratio is larger (that is, when all jumps of $s$ are much larger than the noise-level, and the number of jumps of $s$ is small compared to the sample size), the change-point detection problem is of different nature.
In particular, the number of change-points would be better estimated with procedures targeting identification (such as BIC, or even larger penalties) than efficiency (such as VFCV).
\end{rk}

\subsection{Framework A}
\begin{figure}
\hspace*{-.5cm}
\begin{center}
\includegraphics[width=.45\linewidth]{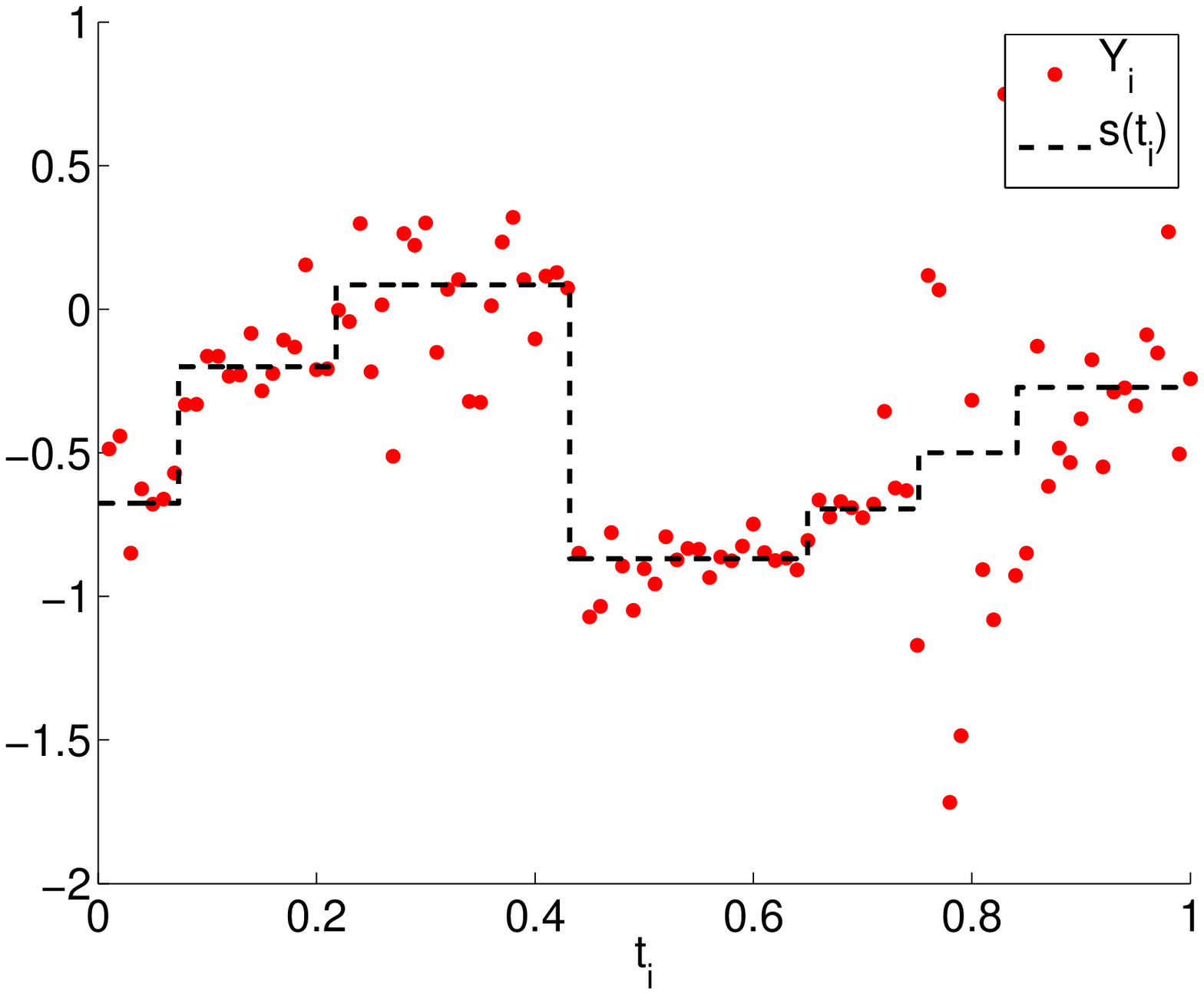} \hspace{.015\linewidth}
\includegraphics[width=.45\linewidth]{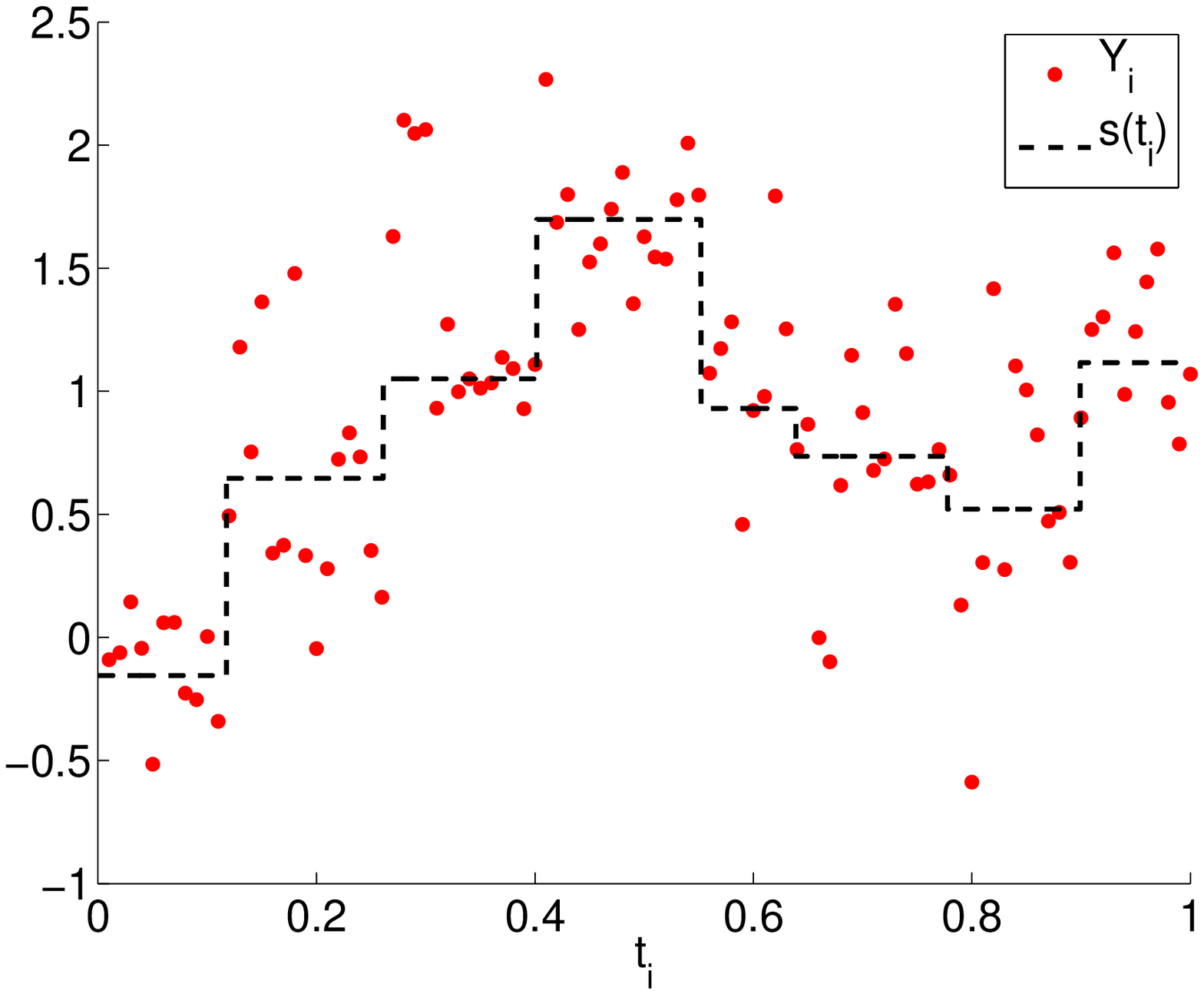}
\end{center}
\caption{Random framework A: two examples of a sample $(t_i,Y_i)_{1 \leq i \leq 100}$ and the corresponding regression function~$s$.\label{fig.ex.randA}}
\end{figure}

In framework A, $s$ and $\sigma$ are generated as follows:
\begin{itemize}
\item $K_s$, the number of jumps of $s$, has uniform distribution over $\set{3, \ldots, \lfloor \sqrt{n} \rfloor}$.
\item For $0 \leq j \leq K_s$, \[ a_{j+1}-a_j = \Delta_{\min}^s + \frac{ (1 - (K_s + 1) \Delta_{\min}^s) U_j} { \sum_{k=0}^{K_s} U_k} \] with $\Delta_{\min}^s = \min\set{ 5/n, 1/(K_s + 1)}$ and $U_0, \ldots, U_{K_s}$ are i.i.d. with uniform distribution over $[0;1]$.
\item $\alpha_0 = V_0$ and for $1 \leq j \leq K_s$, $\alpha_j = \alpha_{j-1} + V_j$ where $V_0, \ldots, V_{K_s}$ are i.i.d. with uniform distribution over $[-1;-0.1] \cup [0.1;1]$.
\item $K_{\sigma}$, the number of jumps of $\sigma$, has uniform distribution in $\set{5, \ldots, \lfloor \sqrt{n} \rfloor}$.
\item For $0 \leq j \leq K_{\sigma}$, \[ b_{j+1}-b_j = \Delta_{\min}^{\sigma} + \frac{ (1 - (K_{\sigma} + 1) \Delta_{\min}^{\sigma}) U_j^{\prime}} { \sum_{k=0}^{K_s} U_k^{\prime}} \] with $\Delta_{\min}^{\sigma} = \min\set{ 5/n, 1/(K_{\sigma} + 1)}$ and $U_0^{\prime}, \ldots, U_{K_{\sigma}}^{\prime}$ are i.i.d. with uniform distribution over $[0;1]$.
\item $\beta_0, \ldots, \beta_{K_{\sigma}}$ are i.i.d. with uniform distribution over $[0.05;0.5]$.
\end{itemize}
Two examples of a function $s$ and a sample $(t_i,Y_i)$ generated in framework~A are plotted on Figure~\ref{fig.ex.randA}.

\subsection{Framework B}
\begin{figure}
\hspace*{-.5cm}
\begin{center}
\includegraphics[width=.45\linewidth]{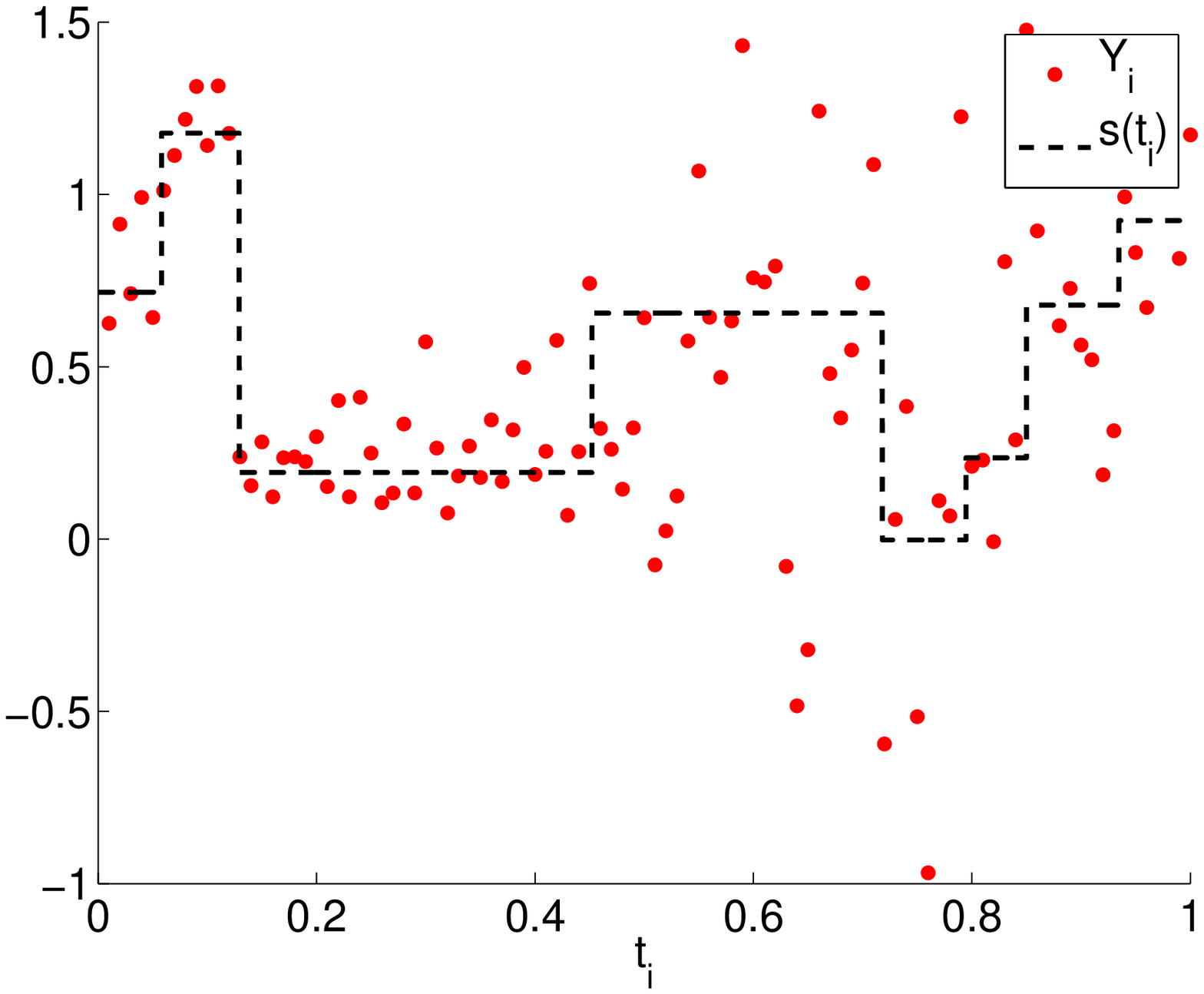} \hspace{.015\linewidth}
\includegraphics[width=.45\linewidth]{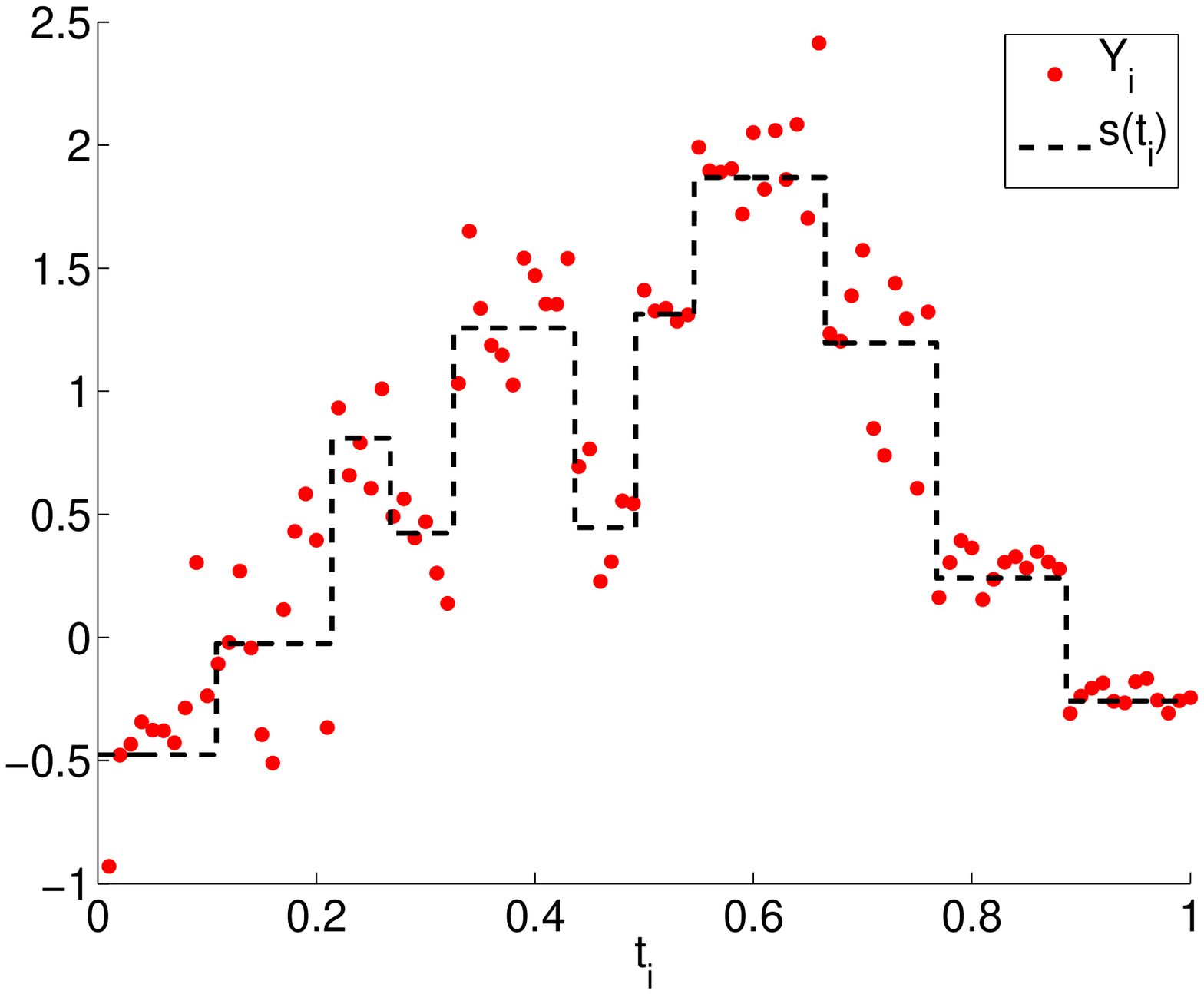}
\end{center}
\caption{Random framework B: two examples of a sample $(t_i,Y_i)_{1 \leq i \leq 100}$ and the corresponding regression function~$s$. \label{fig.ex.randB}}
\end{figure}

The only difference with framework A is that $U_0, \ldots, U_{K_s}$ are i.i.d. with the same distribution as $Z = \abs{10 Z_1 + Z_2 }$ where $Z_1$ has Bernoulli distribution with parameter $1/2$ and $Z_2$ has a standard Gaussian distribution.
%
%
Two examples of a function $s$ and a sample $(t_i,Y_i)$ generated in framework B are plotted on Figure~\ref{fig.ex.randB}.

\subsection{Framework C}
\begin{figure}
\hspace*{-.5cm}
\begin{center}
\includegraphics[width=.45\linewidth]{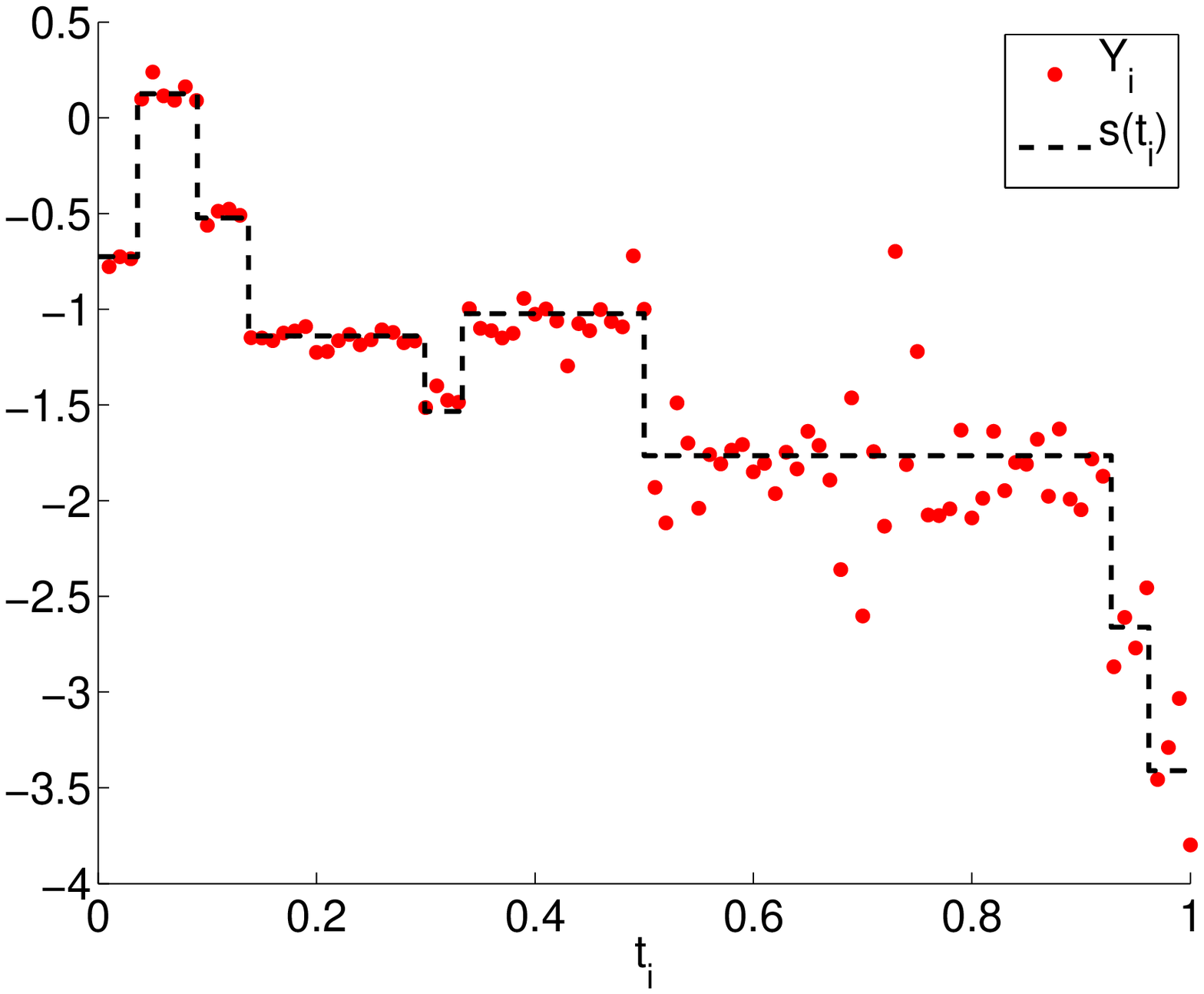} \hspace{.015\linewidth}
\includegraphics[width=.45\linewidth]{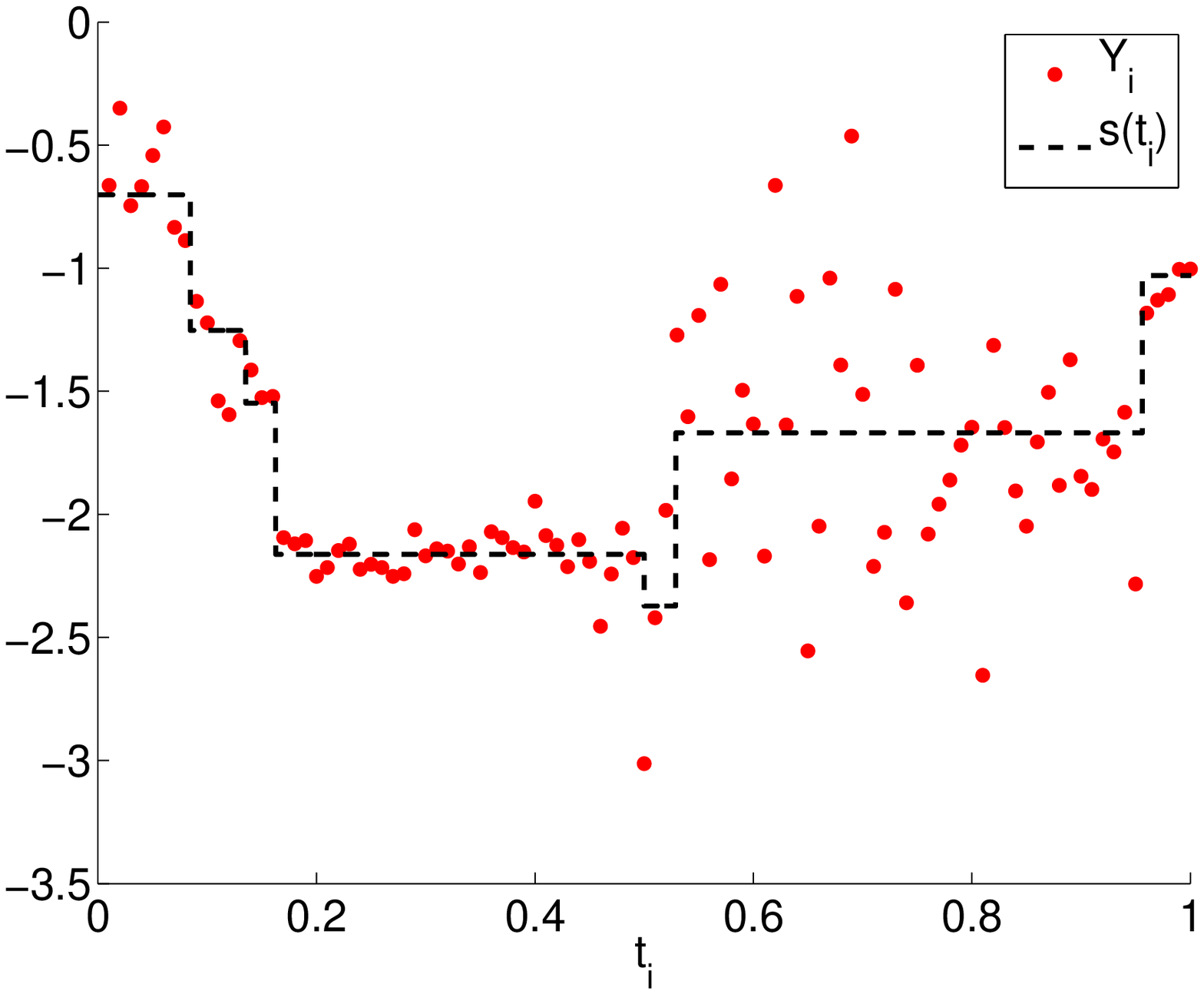}
\end{center}
\caption{Random framework C: two examples of a sample $(t_i,Y_i)_{1 \leq i \leq 100}$ and the corresponding regression function~$s$. \label{fig.ex.randC}}
\end{figure}

The main difference between frameworks~C and B is that $[0;1]$ is split into two regions:
$a_{K_{s,1}+1} = 1/2$ and $K_s = K_{s,1} + K_{s,2} + 1$ for some positive integers $K_{s,1}, K_{s,2}$, and the bounds of the distribution of $\beta_j$ are larger when $b_j \geq 1/2$ and smaller when $b_j<1/2$. 
Two examples of a function $s$ and a sample $(t_i,Y_i)$ generated in framework~C are plotted on Figure~\ref{fig.ex.randC}.
More precisely, $s$ and $\sigma$ are generated as follows:
\begin{itemize}
\item $K_{s,1}$ has uniform distribution over $\set{2, \ldots, K_{\max,1} }$ with $K_{\max,1} = \lfloor \sqrt{n} \rfloor - 1 - \lfloor (\lfloor \sqrt{n} - 1 \rfloor )/3 \rfloor $.
\item $K_{s,2}$ has uniform distribution over $\set{0, \ldots, K_{\max,2} }$ with $K_{\max,2} = \lfloor (\lfloor \sqrt{n} - 1 \rfloor )/3 \rfloor $.
\item Let $U_0, \ldots, U_{K_s}$ be i.i.d. random variables with the same distribution as $Z = \abs{10 Z_1 + Z_2 }$ where $Z_1$ has Bernoulli distribution with parameter $1/2$ and $Z_2$ has a standard Gaussian distribution.
\item For $0 \leq j \leq K_{s,1}$, \[ a_{j+1}-a_j = \Delta_{\min}^{s,1} + \frac{ (1 - (K_{s,1} + 1) \Delta_{\min}^{s,1}) U_j} { \sum_{k=0}^{K_{s,1}} U_k} \] with $\Delta_{\min}^{s,1} = \min\set{ 5/n, 1/(K_{s,1} + 1)}$.
\item For $K_{s,1} +1 \leq j \leq K_{s} $, \[ a_{j+1}-a_j = \Delta_{\min}^{s,2} + \frac{ (1 - (K_{s,2} + 1) \Delta_{\min}^{s,2}) U_j} { \sum_{k=K_{s,1} +1}^{K_{s}} U_k} \] with $\Delta_{\min}^{s,2} = \min\set{ 5/n, 1/(K_{s,2} + 1)}$.
\item $\alpha_0 = V_0$ and for $1 \leq j \leq K_s$, $\alpha_j = \alpha_{j-1} + V_j$ where $V_0, \ldots, V_{K_s}$ are i.i.d. with uniform distribution over $[-1;-0.1] \cup [0.1;1]$.
\item $K_{\sigma}$, $(b_{j+1}-b_j)_{0 \leq j \leq K_{\sigma}}$ are distributed as in frameworks~A and B.
%
\item $\beta_0, \ldots, \beta_{K_{\sigma}}$ are independent. \\
When $b_j < 1/2$, $\beta_j$ has uniform distribution over $[0.025;0.2]$. \\
When $b_j \geq 1/2$, $\beta_j$ has uniform distribution over $[0.1;0.8]$.
\end{itemize}

\section{Additional results from the simulation study}

In the next pages are presented extended versions of the Tables of the main paper, as well as an extended version of Table~\ref{table.BM} (Table~\ref{table.BM.ext}).

\bibliographystyle{plain}
\bibliography{biblio}

\clearpage


%
\begin{table}
\begin{center}
{\small
\begin{tabular}{cc|r@{ $\pm$ }lr@{ $\pm$ }lr@{ $\pm$ }lr@{ $\pm$ }l}
$s_{\cdot}$ &   $\sigma_{\cdot}$   & \multicolumn{2}{c}{$\Emp$} & \multicolumn{2}{c}{$\Loo$} & \multicolumn{2}{c}{$\Lpo_{20}$} & \multicolumn{2}{c}{$\Lpo_{50}$} \\
\hline
 1 & c            & {\bf 1.59} & 0.01 & 1.60       & 0.02 & {\bf 1.58} & 0.01 & {\bf 1.58} & 0.01 \\
   & pc,1         & {\bf 1.04} & 0.01 & 1.06       & 0.01 & 1.06       & 0.01 & 1.06       & 0.01 \\
   & \pctroisOLD\ & {\bf 1.89} & 0.02 & {\bf 1.87} & 0.02 & {\bf 1.87} & 0.02 & {\bf 1.87} & 0.02 \\
   & \pcdeuxOLD\  & {\bf 2.05} & 0.02 & {\bf 2.05} & 0.02 & {\bf 2.05} & 0.02 & {\bf 2.07} & 0.02 \\
   & s            & 1.54       & 0.02 & {\bf 1.52} & 0.02 & {\bf 1.52} & 0.02 & {\bf 1.51} & 0.02 \\
\hline
 2 & c            & {\bf 2.88} & 0.01 & 2.93       & 0.01 & 2.93       & 0.01 & 2.94       & 0.01 \\
   & pc,1         & 1.31       & 0.02 & 1.16       & 0.02 & 1.14       & 0.02 & {\bf 1.11} & 0.01 \\
   & \pctroisOLD\ & 2.88       & 0.02 & 2.24       & 0.02 & 2.19       & 0.02 & {\bf 2.13} & 0.02 \\
   & \pcdeuxOLD\  & 3.09       & 0.03 & 2.52       & 0.03 & 2.48       & 0.03 & {\bf 2.32} & 0.03 \\
   & s            & {\bf 3.01} & 0.01 & {\bf 3.03} & 0.01 & 3.05       & 0.01 & 3.13       & 0.01 \\
\hline
 3 & c            & {\bf 3.18} & 0.01 & 3.25       & 0.01 & 3.29       & 0.01 & 3.44       & 0.01 \\
   & pc,1         & 3.00       & 0.01 & {\bf 2.67} & 0.02 & {\bf 2.68} & 0.02 & 2.77       & 0.02 \\
   & \pctroisOLD\ & 4.06       & 0.02 & {\bf 3.63} & 0.02 & {\bf 3.64} & 0.02 & 3.78       & 0.02 \\
   & \pcdeuxOLD\  & 4.41       & 0.02 & {\bf 3.97} & 0.02 & 4.00       & 0.02 & 4.11       & 0.02 \\
   & s            & 4.02       & 0.01 & {\bf 3.82} & 0.01 & 3.85       & 0.01 & 3.98       & 0.01 \\
\end{tabular}}
\caption{\label{table.losses.1star2id} Average performance
$\Cora{\Proc}$ for change-point detection procedures $\Proc$ among $\Emp$, $\Loo$
and $\Lpo_p$ with $p=20$ and $p=50$. Several regression functions $s$ and
noise-level functions $\sigma$ have been considered, each time with
$N=10\,000$ independent samples. 
Next to each value is indicated the corresponding empirical standard
deviation divided by $\sqrt{N}$, measuring the uncertainty of the
estimated performance.}
\end{center}
\end{table}


%

\begin{table}
\begin{center}
\begin{tabular}{cc|r@{ $\pm$ }lr@{ $\pm$ }lr@{ $\pm$ }l}
$s_{\cdot}$ & $\sigma_{\cdot}$ & \multicolumn{2}{c}{Oracle} & \multicolumn{2}{c}{$\VF_5$} & \multicolumn{2}{c}{BM}  \\
\hline
 1& c            & 1.59 & 0.01 & 5.40        & 0.05 & {\bf 3.91} & 0.03 \\
  & pc,1         & 1.04 & 0.01 & {\bf 11.96} & 0.03 & 12.85      & 0.04 \\
  & \pctroisOLD\ & 1.89 & 0.02 & {\bf 6.43}  & 0.05 & 13.03      & 0.04 \\
  & \pcdeuxOLD\  & 2.05 & 0.02 & {\bf 4.96}  & 0.05 & 13.08      & 0.04 \\
  & s            & 1.54 & 0.02 & {\bf 7.33}  & 0.06 &  9.41      & 0.04 \\
\hline
 2& c            & 2.88 & 0.01 & {\bf 4.51}  & 0.03 &  5.27      & 0.03 \\
  & pc,1         & 1.31 & 0.02 & {\bf 11.67} & 0.09 & 19.36      & 0.07 \\
  & \pctroisOLD\ & 2.88 & 0.02 & {\bf 6.58}  & 0.06 & 19.82      & 0.07 \\
  & \pcdeuxOLD\  & 3.09 & 0.03 & {\bf 6.66}  & 0.06 & 20.12      & 0.07 \\
  & s            & 3.01 & 0.01 & {\bf 5.21}  & 0.04 &  9.69      & 0.40 \\
\hline
 3& c            & 3.18 & 0.01 &  4.41       & 0.02 & {\bf 4.39} & 0.01 \\
  & pc,1         & 3.00 & 0.01 & {\bf 4.91}  & 0.02 & 6.50       & 0.02 \\
  & \pctroisOLD\ & 4.06 & 0.02 & {\bf 5.99}  & 0.02 & 7.86       & 0.03 \\
  & \pcdeuxOLD\  & 4.41 & 0.02 & {\bf 6.32}  & 0.02 & 8.47       & 0.03 \\
  & s            & 4.02 & 0.01 & {\bf 5.97}  & 0.03 & 7.59       & 0.03 \\
\end{tabular}
\caption{Performance $\Corb{\Proc}$ for $\Proc=\Id$ (that is, choosing the dimension $\Do \defegal \argmin_{D \in \D}\set{
\perte{\ERM_{\mhERM(D)}}}$), $\Proc=\VF_V$ with $V=5$ or $\Proc = \BM$.
Several regression functions $s$ and noise-level functions $\sigma$
have been considered, each time with $N=10\,000$ independent samples.
Next to each value is
indicated the corresponding empirical standard deviation divided by
$\sqrt{N}$, measuring the uncertainty of the estimated performance.
\label{table.losses.capacity}}
\end{center}
\end{table}


%

\begin{table}
\begin{center}
\begin{tabular}{cc|r@{ $\pm$ }lr@{ $\pm$ }lr@{ $\pm$ }lr@{ $\pm$ }lr@{ $\pm$ }l}
$s_{\cdot}$ &   $\sigma_{\cdot}$   & \multicolumn{2}{c}{$\alg{\Emp}{\VF_5}$}  & \multicolumn{2}{c}{$\alg{\Loo}{\VF_5}$} & \multicolumn{2}{c}{$\alg{\Lpo_{20}}{\VF_5}$} & \multicolumn{2}{c}{$\alg{\Lpo_{50}}{\VF_5}$} & \multicolumn{2}{c}{$\alg{\Emp}{\BM}$} \\
\hline
 1  & c            &  5.40      & 0.05 & 5.03        & 0.05 &  5.10       & 0.05 &  5.24      & 0.05 & {\bf 3.91} & 0.03 \\
    & pc,1         & 11.96      & 0.03 & {\bf 10.25} & 0.03 & 10.28       & 0.03 & 10.66      & 0.04 & 12.85      & 0.04 \\
    & \pctroisOLD\ &  6.43      & 0.05 & {\bf 5.83}  & 0.05 &  5.99       & 0.05 &  6.20      & 0.05 & 13.03      & 0.04 \\
    & \pcdeuxOLD\  &  4.96      & 0.05 & {\bf 4.82}  & 0.04 & {\bf 4.79}  & 0.05 &  5.02      & 0.05 & 13.08      & 0.04 \\
    & s            &  7.33      & 0.06 & {\bf 6.82}  & 0.05 &  6.99       & 0.06 & {\bf 6.91} & 0.06 &  9.41      & 0.04 \\
\hline
 2  & c            & {\bf 4.51} & 0.03 & {\bf  4.55} & 0.03 & {\bf  4.50} & 0.03 &  4.73      & 0.03 &  5.27      & 0.03 \\
    & pc,1         & 11.67      & 0.09 & {\bf 10.26} & 0.08 & {\bf 10.29} & 0.08 & 10.45      & 0.09 & 19.36      & 0.07 \\
    & \pctroisOLD\ &  6.58      & 0.06 & 5.85        & 0.06 & 5.85        & 0.06 & {\bf 5.49} & 0.06 & 19.82      & 0.07 \\
    & \pcdeuxOLD\  &  6.66      & 0.06 & 5.81        & 0.06 & {\bf  5.74} & 0.06 & {\bf 5.66} & 0.06 & 20.12      & 0.06 \\
    & s            & {\bf 5.21} & 0.04 & {\bf  5.19} & 0.03 & {\bf  5.17} & 0.03 &  5.51      & 0.04 &  9.69      & 0.04 \\
\hline
 3  & c            & 4.41       & 0.02 & 4.54        & 0.02 & 4.62        & 0.02 &  4.94      & 0.02 & {\bf 4.39} & 0.01 \\
    & pc,1         & 4.91       & 0.02 & {\bf 4.40}  & 0.02 & {\bf 4.44}  & 0.02 &  4.69      & 0.02 & 6.50       & 0.02 \\
    & \pctroisOLD\ & 5.99       & 0.02 & {\bf 5.34}  & 0.02 & 5.42        & 0.02 &  5.75      & 0.02 & 7.86       & 0.03 \\
    & \pcdeuxOLD\  & 6.32       & 0.02 & {\bf 5.74}  & 0.02 & 5.81        & 0.02 &  6.24      & 0.02 & 8.47       & 0.03 \\
    & s            & 5.97       & 0.02 & {\bf 5.72}  & 0.02 & 5.86        & 0.02 &  6.07      & 0.02 & 7.59       & 0.03 \\
\end{tabular}
\caption{\label{table.1*2VF} Performance
$\Cor(\Proc)$ for several change-point detection procedures $\Proc$. Several regression functions $s$ and noise-level
functions $\sigma$ have been considered, each time with $N=10\,000$
independent samples. Next to each value is indicated the
corresponding empirical standard deviation.}
\end{center}
\end{table}

%

\begin{table}
\begin{center}
\begin{tabular}{c|r@{ $\pm$ }lr@{ $\pm$ }lr@{ $\pm$ }l}
Framework       & \multicolumn{2}{c}{A}   & \multicolumn{2}{c}{B}   & \multicolumn{2}{c}{C} \\
\hline 
 $\alg{\Emp}{\BM}$                    & 6.82       & 0.03 & 7.21       & 0.04 & 13.49      & 0.07 \\
 $\alg{\Emp}{\VF_5}$                  & 4.78       & 0.03 & 5.09       & 0.03 &  7.17      & 0.05 \\
 $\alg{\Loo}{\VF_5}$                  & {\bf 4.65} & 0.03 & {\bf 4.88} & 0.03 &  6.61      & 0.05 \\
 $\alg{\Lpo_{20}}{\VF_5}$             & 4.78       & 0.03 & {\bf 4.91} & 0.03 & {\bf 6.49} & 0.05 \\
 $\alg{\Lpo_{50}}{\VF_5}$             & 4.97       & 0.03 & 5.18       & 0.04 &  6.69      & 0.05 \\
\end{tabular}
\caption{\label{table.random} 
Performance $\CorR(\Proc)$ of several model selection procedures $\Proc$ in  frameworks A, B, C with sample size $n=100$. 
In each framework, $N=10,\,000$ independent samples have been considered. 
Next to each value is indicated the corresponding empirical standard deviation divided by $\sqrt{N}$.}
\end{center}
\end{table}

\begin{table}
\begin{center}
\begin{tabular}{c|r@{ $\pm$ }lr@{ $\pm$ }lr@{ $\pm$ }l}
Framework       & \multicolumn{2}{c}{A}   & \multicolumn{2}{c}{B}   & \multicolumn{2}{c}{C} \\
\hline 
 $\alg{\Emp}{\BM}$                    &  9.04      & 0.12 & 11.62      & 0.14 & 21.21      & 0.31 \\
 $\alg{\Emp}{\BM_{\widehat{\sigma}}}$ &  5.34      & 0.10 &  6.24      & 0.11 & 11.48      & 0.22 \\
 $\alg{\Emp}{\VF_5}$                  &  5.10      & 0.11 &  5.92      & 0.11 &  7.31      & 0.14 \\
 $\alg{\Loo}{\VF_5}$                  & {\bf 4.90} & 0.11 & {\bf 5.63} & 0.11 & {\bf 6.89} & 0.16 \\
 $\alg{\Lpo_{20}}{\VF_5}$             & {\bf 4.88} & 0.10 & {\bf 5.55} & 0.10 & {\bf 6.82} & 0.15 \\
 $\alg{\Lpo_{50}}{\VF_5}$             &  5.11      & 0.11 & {\bf 5.49} & 0.10 &  7.14      & 0.15 \\
\end{tabular}
\caption{\label{table.random.n200} 
Performance $\CorR(\Proc)$ of several model selection procedures $\Proc$ in  frameworks A, B, C with sample size $n=200$. 
In each framework, $N=1,\,000$ independent samples have been considered. 
Next to each value is indicated the corresponding empirical standard deviation divided by $\sqrt{N}$.}
\end{center}
\end{table}



%

\begin{table}
\begin{center}
\begin{tabular}{cc|r@{ $\pm$ }lr@{ $\pm$ }lr@{ $\pm$ }lr@{ $\pm$ }l}
$s_{\cdot}$ &   $\sigma_{\cdot}$   & \multicolumn{2}{c}{$2 \widehat{K}_{\mathrm{max. jump}}$}  & \multicolumn{2}{c}{$2 \widehat{K}_{\mathrm{thresh.}}$} & \multicolumn{2}{c}{$\sighsq$} & \multicolumn{2}{c}{$\sigtrue^2$} \\
\hline
 1  & c                 &  6.85 & 0.12 &  3.91 & 0.03 &  1.74 & 0.02 &  2.05 & 0.02 \\
    & pc,1              & 70.97 & 1.18 & 12.85 & 0.04 &  1.13 & 0.02 & 10.20 & 0.05 \\
   & \pctroisOLD\      & 23.74 & 0.26 & 13.03 & 0.04 &  3.55 & 0.04 & 10.43 & 0.05 \\
    & \pcdeuxOLD\       & 17.56 & 0.15 & 13.08 & 0.04 &  4.42 & 0.04 & 10.43 & 0.05 \\
    & s                 & 20.07 & 0.31 &  9.41 & 0.04 &  2.18 & 0.03 &  1.66 & 0.02 \\
\hline
2  & c                 &  6.02 & 0.03 &  5.27 & 0.03 &  3.58 & 0.02 &  3.54 & 0.02 \\
   & pc,1              & 17.83 & 0.10 & 19.36 & 0.07 &  8.52 & 0.06 & 15.62 & 0.08 \\
   & \pctroisOLD\      & 17.63 & 0.10 & 19.82 & 0.07 & 10.77 & 0.07 & 16.56 & 0.08 \\
   & \pcdeuxOLD\       & 17.76 & 0.10 & 20.12 & 0.07 & 10.58 & 0.07 & 16.64 & 0.08 \\
   & s                 & 10.17 & 0.05 &  9.69 & 0.04 &  5.28 & 0.03 & 10.95 & 0.02 \\
\hline
 3  & c                 &  4.97 & 0.02 &  4.39 & 0.01 &  4.62 & 0.01 &  4.21 & 0.01 \\
    & pc,1              &  7.18 & 0.03 &  6.50 & 0.02 &  4.52 & 0.02 &  6.70 & 0.03 \\
   & \pctroisOLD\      &  8.14 & 0.03 &  7.86 & 0.03 &  6.22 & 0.02 &  7.55 & 0.03 \\
    & \pcdeuxOLD\       &  8.66 & 0.03 &  8.47 & 0.03 &  6.64 & 0.02 &  8.00 & 0.03 \\
    & s                 &  8.50 & 0.04 &  7.59 & 0.03 &  5.94 & 0.02 & 15.50 & 0.04 \\
\hline
 \multicolumn{2}{c|}{A} &  7.52 & 0.04 &  6.82 & 0.03 &  4.86 & 0.03 &  5.55 & 0.03 \\
 \multicolumn{2}{c|}{B} &  7.89 & 0.04 &  7.21 & 0.04 &  5.18 & 0.03 &  5.77 & 0.03 \\
 \multicolumn{2}{c|}{C} & 12.81 & 0.08 & 13.49 & 0.07 &  8.93 & 0.06 & 12.44 & 0.07 \\
 \end{tabular}
\caption{\label{table.BM.ext} Performance
$\Cor(\BM)$ with four different definitions of $\Ch$ (see text), in some of the simulation settings considered in the paper. 
In each setting, $N=10\,000$ independent samples have been generated. 
Next to each value is indicated the corresponding empirical standard deviation divided by $\sqrt{N}$.}
\end{center}
\end{table}